\documentclass[a4paper,twoside]{article}
\usepackage{deluxetable}
\usepackage{amssymb}

\usepackage{times,epsfig}
\usepackage{graphics,graphicx}
\usepackage{amsmath}
\usepackage{amsfonts}
\usepackage{psfig}

\usepackage{mathrsfs}
\usepackage{amssymb}
\usepackage{bm}
\newcommand{\affil}[1]{$^{\rm #1}$}
\usepackage[authoryear]{natbib}
\bibpunct{ (}{)}{;}{a}{}{,}
\usepackage{graphicx}
\date{Accepted by PASA,  22 October 2012} 
\newcommand{\kms}{\mbox{km\,s$^{-1}$}}
\def\kms {\ifmmode{{\rm ~km~s}^{-1}}\else{~km~s$^{-1}$}\fi}
\def\etal {~{et~al.}~}
\def\lsun {\ifmmode{{\rm ~L}_\odot}\else{~L$_\odot$}\fi}
\def\sun {\odot}
\def\deg {^{\circ} }
\def\sqdeg {\,deg$^2$}
\def\arcsec {\,arcsec}
\def\arcmin {\,arcmin}
\def\ujybm {\,$\mu$Jy/beam}
\def\ujy {\,$\mu$Jy}

\newbox\grsign \setbox\grsign=\hbox{$>$} \newdimen\grdimen \grdimen=\ht\grsign
\newbox\simlessbox \newbox\simgreatbox
\setbox\simgreatbox=\hbox{\raise.5ex\hbox{$>$}\llap
 {\lower.5ex\hbox{$\sim$}}}\ht1=\grdimen\dp1=0pt
\setbox\simlessbox=\hbox{\raise.5ex\hbox{$<$}\llap
 {\lower.5ex\hbox{$\sim$}}}\ht2=\grdimen\dp2=0pt

\def\lsim{\mathrel{\rlap{\lower4pt\hbox{\hskip1pt$\sim$}}
    \raise1pt\hbox{$<$}}}                
\def\gsim{\mathrel{\rlap{\lower4pt\hbox{\hskip1pt$\sim$}}
    \raise1pt\hbox{$>$}}}                

\def\apj {{\it Ap.~J.}}
\def\apjl {{\it Ap.~J.\ (Letters)}}
\def\apjs {{\it Ap.~J.\ Suppl.}}
\def\aj {{\it A.~J.}}
\def\aa {{\it Astr.~Ap.}}
\def\aap {{\it Astr.~Ap.}}

\def\aapr {{\it Astr.~Ap.\ Rev.}}

\def\apss {{\it Ap. Sp. Sci.}}
\def\araa {{\it Ann.\ Rev.\ Astr.\ Ap.}}
\def\aspc {{\it Ast.\ Soc.\ Pacific\ Conference Series}}

\def\jcap {{\it J. Cosm. Astroparticle Phys.}}
\def\mnras {{\it MNRAS}}

\def\nat {{\it Nature}}
\def\pasa {{\it PASA}}
\def\pasp {{\it PASP}}
\def\prd {{\it Phys. Rev. D}}
\def\procspie {{\it Proc. SPIE}}

\def\aj{AJ}                   
\def\araa{ARA\&A}             
\def\apj{ApJ}                 
\def\apjl{ApJ}                
\def\apjs{ApJS}               
\def\apss{Ap\&SS}             
\def\aap{A\&A}                
\def\aapr{A\&A~Rev.}          
\def\aaps{A\&AS}              
\def\mnras{MNRAS}             
\def\nar{New Astr. Rev}
\def\prd{Phys.~Rev.~D}        
\def\pasp{PASP}               
\def\nat{Nature}              

\def\procspie{Proc.~SPIE}   

\def\grtsim{\mathrel{\hbox{\rlap{\hbox{\lower2pt\hbox{$\sim$}}}\raise2pt\hbox{$>$}}}}
\def\lesssim{\mathrel{\hbox{\rlap{\hbox{\lower2pt\hbox{$\sim$}}}\raise2pt\hbox{$<$}}}}

\def\lsim{\,\lower2truept\hbox{${<\atop\hbox{\raise4truept\hbox{$\sim$}}}$}\,}
\def\gsim{\,\lower2truept\hbox{${> \atop\hbox{\raise4truept\hbox{$\sim$}}}$}\,}
\def\simlt{\mathrel{\rlap{\lower 3pt\hbox{$\sim$}}
        \raise 2.0pt\hbox{$<$}}}
\def\simgt{\mathrel{\rlap{\lower 3pt\hbox{$\sim$}}
        \raise 2.0pt\hbox{$>$}}}

\title{\large\bf\flushleft {Radio Continuum Surveys with Square Kilometre Array Pathfinders}}
\author{\parbox{\textwidth}{\flushleft
\vspace{-0.5cm}
{\it
Ray P.\ Norris\affil{1,2},
J.\ Afonso\affil{3},
D.Bacon\affil{4},
Rainer Beck\affil{5},
Martin Bell\affil{2, 6, 15},
R. J. Beswick\affil{7},
Philip Best\affil{8},
Sanjay Bhatnagar\affil{9},
Annalisa Bonafede\affil{10},
Gianfranco Brunetti\affil{11},
Tam\'as Budav\'ari\affil{12},
Rossella Cassano\affil{11},
J.\,J.\ Condon\affil{9},
Catherine Cress\affil{13},
Arwa Dabbech\affil{14},
I.\ Feain\affil{1, 15},
Rob Fender\affil{6},
Chiara Ferrari\affil{14},
B.M. Gaensler\affil{2, 15},
G. Giovannini\affil{11},
Marijke Haverkorn\affil{30,34},
George Heald\affil{16},
Kurt van der Heyden\affil{17},
A.\,M.\ Hopkins\affil{18},
M.\ Jarvis\affil{13,19 ,36},
Melanie Johnston-Hollitt\affil{35},
Roland Kothes\affil{32},
Huib van Langevelde\affil{20, 30},
Joseph Lazio\affil{21},
Minnie Y. Mao\affil{1,18,22,23},
Alejo Mart\'\i nez-Sansigre\affil{4},
David Mary\affil{14},
Kim McAlpine\affil{19,24},
E.\ Middelberg\affil{25},
Eric Murphy\affil{26},
P.\ Padovani\affil{27},
Zsolt Paragi\affil{20},
I.\ Prandoni\affil{11},
A. Raccanelli\affil{4,21, 33},
Emma Rigby\affil{28},
I. G. Roseboom\affil{8}
H. R\"ottgering\affil{30},
Jose Sabater\affil{8},
Mara Salvato\affil{10},
Anna M. M. Scaife \affil{6},
Richard Schilizzi \affil{7},
N.\ Seymour\affil{1},
Dan J. B. Smith\affil{19},
Grazia Umana\affil{31},
G.-B. Zhao\affil{4},
Peter-Christian Zinn\affil{25,1}
 \\
\vspace{0.4cm}
}}}
\begin{document}
\twocolumn[
\begin{changemargin}{.8cm}{.5cm}
\begin{minipage}{.9\textwidth}
\vspace{-1cm}
\maketitle
%
%

{\bf Abstract:
In the lead-up to the Square Kilometre Array (SKA) project,  several next-generation radio telescopes and upgrades are already being built around the world. These include
APERTIF (The Netherlands), ASKAP (Australia),  eMERLIN (UK), VLA (USA), e-EVN (based in Europe),
 LOFAR (The Netherlands),  Meerkat (South Africa), and the Murchison Widefield Array (MWA). Each of these new instruments has different strengths, and coordination of surveys between them can help maximise the science from each of them.  A radio continuum survey is being planned on each of them with the primary science objective of understanding the formation and evolution of galaxies over cosmic time, and the cosmological parameters and large-scale structures which drive it. In pursuit of this objective, the different teams are developing a variety of new techniques, and refining existing ones.
To achieve these exciting scientific goals, many technical challenges must be addressed by the survey instruments. Given the limited resources of the global radio-astronomical community, it is essential that we pool our skills and knowledge. We do not have sufficient resources to enjoy the luxury of re-inventing wheels. We face significant challenges in calibration, imaging, source extraction and measurement, classification and cross-identification, redshift determination, stacking, and data-intensive research. As these instruments extend the observational parameters, we will face further unexpected challenges in calibration, imaging, and interpretation. If we are to realise the full scientific potential of these expensive instruments, it is essential that we devote enough resources and careful study to understanding the instrumental effects and how they will affect the data. We have established an SKA Radio Continuum Survey working group, whose prime role is to maximise science from these instruments by ensuring we share resources and expertise across the projects. Here we describe these projects, their science goals, and the technical challenges which are being addressed to maximise the science return.
}


\medskip
\medskip
\end{minipage}
\end{changemargin}

]

\section{Introduction \& Background}
The Square Kilometre Array (SKA) is a proposed major internationally-funded radio telescope  \citep{Dewdney09} which is expected to be completed in the next decade. It will be many times more sensitive than any existing radio telescope, covering centi\-metre to metre wavelengths, and will answer fundamental questions about the Universe by surveying the radio sky and studying individual objects in detail \citep{Carilli04}. It will consist of many antennas, constituting an effective collecting area of about one square kilometre, deployed over two sites in Australia and  South Africa.

In the lead-up to the SKA,  several next-generation radio telescopes and upgrades are being constructed around the world, including
APERTIF (The Netherlands), AS\-KAP (Australia),  eMERLIN (UK), e-EVN (based in Europe),
 LOFAR (The Netherlands), Meer\-kat (South Africa), MWA (Australia) and the VLA (Karl G. Jansky Very Large Array, in the USA). Large continuum surveys are being planned for many of these telescopes.

\begin{figure}[h]
\begin{center}
\includegraphics[width=7cm, angle=0]{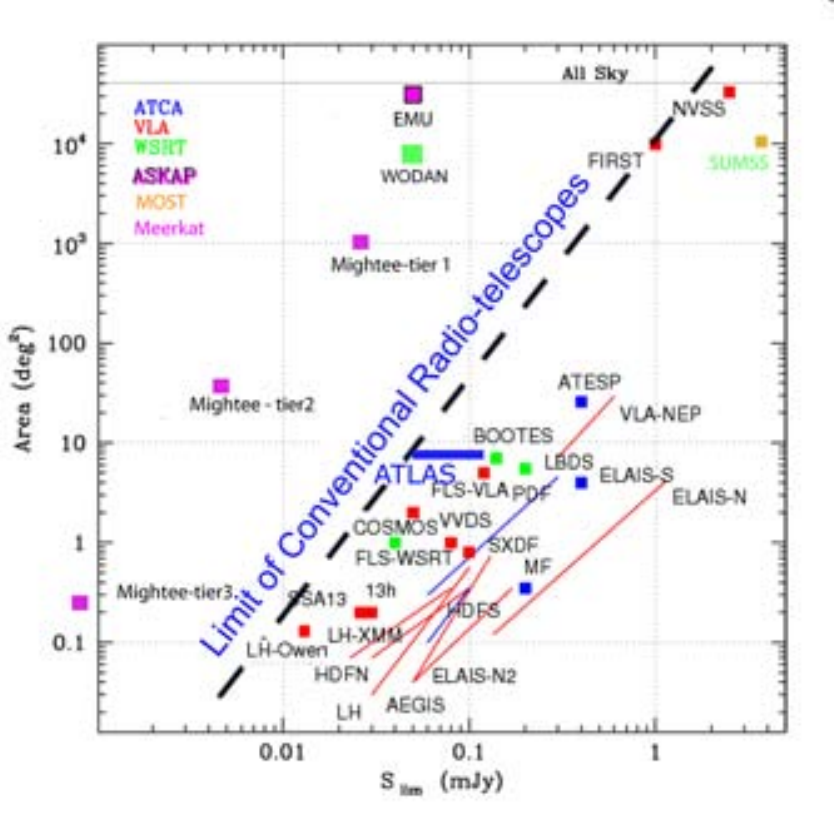}
\caption{Comparison of existing and planned deep 1.4\,GHz radio surveys. The horizontal axis shows the  5-$\sigma$ sensitivity, and the vertical axis shows the sky coverage. The diagonal dashed line shows the approximate envelope of existing surveys, which is largely determined by the availability of telescope time.  
For example, to extend NVSS to the sensitivity of EMU would have required over 600 years of (pre-upgrade) VLA time, so in practice this would not have been possible, and this line therefore represents a hard limit to the sensitivity of traditional surveys.
 The squares in the top-left represent the new radio surveys discussed in this paper.  
Surveys at other frequencies are not shown in this diagram, as their relative sensitivity depends on the assumed spectral index of the sources, although we do include SUMMS at 843 MHz, without making any correction for spectral index.
A similar comparison of low-frequency surveys can be found in \citet{Tingay12}.
}
\label{fig1}
\end{center}
\end{figure}

The predicted sensitivities and areas for these surveys are shown in Figure~\ref{fig1}, alongside existing 1.4\,GHz continuum radio surveys. The largest published radio survey, shown in the top right, is the wide but shallow NRAO VLA Sky Survey (NVSS) \citep{Condon98}.  The most sensitive published radio survey is the deep but narrow Lockman Hole observation  \citep{Owen08} in the lower left, which has recently been overtaken by an even deeper observation of the same field \citep{Condon12}. All current surveys are bounded by a diagonal line that roughly marks the limit of available telescope time of current-generation radio telescopes. The region to the left of this line is currently unexplored, and this area of observational phase space presumably contains as many potential new discoveries as the region to the right.
 
Most of these projects have multiple science goals, but they have one goal in common, which is to survey the radio continuum emission from galaxies, in order to understand the formation and evolution of galaxies over cosmic time, and the cosmological parameters and large-scale structures that drive it. In pursuit of this goal, the different teams are developing techniques such as multi-scale deconvolution, source extraction and classification, and multiwavelength cross-ident\-ification. Furthermore, these projects share specific scientific goals, some of which  require further definition before a well-planned survey can be executed. Finally, each of these new instruments has different strengths, and coordination of surveys between them can help maximise the scientific return from each.
 
 The  radio continuum surveys planned for these instruments typically reach orders of magnitude deeper than traditional surveys. Consequently, the resulting surveys will be far more than just a deep version of existing surveys, but will be qualitatively different, in most cases being dominated by star-forming galaxies.
For example, only a total of about 10\,\sqdeg\  of the sky has been surveyed at 1.4\,GHz to the planned 10\,\ujybm\ rms of many of these surveys, in fields such as the {\it Hubble}, {\it Chandra}, ATLAS, COSMOS and Phoenix deep fields  \citep{Hopkins03, Huynh05, Biggs06, Norris06, Middelberg08a, Miller08, Schinnerer07,   Morrison2010,  Hales12, Grant11, Guglielmino12}.

Surveys at this depth extend beyond the traditional domains of radio astronomy, where sources are predominantly radio-loud galaxies and quasars, into the regime of star-forming galaxies. At this depth, even the most common active galactic nuclei (AGN) are radio-quiet AGNs, which make up most of the X-ray extragalactic sources.
As a result, the role of radio astronomy is changing. Previous radio-astronomical surveys had their greatest impact in the niche area of radio-loud AGNs. For example, the numbers of sources in  NVSS \citep{Condon98} and FIRST\citep{Becker95} are overwhelmingly dominated by AGNs, and only a small fraction of their sources are star-forming galaxies. In contrast, most sources detected by the next-gener\-ation surveys are star-forming galaxies, so that these surveys are dominated by the same galaxies as are studied by optical and infrared (IR) surveys. As a result,  these next-generation radio surveys are an increasingly important component of multiwavelength studies of galactic evolution. 

These surveys, however, will also encounter a new set of technical and scientific challenges. For example, the new continuum surveys such as EMU (see \S\,\ref{askap}) and WO\-DAN (see \S\,\ref{wodan})
should cover the whole sky at sensitivities previously attained only in small fields ($\sim$ 10
\ujybm\ rms) by instantaneously covering wide fields ($\sim$ 30 \sqdeg) with large bandwidths ($\sim$ 300 MHz) and long integration times ($\sim$
12 hours), and not by virtue of larger collecting areas or lower
system noise temperatures.  To reach the sensitivities predicted by the
radiometer equation, these surveys must have exceptionally high
dynamic ranges (e.g., about 42 db for EMU) and low systematic errors.
This requires very high quality hardware, calibration techniques, and
imaging algorithms.  In particular, the primary beams must have $<$ 1\%
uncertainties in reconstructed position, size, and shape.  The ASKAP
dishes, for example, appear to have good surface and pointing accuracies, and their
3-axis mounts maintain a constant footprint of the primary beam on the sky.  The
electronically formed primary beams of phased-array feeds also have to
be very stable, and the survey astronomers need to work closely
with the engineers to ensure that amplitude and phase calibration on
strong sources in the array field-of-view can be transferred from one
primary beam to another.  Finally, the success of a survey in the
astronomical community depends
as much
on the survey results being made
available to everyone in an easily-accessible fashion (such as through
a web portal, or virtual observatory tools, and with useful and intuitive query engines)
as
on whether the
survey actually meets every performance goal, such as sensitivity.

Moreover, some of these projects differ from earlier surveys in that one goal is to cross-identify the detected radio sources with major surveys at other wavelengths, and produce public-domain Virtual Observatory (VO) accessible catalogues as ``value-added'' data products. This is facilitated by the growth in the number of major surveys spanning all wavelengths, discussed below in \S\,\ref{crossid}.

To address these challenges,
a new group, the SKA PAthfinder Radio Continuum Survey (SPARCS) Working Group has been established, with the following goals:
\begin{itemize}
\item To coordinate developments of techniques, to avoid duplication of effort and ensure that each project has access to best practice.

\item To hold cross-project discussions of the specific scientific goals, to ensure cross-fertilisation of ideas and optimum survey strategies.

\item To coordinate the surveys in their choice of area, depth, location on the sky, and other survey parameters, to maximise the scientific return from the surveys.

\item To distil the SKA pathfinder experiences in order to provide the most relevant and up to date input for SKA planning.

\end{itemize}

SPARCS was initially proposed during the SKA meeting in Manchester in March 2010, was formed in May 2010, and
held its first meeting at the Lorentz Center in Leiden, The Netherlands, in February 2011. This review paper documents the outcomes of that workshop, together with related developments since the workshop. The authors are the speakers and facilitators of the workshop, together with other domain experts who were unable to attend the workshop in person.

This paper describes the SKA pathfinder continuum surveys in \S\,2, setting out their scientific goals in  \S\,3, and identifying the challenges to achieve these goals in \S\,4.  \S\,5 concludes by summarising how these challenges are being addressed, and how SPARCS is helping to coordinate and facilitate this process. We refer to the collection of all surveys discussed in this paper as the SPARCS surveys, and the sources detected in those surveys as the SPARCS sources. We assume the convention for spectral index $\alpha$ that $S\propto \nu^{\alpha}$ throughout.

\section{Pathfinders and their Continuum Surveys}
\subsection{APERTIF-WODAN}
\label{wodan}
APERTIF, the new Phased Array Feed (PAF) receiver system for the Westerbork
Synthesis Radio Telescope (WSRT) will dramatically enlarge the
instantaneous field of view (FOV) of the WSRT
\citep{Oosterloo10}  by replacing
the current single front-end feeds by PAFs. Each
of the PAFs consists of 121 Vivaldi elements and will detect the
radiation field (in dual polarisation) in the focal plane of each dish
over an area of about one square metre at an observing frequency of 1.4 GHz. Because of this, many beams
can be formed simultaneously for each dish making it possible to image
an area of about 8 square degrees on the sky, which is an increase of
about a factor of 30 compared to the current WSRT. Its large 300 MHz
bandwidth will not only cater for sensitive continuum imaging, but is also crucial
for  efficient HI and OH emission surveys and for
studies of polarised emission from large areas.

The WODAN  (Westerbork Observations of the Deep APERTIF Northern-Sky) survey  \citep{Rottgering10b} will use APERTIF to survey the northern 25\% of the sky (i.e. North of declination $+30\deg$) that is inaccessible to ASKAP, to a target rms sensitivity of 10 \ujybm\ at a spatial resolution of 15 \arcsec, although confusion noise with a 15-\arcsec beam may increase the observed rms noise level to about 20 \ujybm. 


\subsection{APERTIF-BEOWULF and APERTIF-\\FRIGG}
\label{beowulf}
BEOWULF and FRIGG are both polarisation projects proposed for APERTIF.

The BEOWULF (B-field Estimation and Observational Wide-field Understanding of Large-scale Faraday-structure) survey \citep{Scaife10} will use APERTIF to survey an area of 64 \sqdeg\ within the Perseus-Pisces super-cluster of galaxies in order to construct a finely spaced rotation measure (RM) grid tuned for cosmological and large-scale structure studies of magnetism within both clusters and the filamentary inter-cluster medium. This survey will complement other wide area polarisation surveys: by measuring RMs to 5,000 background sources it will give an RM-grid sampling of 3.5\,arcmin. The BEOWULF survey will provide vital insights into whether magnetic fields were primordial in clusters of galaxies, or whether the galaxies themselves later injected the field. The survey is ideally suited to APERTIF whose wide FOV enables observations of large scale structures, and whose frequency coverage is well matched to the comparatively high RMs expected from such a region.

FRIGG (Faraday Rotation Investigation of Galaxies and Groups) survey \citep{Beck10} will observe large areas around 4 large northern nearby spiral galaxies, the bridge between M\,31 and M\,33, and 5 galaxy groups. Its goal is to measure grids of Faraday rotation measures of background sources and faint diffuse emission, in order to unravel the extent and dynamical importance of cosmic magnetic fields in galaxies and in intergalactic space in galaxy groups.

\subsection{ASKAP-EMU}
\label{askap}
The Australian SKA Pathfinder \cite[ASKAP:][]{Johnston07, Johnston08, Deboer09} is a new radio telescope being built on the Australian SKA site in Western Australia, at the Murchison Radio-astronomy Observatory, with a planned completion date of 2013-14. It will consist of 36 12-metre antennas distributed over a region 6 km in diameter. Each antenna is equipped with a PAF \citep{Bunton10} of 96 dual-polarisation pixels operating in a frequency band of 700--1800 MHz.
As a result, ASKAP will have a  field of view up to 30 \sqdeg.
To ensure good calibration, the antennas are a novel 3-axis design, with the feed and reflector rotating  to mimic  the effect of an equatorial mount,  ensuring a constant position angle of the PAF and sidelobes on the sky.

In continuum mode, ASKAP will observe a 300 MHz band, split into 1 MHz channels, with full Stokes parameters measured in each channel. The data will be processed in a  multi-frequency synthesis mode, in which data from each channel are correctly gridded in the \emph{uv} plane. As well as producing images and source catalogues, the processing pipeline will also measure spectral index, spectral curvature, and all polarisation products across the band.
All data processing steps, from the output of the correlator to science-quality images, spectra, and catalogues, are performed in automated pipelines \citep{Cornwell11} running on a highly distributed parallel processing computer. These steps include flagging bad data, calibration, imaging, source-finding, and archiving.

The Evolutionary Map of the Universe \citep[EMU: ][]{Norris11b} will use ASKAP to make a deep (10 \ujybm\ rms) radio continuum survey of the entire Southern Sky, extending as far North as $+30\deg$. EMU will cover roughly the same fraction (75\%) of the sky as the benchmark NVSS survey \citep{Condon98}, but will be 45 times more sensitive, and will have an angular resolution (10 arcsec) 4.5 times better. Because of the excellent short-spacing \emph{uv} coverage of AS\-KAP, EMU will also have higher sensitivity to extended structures.  Like most radio surveys, EMU will adopt a 5-$\sigma$ cutoff, leading to a source detection threshold of 50\,\ujybm. EMU is expected to generate a catalogue of about 70 million galaxies, and all
radio data from the EMU survey will be placed in the public domain as soon as the data quality has been assured.

Together, EMU and WODAN (see \S \ref{wodan}) will provide full-sky 1.3\,GHz imaging at $\sim$ 10--15 \arcsec\ resolution to an rms noise level of 10 \ujybm, providing an unprecedented sensitive all-sky radio survey as a legacy for astronomers at all wavelengths. The EMU and WODAN surveys will overlap  by a few degrees of declination to provide a comparison and cross-validation, to ensure consistent calibration, and to check on completeness and potential sources of bias between the surveys.

\subsection{ASKAP-POSSUM}
\label{possum}
POSSUM (POlarisation Sky Survey of the Universe's Magnetism)   is an all-sky ASKAP survey of linear polarisation \citep{Gaensler10}. It is expected that POSSUM will be commensal with EMU, and that the two surveys will overlap considerably in their analysis pipe\-lines and source catalogues. POSSUM will provide a catalogue of polarised fluxes and Faraday rotation measures for approximately 3 million compact extragalactic sources. These data will be used to determine the large-scale magnetic field geometry of the Milky Way, to study the turbulent properties of the interstellar medium, and to constrain the evolution of intergalactic magnetic fields as a function of cosmic time. POSSUM will also be a valuable counterpart to EMU, in that it will provide polarisation properties or upper limits to polarisation for all sources detected by EMU.

\subsection{{\em e-}MERLIN}
\label{emerlin}
The {\em e-}MERLIN array \citep{garrington04}, operated by the University of Manchester, is a significant upgrade to the existing telescopes
which form the MERLIN array in the UK. Consisting of seven telescopes, spread across the UK with a maximum baseline of 217\,km, {\em e-}MERLIN  provides high angular
resolution (10-150\,mas) imaging and spectroscopy in three broad cm-wavebands (1.3-1.8\,GHz, 4-8\,GHz and 22-24\,GHz).  The {\em e-}MERLIN project includes a major upgrade to the
existing telescope hardware, including the installation of new receivers, analogue and digital electronics, optical-fibre links to each telescope and a powerful new digital
correlator at Jodrell Bank Observatory (JBO). This upgrade increases the usable bandwidth by more than two orders of magnitude, compared to the old MERLIN system, so that the
continuum sensitivity is increased by a factor of between 10 and 30. In addition, the increase in bandwidth will enable multi-frequency synthesis, dramatically improving the \emph{uv} coverage for continuum
observations and enabling simultaneous spectral-index imaging.

The {\em e-}MERLIN upgrade is well underway with new receivers in service, the
dedicated optical fibre network and the digital transmission equipment operational and returning 30\,Gbit/s of data from each telescope back to a new correlator. Currently {\em
e-}MERLIN is undergoing the final stages of instrument commissioning and will be fully operational during 2012.

These upgrades to the system allow {\em e-}MERLIN to typically provide \ujy\ sensitivity with up to 4\,GHz of instantaneous bandwidth. This sensitive continuum imaging can be
simultaneously combined with powerful spectral line and polarisation capabilities provided by the new highly flexible WIDAR correlator which is a twin of the new VLA correlator
and has been developed by DRAO in Canada.

The {\em e-}MERLIN array is an important SKA path\-fi\-nder instrument covering
a unique scientific niche due to its baseline lengths, which are
intermediate between those provided by the VLA and VLBI arrays. Technically this provides a test-bed for the development of fibre-based data transport and time distribution over
distances spanning many hundreds of kilometres. Scientifically, results stemming from its high angular resolution will provide significant input into the eventual locations of dish
receptors as part of SKA Phase 1.

{\em e-}MERLIN is an open-user instrument with twice yearly proposal deadlines. As with
many other new radio facilities, however, it has pre-allocated a large fraction of observing time
to large key-science projects. Eleven large {\em e-}MERLIN `legacy' projects\footnote{http://www.e-merlin.ac.uk/legacy/} have been allocated a total of 5000 hours (approximately
50\% of the available observing time) over the first 5 semesters of operation. These key projects span a wide range of astrophysics from studies of planet formation, pulsar proper
motions, stars, nearby normal and star-forming galaxies, classical radio galaxies, strong gravitational lenses, galaxy clusters, and the physics of high redshift star-formation and AGN
dominated galaxies.  In each of these programmes the strengths of {\em e-}MERLIN, in terms of its sensitivity and critically high angular resolution at centimetre wavelengths, are

being exploited  to provide a spatially resolved view of of these sources, complementing studies planned or underway with other SKA
pathfinder instruments, and new facilities at other wavelengths.

\subsection{VLA}

The Karl G. Jansky Very Large Array \citep[VLA:][]{napier06} represents a
major upgrade to the VLA. Although the upgraded VLA was not designed to be a survey telescope or an SKA
pathfinder, its high sensitivity, continuous frequency coverage,
flexible wideband correlator, and scaled array configurations make the
VLA a useful complement to the dedicated pathfinders, and
to the all-sky continuum surveys that are being planned.  For example,
the VLA could observe ``reference fields" covering a few small areas
of sky with higher angular resolution (e.g. 5\,arcsec) and sensitivity
(e.g. 5\,\ujybm\ rms) than the 1.4\,GHz EMU and WODAN surveys in
order to join those surveys seamlessly and provide a reference for
resolution corrections, flux densities, positions, etc. of the
faintest survey sources.

The VLA is a general-purpose user
instrument, open to researchers around the world.
The VLA upgrade is approaching completion, and algorithms for making
20\,cm images with large fractional bandwidth and high dynamic range
are now being tested on real data.  The ultimate continuum
sensitivity limits (source confusion or dynamic range) of the full SKA
depend on the surface density on the sky of sub-$\mu$Jy sources.  If dynamic range permits, by imaging a single
field at 20\,cm to the confusion limit with 5\,\arcsec\ resolution, the
VLA in its B configuration will place a tight statistical constraint
on the source density of objects fainter than 1\,\ujy.

\subsection{eVLBI}

The SKA will be a real-time instrument with preferably
25\% of its collecting area forming long baselines that
extend up to to 3000 km. Data caching and transportation,
distribution of the clock signal, and operations of this
array will present a great
technical challenge. The European VLBI Network (EVN) as
an SKA pathfinder is addressing these issues within
the framework of the NEXPReS project (http://www.nexpres.eu/).
The primary goal of NEXPReS is to support and develop
real-time electronic-VLBI (e-VLBI) operations in the EVN
and on global scales.

Science observations with the e-EVN have routinely been
carried out since 2006. The possibility of operating a
real-time VLBI array with baselines exceeding 12,000 km was
recently demonstrated by \cite{giroletti11}. Relaxing the
data storage limitations at the telescopes creates an obvious
advantage for the EVN. This allows for more flexible
operations to carry out transient science (see \S\,\ref{transients}),
it may provide seamless data transport at $>$1~Gbps data rates
for superior sensitivity, and makes it possible to
conduct automated observations
for efficient VLBI surveys, or to respond to external triggers.

The importance of long baselines (and even the possibility of
baselines to space) for the SKA have been discussed by
\citet{fomalont04} and \citet{gurvits04}. Present day
observations of faint, compact radio sources are beginning
to shed light on the science that could be done with
one or two orders of magnitude better sensitivity. The mJy sky
is thought to be dominated by AGN, compact on milliarcsec
scales, while in the sub-mJy regime star-forming galaxies become
increasingly important, and a number of objects are now known
which show both strong AGN activity and vigorous star-formation
activity \citep[e.g.][]{Afonso01,perez10, norris12}. VLBI, and the long
baselines of SKA, will play a key role in distinguishing between
AGN and SF galaxies.

The long baselines of the SKA will be particularly important
for studying
the details of feedback processes between AGN and star-formation \citep{Croton06},   
the role of binary AGN mergers in galaxy evolution and feedback \citep[e.g.][]{dotti12},
the first supermassive black holes at the highest redshifts \citep{falcke04,frey10,frey11}
and several other applications (see \citet{godfrey12} for a review).

In traditional VLBI observations, the field of view was severely limited
by time- and bandwidth smearing. Since the sub-mJy sky is densely populated with radio sources, VLBI
observations are moving from the traditional mode of
observing single, bright, widely-separated objects to a mode which
allows observations of wide fields. This transition poses a
significant technological challenge, because the combination of high
resolution and wide fields requires high spectral and temporal
resolution in correlation, to prevent loss of coherence arising from
averaging effects (bandwidth and time smearing). Early wide-field VLBI experiments
\citep[e.g.][]{garrett05},
were limited by the hardware
VLBI correlators, and by limited computing resources.
Recently, software
correlators running on cheap commodity computers have provided the
power and flexibility needed for VLBI observations of wide fields \citep{deller11}.

Since sources detectable with VLBI, even in the sub-mJy regime, are
relatively sparsely distributed, it has proven to be more
efficient to generate multiple phase centres with small fields of view
directed towards known locations of radio emission, rather than to
image huge blank portions of the sky. Alternative, equivalent
approaches have also been proven to be successful \citep{morgan11}.
The multi-phase centre technique has only
recently been brought to production \citep{middelberg11} but has already
demonstrated that galaxies can contain a radio-active AGN even when an AGN is not evident in comprehensive radio, optical, and X-ray
observations  \citep{middelberg12}.

Wide-field VLBI observations require new calibration
steps to reach the full sensitivity of the data, such as primary beam
corrections and multi-phase centre self-calibration, which help to
make the technique useful for a wide range of applications. For
example, the multi-phase centre technique can be used to image many
sources around a faint target to aid in self-calibration. While no single source may be sufficiently strong for self-calibration, the ensemble of sources can provide a robust solution. The technique may also be used to provide an
astrometric reference, or to image many background sources to probe
a foreground object.

\subsection{ LOFAR}
\label{lofar}

LOFAR, the Low Frequency Radio Array, is a pan-European radio
phased-array telescope that is currently being commissioned.   The two types of
antenna, one optimised for 30 -- 80 MHz and the other for 110 -- 240 MHz, are grouped together in stations the
size of soccer fields. 40 stations are distributed over an area of diameter of 100 km in The Netherlands, and a further eight stations are located in Germany, UK,
Sweden, and France. The signals from the antennas are digitised
to form many beams on the sky, making LOFAR an extremely
efficient survey instrument.   LOFAR has already generated images that are the deepest ever at these low frequencies.

A key
motivation of LOFAR is to provide the entire international
astronomical community with surveys of the radio sky that have
a long-lasting legacy value for a broad range of astrophysical
research.
The LOFAR continuum survey  \citep{Rottgering10a} will cover the northern half of the sky
(i.e.\ North of declination $0\deg$).
LOFAR will be especially complementary to WOD\-AN and EMU in surveying the sky at high sensitivity and resolution but at a much lower frequency.

The three fundamental areas of astrophysics that have driven the
design of the planned LOFAR surveys are: (i) the formation of massive galaxies at the epoch of
reionisation, (ii) magnetic fields and shocked hot gas associated with
the first bound clusters of galaxies, and (iii) star formation
processes in distant galaxies.
The areas, depths and
frequencies of the surveys have been chosen to
contain: (i) $\sim$100 powerful radio galaxies close to or at the epoch of
reionisation, (ii) $\sim$100 radio halos at the epoch when the first
massive bound galaxy clusters appeared, and (iii) $\sim$100 protoclusters.
The resulting  survey parameters are based on estimates of luminosity functions
for powerful radio galaxies by \citet{Wilman08}, for
radio halos by \citet{ens02b} and \citet{cas10a},
and for protoclusters by \citet{Venemans07}.

To achieve the goals of the LOFAR surveys,
a three-tiered approach has been adopted  \citep{Rottgering10a}. Tier 1 is a 2$\pi$ steradian
survey reaching an rms of 0.07\,mJy at
15--65 and 120--180\,MHz, and is designed to  detect $\sim 100$ cluster halos at $z>0.6$ and
$\sim$ 100 $z>6$ radio galaxies. Tier 2 is a deep survey
over 500\,\sqdeg\ at 30, 60, and 150\,MHz. At 150\,MHz, this will require 55 pointings, to be centred on
the following science targets:\\
\hspace*{1em} $\bullet$ 25 well-studied extragalactic-fields with superb\\
\hspace*{2em} existing multiwavelength data;\\
\hspace*{1em}  $\bullet$ 15 fields centred on clusters or super-clusters; \\
\hspace*{1em} $\bullet$ 15 fields centred on nearby galaxies.\\

\begin{figure}[h!]
\centerline{
\includegraphics[width=8cm,angle=0]{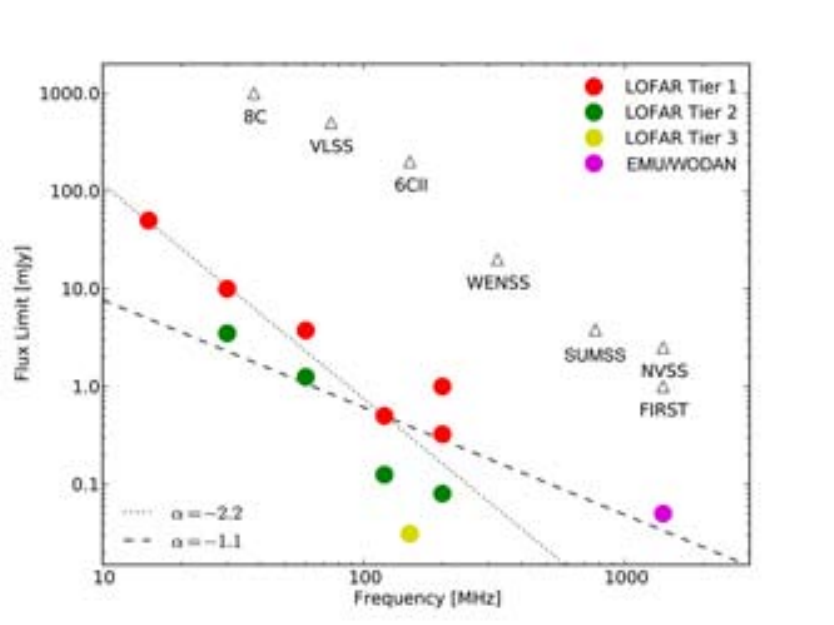}}
\caption{\label{radiosurveys} Flux limits ($5\,\sigma$) of the proposed LOFAR
surveys compared to other radio surveys.  The triangles represent
existing surveys.  The lines represent different
power-laws (with $\alpha=-1.6$ and $-0.8$) to
illustrate how, depending on the spectral indices of the sources, the
LOFAR surveys will compare to others.}
\end{figure}

Tier 3 is an ``ultra-deep'' survey at 150\,MHz covering a single pointing of 100\,\sqdeg\
reaching the confusion level of 7\,\ujybm\ rms. Figure~\ref{radiosurveys}
shows the resulting depth versus frequency.

WODAN and EMU have an enormous synergy with the
LOFAR surveys: virtually all the $5 \times 10^7$ radio sources from
the LOFAR all sky surveys will have their flux density at 1400\,MHz
measured (i.e.\ Figure~\ref{fig1}). Together they will yield data for all radio loud AGN in the northern hemisphere, and most
luminous starbursts up to $z=2$.
The resulting densely populated radio colour-colour diagrams
will be a powerful tool to spectrally discriminate between very rare
radio sources with extreme radio spectra, such as diffuse emission from
clusters and very distant radio galaxies. Nearby resolved sources will have
spectral index and spectral curvature maps, a
very rich source of information to constrain many physical
parameters. As the combined surveys will cover the entire sky, measurements of  the Integrated
Sachs-Wolfe effect,  galaxy auto-correlation functions and
cosmic magnification will substantially tighten cosmological model parameters
\citep{Raccanelli11}.

LOFAR will also support a Key Science Project on Cosmic Magnetism
(MKSP\footnote{http://www.mpifr-bonn.mpg.de/staff/rbeck/MKSP/mksp.html}),
which is likely to make significant contributions to our knowledge of the magnetic sky (see \S\,\ref{magnet}).

\subsection{Meerkat/Mightee}
\label{mightee}

MeerKAT \citep{Jonas10} is the South African SKA path\-fi\-nder telescope. Meerkat Phase 1 will consist of 64 dishes, each 13.5\,m
in diameter, equipped with single-pixel receivers. An offset Gregorian dish configuration has been chosen because the unblocked aperture provides good optical performance, sensitivity  and imaging quality. It also facilitates the installation of multiple receiver systems in the primary and secondary focal areas. MeerKAT's 64 dish array layout will be distributed over two components. A dense inner component will contain 70\% of the dishes with a Gaussian uv-distribution with a dispersion of 300m. The outer component will contain the remaining 30\% of the dishes, having a Gaussian
\emph{uv}-distribution with a dispersion of 2500\,m and a longest baseline of 8\,km. A potential future extension   (Meerkat Phase 2) could see 7 additional dishes being added to extend the longest baselines to about 20\,km. MeerKAT will support a wide range of observing modes, including deep continuum, polarisation and spectral line imaging, pulsar timing, and transient searches. The plan is to provide standard data products, including an imaging pipeline. A number of ``data spigots" will also be available to support user-provided instrumentation.  

70\% of observing time on MeerKAT, for the first few years, is allocated for large survey projects of 1000 hours or more, while the remaining 30\% is reserved for smaller PI driven proposals (of which 5\% will be director's discretionary time).
Proposals for key projects were solicited in 2010, and 10 projects were selected. Two (a pulsar timing and a deep HI survey) were chosen as top priority, and a further eight as second priority, one of which is MIGHTEE.

The MIGHTEE survey \citep{Heyden10} aims to probe to much fainter flux densities (0.1-1\,\ujybm\ rms) than the EMU/WODAN surveys, over smaller areas ($\sim 35$\,\sqdeg) at higher angular resolution, at an observing frequency of 1.4 GHz. The higher sensitivity and resolution will enable exploration of AGN, star-forming galaxies and galaxy clusters from the epoch of reionisation through to the present day. The MIGHTEE survey strategy will follow a 3-tiered approach both in survey area and in sensitivity,  using the longer baselines to be able to probe below the confusion limit of the shorter baseline surveys such as EMU and WODAN.

Tier 1 will cover 1000\,\sqdeg\ using Meerkat Phase 1 at 1.4 GHz to a target rms flux-density of 5\,\ujybm\. This will allow detailed studies of the distant Universe and the evolution of the lower-luminosity radio source populations into the epoch of reionisation. Tier 2 will exploit the longer baselines of Meerkat Phase 2 and increased bandwidth to observe a
single 35\,\sqdeg\  pointing down to 1\,\ujybm\ rms. This will enable studies of the evolution of Milky Way type galaxies up to $z\sim 4$  and pioneering weak lensing analyses at radio wavelengths. These observations will also be used to test techniques to increase dynamic range in preparation for the full SKA. Tier 3 will cover a much smaller area (1\,\sqdeg) to 0.1\,\ujybm\ rms. This  will allow us to isolate the level of AGN activity in star-forming galaxies and to investigate the morphological properties of AGN from $z\sim 0.5$ to $z\sim 6$.

\subsection{Murchison Widefield Array}
\label{mwa}

The Murchison Widefield Array \citep[MWA; ][]{Lonsdale09, Tingay12} is a low-frequency synthesis telescope under construction at Boolardy, Western Australia, adjacent to ASKAP. It is due for completion in November 2012, making it the first of the SKA precursor telescopes to commence full operation. It consists of 2048 dual-polarisation dipole antennas, arranged as 128 ``tiles'', each consisting of a 4 $\times$ 4 array of dipoles designed to operate in the 80-300 MHz frequency range.  Each tile performs an analogue beam-forming operation, narrowing the field of view to an electronically steerable ~25 degrees at 150 MHz.
The majority of the tiles (112) are scattered across a roughly 1.5 km core region, forming an array with very high imaging quality, and a field of view of several hundred square degrees at a resolution of several arcmin. The remaining 16 tiles are located outside the core, yielding baselines up to 3 km to allow higher angular resolution. The MWA calibration and imaging is undertaken on 24 IBM iDataPlex dual Xeon servers, each housing 2  NVIDIA Tesla M2070 Graphics Processing Units. The correlated data are calibrated in real time using novel position-dependent self-calibration algorithms.

MWA is focused on four key science projects \citep{Bowman12}: (a) the detection and characterisation of 3-dimensional brightness temperature fluctuations in the 21cm line of neutral hydrogen during the Epoch of Reionisation (EoR) at redshifts from 6 to 10, (b) solar imaging and remote sensing of the inner heliosphere via propagation effects on signals from distant background sources, (c) high-sensitivity exploration of the variable radio sky, and (d) wide-field galactic and extragalactic astrophysics including a deep all-sky survey over the MWA frequency range with full polarimetry and spectral resolution. An initial survey of 2400 square degrees on the prototype `32T' system has recently been completed \citep{Bernardi12} and a full sky survey south of $+30\deg$ is being planned for the full array.

\subsection{Other survey telescopes}
We acknowledge that many other radio telescopes, such as the Giant Metre Wave Radio Telescope \citep[GMRT:][]{Anantha01} and the Very Long Baseline Array \citep[VLBA:][]{Napier94} are also making very significant contributions both to the continuum survey science and to the survey techniques necessary for the SKA, but are outside the scope of this paper.

\section{Science Goals}
\label{science}

\subsection{The Radio Sky}

The goal of most SKA pathfinder radio continuum surveys  is to make a deep  radio continuum survey of a significant fraction of the sky. For example, EMU and WODAN together will cover the entire sky at 1.4\,GHz to a depth of about
10--15\,\ujybm\ rms with an angular resolution of 10--15\,arcsec. LOFAR will complement this with a  low-frequency survey of comparable intrinsic sensitivity, and VLA and Meerkat will be able to probe individual objects, or survey smaller areas, to a much greater depth. The largest comparable survey currently existing is the benchmark NVSS survey \citep{Condon98}, compared to which EMU plus WODAN will be about 40 times more sensitive, with an angular resolution about four times better, and with higher sensitivity to extended structures. EMU and WODAN together
are expected to detect and catalogue about 100 million sources.

\begin{figure}[h]
\begin{center}
\includegraphics[width=8cm, angle=0]{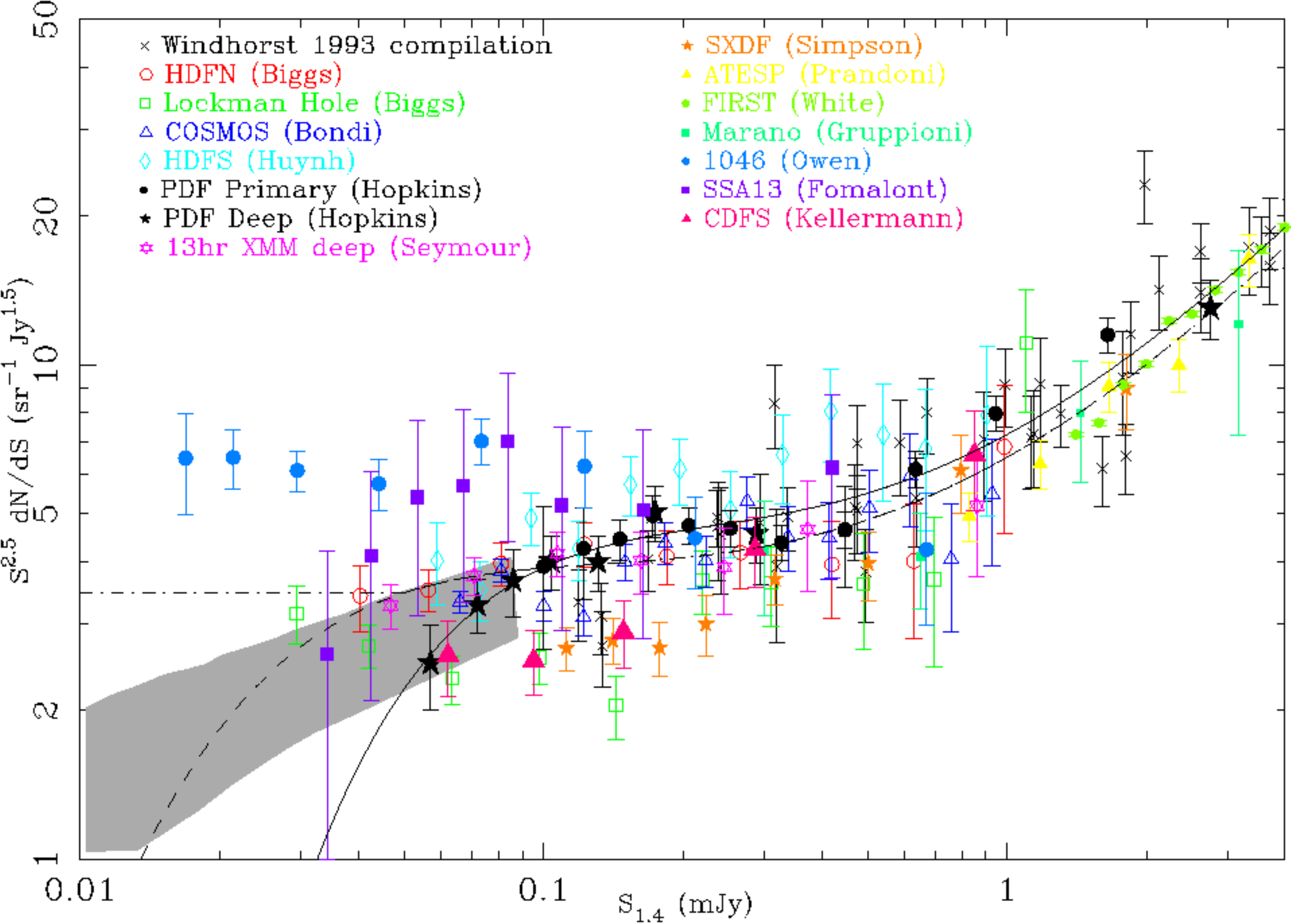}
\caption{The Euclidean normalised differential radio source counts at 1.4 GHz, based on and updated from the distribution shown in \cite{Hopkins03}.  The solid curve is the polynomial fit from \cite{Hopkins03},
and the dashed curve is an updated polynomial fit. The horizontal dot-dashed line  represents a non-evolving population in a Euclidean universe.
The shaded region shows the prediction based on fluctuations due to weak confusing sources ( a ``P(D) analysis'') from
\citet{Condon74} and \citet{Mitchell85}.
 }
\label{srccnt}
\end{center}
\end{figure}

\begin{figure}[h]
\begin{center}
\includegraphics[width=8cm, angle=0]{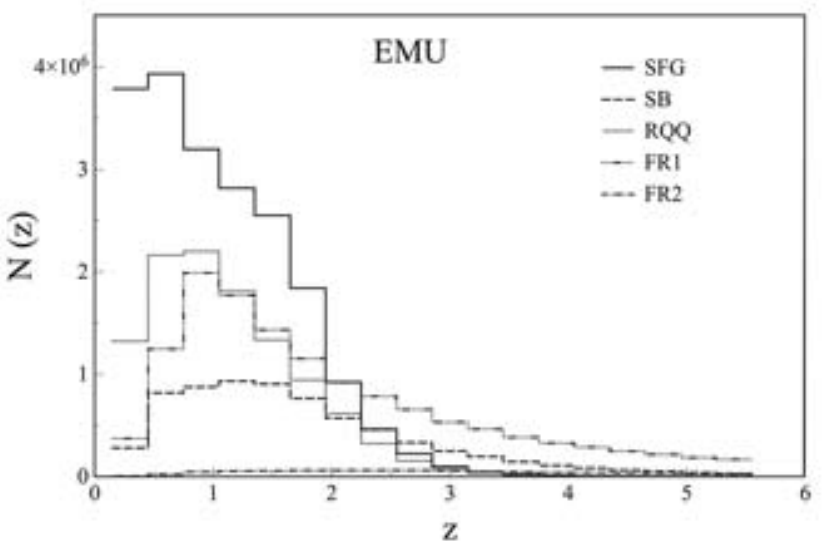}
\caption{Expected redshift distribution of  sources with $S_{1.4} > 10$\,\ujybm, based on the SKADS simulations  \citep{Wilman08, Wilman10}. The five lines show the distributions for star-forming galaxies (SFG), starburst galaxies(SB), radio-quiet quasars (RQQ), and radio-loud galaxies of Fanaroff-Riley types I and II \citep[FRI \& FR2; ][]{FRI}. The vertical scale shows the total number of sources expected to be detected.
}
\label{nz}
\end{center}
\end{figure}

At the sensitivity level of these surveys, the radio sky has already been studied by a number of deep surveys, and
the flux-density--dependent surface-density statistics have been extensively measured (Figure~\ref{srccnt}).
However, these surveys cover a total  of only a few square degrees
\citep{Hopkins03, Huynh05, Biggs06, Norris06, Middelberg08a, Miller08, Schinnerer07,   Morrison2010,  Hales12, Grant11,  Guglielmino12}. We know broadly what types of galaxy we will detect, and their approximate redshift distribution (see
Figure~\ref{nz}) but a number of significant questions remain:
\begin{itemize}
\item What types of object dominate the source counts at low fluxes?
\item Are there undiscovered populations?
\item What causes the scatter between different surveys at low fluxes?
\item Do we need better simulations to guide research in this area?
\item How do environment and large-scale structure affect the evolution of radio sources?
\end{itemize}

At high flux densities, the source counts  (Figure~\ref{srccnt}) are dominated by AGN, following a smooth power law distribution from Jansky levels down to about 1\,mJy. Below 1\,mJy, the normalised source counts flatten, suggesting an additional population of radio sources. The source counts below 1\,mJy are less well-constrained, with a scatter between different surveys which can reach a factor of
$\sim 2-3$.

While some fraction of this scatter may be due to cosmic variance \citep[e.g.][]{Pra12}, that cannot be the only cause. For comparison, in the X-ray band,
the scatter between the GOODS North and South fields is only $\sim 25\%$ ($\sim 3 \sigma$). Furthermore, \citet{Condon12} have shown that some of the largest variations can be accounted for by errors in the  many necessary corrections for effects such as resolution bias, clean bias, Eddington bias, and completeness corrections, together with imaging errors such as  excessive deconvolution, bandwidth smearing, and insufficient beam sampling in the image plane.
 There is also an inconsistency in that some authors measure the numbers of radio components, while other measure the numbers of sources, each of which may consist of several
components, the numbers of which in turn may vary as a function of flux density.

Resolving this discrepancy is critical, since this factor of $2 - 3$ scatter introduces large uncertainties
 in the comparison
of observed number counts with detailed, model-based predictions. Obviously, radio surveys need to move beyond the small-area surveys that have previously dominated the literature. It is also essential that, before the large radio surveys start, we obtain consensus on how to image and analyse radio survey data to produce reliable and consistent source flux densities and source counts.

\subsubsection{The radio sky at $\mu$Jy levels: AGN or star formation?}

\begin{figure}[h]
\begin{center}
\includegraphics[width=7cm, angle=0]{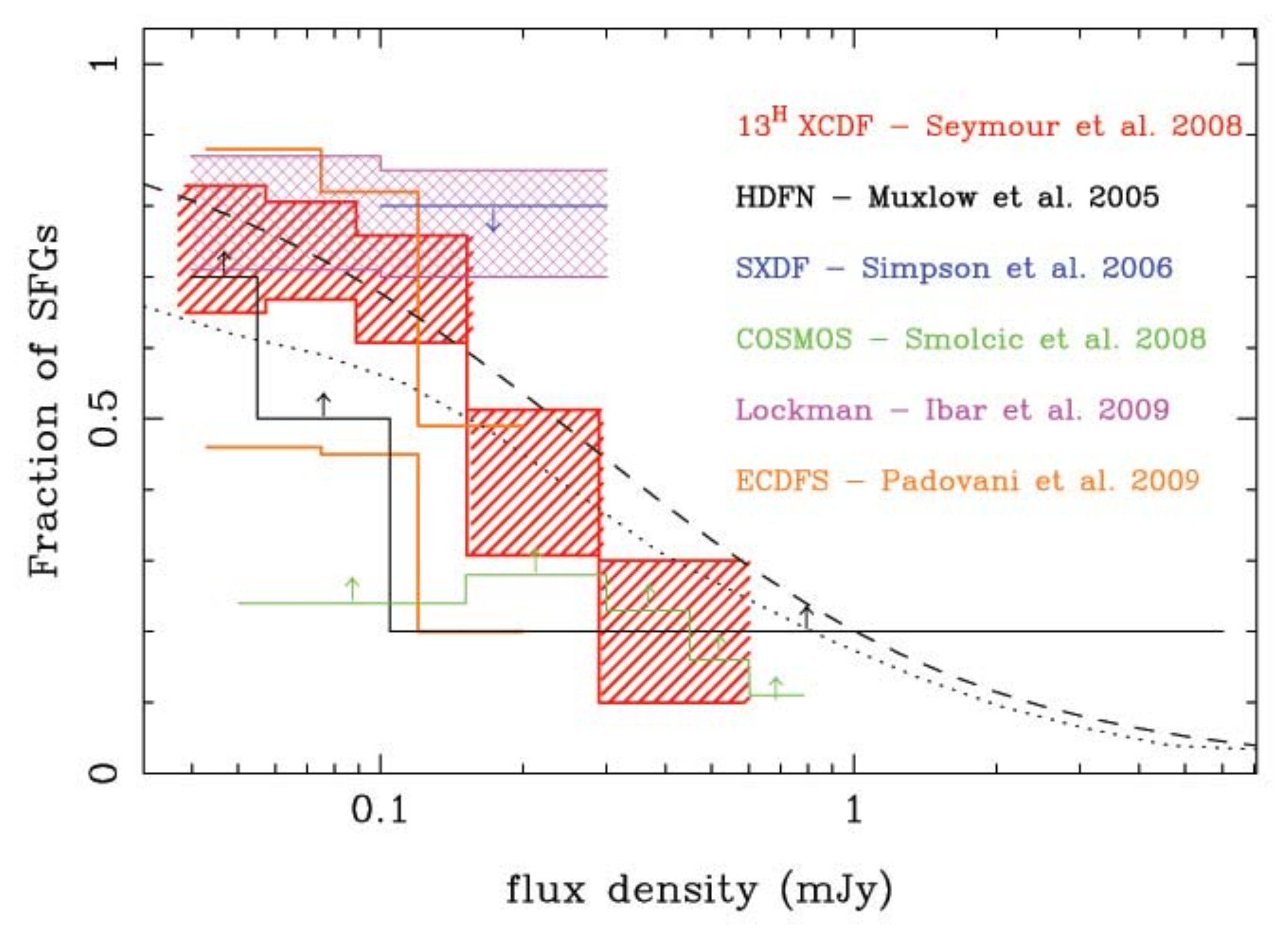}
\caption{Differential fraction of star-forming galaxies as a function of 1.4 GHz
flux density, taken from \cite{Norris11b}. Shaded boxes,
and the two lines for Padovani et al.,  show the range of uncertainty
in the survey results. Arrows indicate constraints from other surveys.
These results show that the fraction of star-forming galaxies
increases rapidly below 1 mJy and, at the $50\,\mu$Jy survey limit of EMU/WODAN,
about 75\% of sources will be star-forming galaxies.
}
\label{SFfraction}
\end{center}
\end{figure}

After years of intense debate, it is now clear that while star-forming (SF) galaxies contribute significantly to
the sub-mJy counts, they do not dominate it above
$\sim\,200~\mu$Jy. Recent results \citep[e.g.\ ][]{Seymour08,Smolcic08,Padovani09}, which are summarised in
Figure~\ref{SFfraction}, show that both SF galaxies and AGN contribute significantly to the source counts. About half of these AGN are radio-quiet, characterised by relatively low radio-to-optical flux density ratios and
radio powers, as compared to radio quasars \citep[see also ][]{pra10}.
Above 1\,mJy, the contribution of radio-quiet objects to the source counts  is insignificant, and they
are dominated by radio-loud AGN.

To probe the radio-quiet AGN component, it is therefore necessary to analyse the deepest radio fields
($S\leq 100-200\,\mu$Jy),
where most sources are typically associated with star-forming galaxies. Radio-quiet AGN share many properties with star-forming galaxies: they have similar radio luminosities ($10^{22-24}$ W/Hz), steep radio spectra ($\alpha<-0.5$), and similar infrared/radio flux ratios\citep[e.g.][]{Roy98}.

In addition radio-quiet AGN are often associated with faint optical
galaxies characterised by Seyfert-2-like spectra, containing both star-forming and AGN components. Even when available, if
optical spectroscopy is not of sufficient quality, it can be difficult to
derive reliable emission line ratios, and distinguish between starburst
and Seyfert 2 spectra, through BPT \citep{BPT} diagrams.
\citep[see e.g.\ ][]{Pra09}.
This makes it difficult
to distinguish radio-quiet AGN from star-forming galaxies, even with multiwavelength information,
although the availability of X-ray data can help \citep{Padovani11}.

\begin{figure}
\begin{center}
\includegraphics[width=7cm]{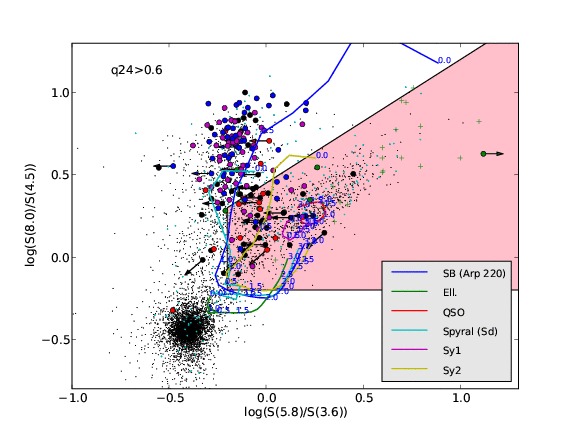}
\caption[]{IRAC colour-colour plot of the FLS radio sources with $q_{24}=log(S_{24  \mu}/S_{1.4  \rm{GHz}}>0.6$ (filled symbols), from \cite{Pra09}.
Colours refer to optical spectral classification: star-forming galaxies
(blue); narrow/broad line AGN (green); early-type galaxies (red); galaxies
with narrow emission line, which do not have a secure optical
classification (magenta); sources with no optical spectroscopy available
(black).
Arrows indicate upper/lower limits.
The lines indicate the expected IRAC colours as a function of redshift for different source types
(see legend). The expected location for AGN is highlighted in pink.
For reference we also show IRAC colours of: {\it (a)} all FLS IRAC-identified
radio sources (no optical identification selection applied, cyan dots);
{\it (b)} the entire FLS IR-selected star/galaxy population
(no radio selection applied, black dots); and {\it (c)} a sample of  
high redshift obscured (type-2) quasars
\citep{Mar06}, green crosses.
}
\label{fig:IRACplot}
\end{center}
\end{figure}

Spitzer IRAC colours, when available, can be very effective
in separating radio quiet AGN from star-forming galaxies as shown in Figure~\ref{fig:IRACplot}, where the well-known IRAC colour-colour plot
\citep{Lac04}
is exploited for the First Look Survey (FLS) deep radio field.
A fraction of sources with typical AGN IRAC colours
appear to be genuine radio-quiet AGN, and account for $\sim 45\%$ of the overall AGN component in the FLS.

A number of other methods can also be used to distinguish AGN from SF galaxies in radio surveys\citep[e.g.][]{Norris08},
including

\begin{itemize}
\item Radio morphology \citep[e.g.][]{Biggs08, Biggs10},
\item Radio spectral index \citep[e.g.][]{Ibar09, Ibar10},
\item Radio--far-infrared ratio \citep[e.g.][]{Norris06, Middelberg08a},
\item Radio--near-infrared ratio \citep[e.g.][]{Willott03},
\item Radio polarisation  \citep[e.g.][]{Hales12},
\item Radio variability  \citep[e.g.][]{Norris08,Vast10, Murphy12},
\item Optical and IR colours, including the use of spectral energy distribution (SED) templates \citep[e.g.][]{Lac04},
\item Optical line ratios \citep[e.g.][]{BPT},
\item X-ray hardness ratio \citep[e.g.][]{Rosati02}
\item X-ray power \citep[e.g.][]{Padovani11},
\item Radio source brightness measured by VLBI \citep[e.g.][]{kewley00, parra10,alexandroff12}.
\end{itemize}

None of these techniques is foolproof and universal, and a combination of techniques is necessary to provide unambiguous classification.
 In preparation for the daunting task
of classifying tens of millions of radio sources, we need to define reliable methods for classifying radio sources.

\subsubsection{Composite Galaxies}
It is well-established \citep[e.g.][]{Roy98} that the radio emission from Seyfert galaxies contains significant contributions from both AGN and star forming activity, and there is growing recognition \citep{Norris08, Lutz10, Shao10, Seymour11} that this may also be true of high-luminosity galaxies, particularly at high redshift. Many of these galaxies are not simply ``star-forming'' or ``AGN'' but include a significant contribution from both. In extreme cases \citep{norris12} a galaxy may appear to be ``pure AGN'' at one wavelength and ``pure SF'' at another.

Particularly at high redshifts,  such composite AGN/SF systems constitute a significant fraction of radio sources, and so  a simple classification into AGN or SF galaxy is inadequate. Instead  the  contribution to the galaxy's luminosity from both AGN and SF activity must be assessed, to measure the relative contributions from underlying physical
properties such as black hole and galaxy mass, star-formation rate, environment, etc. However, the relative contribution will vary
depending on the observing band, and so comprehensive multiwavelength data are required.
However, such
detailed information will be hard to obtain for very large samples, perhaps necessitating  a simpler
classification scheme, which may lead to oversimplifications or even to incorrect interpretation of the data.

It is therefore essential to
develop methods which fit the SED of a given radio source with both a
starburst and AGN component \citep[e.g.\ ][]{Afonso01}. This
approach is clearly only possible for fields with good multiwavelength
data,  but will inform analysis
of fields with lower quality ancillary data. Observations from Spitzer and Herschel cover
the bulk of the energy output of most galaxies and can
be used to distinguish the starburst and AGN components. For example, the far-IR
is directly related to the star formation rate,
and, if the FIR-radio correlation remains true at all redshifts \citep[e.g.][and references therein]{Mao11a}, can be
used to determine what fraction of the radio
emission is due to star formation. Hence, the remaining radio
emission would be due to AGN activity. The AGN bolometric luminosity can
be estimated from the mid-IR component which is not necessarily proportional
to the AGN emission at radio wavelengths. Once the starburst
and AGN fractional contribution to individual sources are determined, it is possible to construct
derived results such as luminosity functions and relative contributions by
type to the source counts.

\subsubsection{Are there undiscovered populations of radio sources?}
The cosmic radio background (CRB) has been well-studied at high frequencies
by instruments such as  COBE and WMAP, but is less well-studied at
 low frequencies.


\cite{fixsen11} measured the background sky temperature at five frequencies between 3 and 90 GHz
using an in situ calibrator on board the balloon-borne ARCADE2 experiment.
These authors found an excess radio sky temperature above that due to the
Cosmic Microwave Background (CMB) below 10\,GHz. The CRB was measured to have
an excess of $50\pm7$\,mK at 3.3\,GHz above a CMB temperature of $2.730\pm0.004$\,K, which is a factor of five higher than expected from known source populations.

The ARCADE2 result is either due to an instrumental or calibration problem (e.g. incorrectly subtracting the Galactic foreground emission), or it is a startling result which will necessitate a drastic revision of our models of extragalactic radio sources. Given that \citet{fixsen11} appear to have taken careful steps to avoid errors,
there is a \emph{prima facie} case that this result is correct, and the radio-astronomical community
has been galvanised to search for this putative new population. Since it is inconsistent
with known radio source populations  \citep[e.g.][]{zotti09,pad11}, it must, if confirmed,  be caused by another
population.
For example, the radio emission could be caused by dark matter annihilation \citep{Fornengo11}, in  which case the emission would trace the dark matter distribution of cluster galaxies, resulting in a scale size of  $\sim$~arcmin. Other mechanisms are also possible, such as diffuse emission from clusters or halos, or very low surface brightness emission from extended AGN radio lobes, or from a population of dwarf galaxies.

\citet{Condon12} have conducted a deep survey with the VLA which shows that the ARCADE2 result cannot be caused by a population of compact objects spatially associated with galaxies. It is therefore either due to a diffuse population of objects which cannot be detected with the shortest baselines of the VLA observations, or they are due to an instrumental or calibration error. A number of  observational programs continue with the VLA and ATCA to determine the cause of this result.

\subsubsection{Simulations}
There is widespread agreement that the SKADS continuum sky
\citep{Wilman08} represents the current state of the art in simulations of the
radio sky. Such simulations have two distinct functions.
\begin{itemize}
\item They represent useful approximations to the real sky which can be used  to test
algorithms and design surveys.
\item They can be viewed as an attempt to
encapsulate the current ``best knowledge'' of the sky, and can  therefore be used to test models of radio source evolution.
\end{itemize}

These two functions have different implications for future updates of the
SKADS continuum sky.  For example, the SKADS continuum sky was
constructed from the best information then available of the
radio luminosity functions of different populations of radio objects, and deeper and wider surveys have since become available.  The SKADS simulation also assumes that
sources are members of distinct populations, omitting the
composite sources in which
both star formation and AGN activity  contribute to their radio luminosity.  
Such small differences from the real sky are unlikely to be important for the design of a survey, but will be important if SKADS is to be used to test theoretical models.

A useful, and relatively simple, modification to the
SKADS continuum sky would be to change the assumed luminosity function
for one or more of the populations in order to assess how that might
influence the results of various continuum surveys.  This change
might be implemented in a parametric way, allowing users to specify
parameters of a luminosity function.  

More difficult would be an attempt to incorporate ``best
knowledge,'' with the aim of producing ``realistic'' sky images.  
Experience from the
\textit{Herschel} mission bolsters the concern that identifying and
incorporating information can be extremely difficult.
Similarly,
incorporating polarisation information would be useful, so that one
could calculate Faraday rotation measures (RMs), but would be quite
difficult in practice because of the need to trace the geodesic of a
photon in order to determine the \hbox{RM}.  
Nevertheless, there are efforts underway to insert realistic source
shapes into the SKADS sky, and the SKADS
team is investigating how to add the signature of the cosmic dipole.

It would be very useful to include models of more complex structures
in the SKADS sky, such as those of large radio sources, diffuse cluster sources, and   structures  near the
Galactic plane, which could potentially limit the lowest Galactic
latitude to which an ``all-sky'' survey could probe.  However, as
described in Section \ref{calibration},
imaging a field containing many compact sources may have
similar issues as to imaging a field containing large scale structure.

\subsection{The evolution of star formation}
\label{SFAGN}

The cosmic star formation history has been studied thoroughly
over the past decade and a half, building up a remarkably consistent
picture.
As shown in Figure~\ref{SFfraction},
only a small fraction of radio sources above 1
mJy are star-forming galaxies. Below 1 mJy, a much larger volume of
the star forming galaxy population becomes detectable, and the population
becomes significant at flux densities below about 0.2\,mJy.
Star-forming galaxies will therefore dominate the deep and wide area continuum
surveys proposed with the ASKAP, APERTIF and MeerKAT telescopes, probing
the entire sky to depths of about 50\,\ujybm.

The star formation history is now well established up to
almost the epoch of reionisation. The space density of star formation rate  declines
by an order of magnitude
between a redshift of unity and zero, and is almost flat at
higher redshifts, with the suggestion of a decline above a redshift
of $z\sim 5$ \citep{Hopkins06}. At higher redshifts there is
still some discrepancy, with evidence from Lyman dropout sources
suggesting a dramatic decline above z=6 \citep{Bouwens08},
and evidence from gamma-ray bursts suggesting a much slower decline
\citep{Kistler09}. The mass-dependence of the star formation rate
density has also begun to be constrained, with the contribution of
more massive galaxies to the star formation rate density increasing
at higher redshift \citep[e.g.][]{Juneau05, Feulner05,
Mobasher09}. Environmental effects as well have been shown
to change with redshift, with star formation in high density, cluster
like environments being enhanced at early times \citep{Elbaz07}
but suppressed at later times \citep{Lewis02, Gomez03}.

\begin{figure}[h]
\begin{center}
\includegraphics[width=7cm, angle=0]{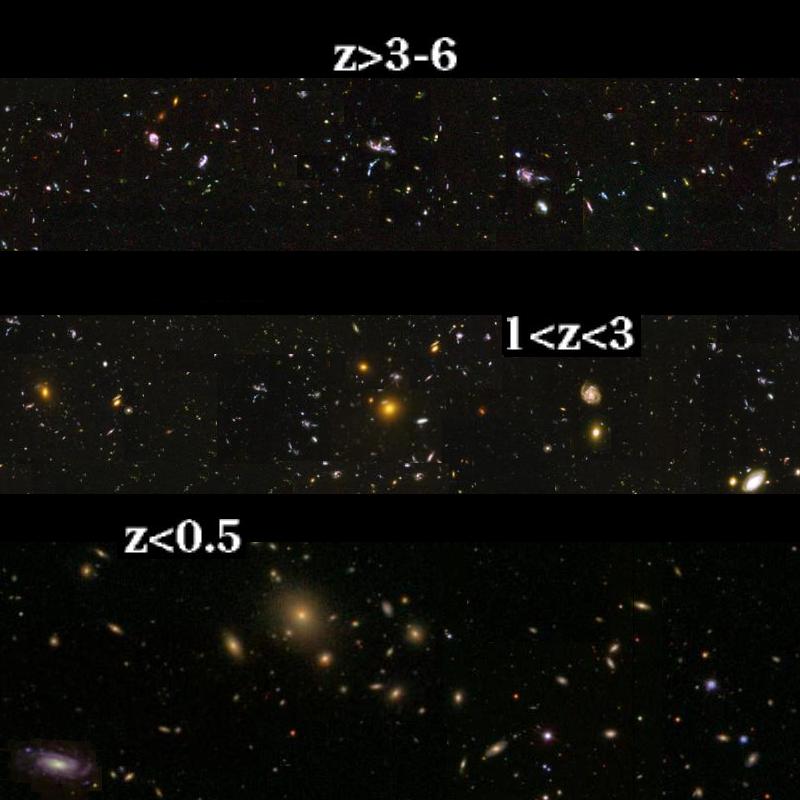}
\caption{An illustration of the density-dependence of the downsizing
of galaxy star formation rates, using optical images from a variety of sources.  At the highest redshifts
(z$>$3), star formation is occurring predominantly in massive galaxies
(those that become local massive ellipticals) that live
in the most overdense regions (that evolve into today's massive
clusters). At lower redshifts ($1<z<3$), where EMU/WODAN will be
sensitive to the most extreme star forming systems, star formation
is dominated by lower mass systems, in  less
dense environments. By the current epoch, star formation is
limited primarily to low-mass galaxies in the outskirts of
clusters, and in the lowest-density environments.}
\label{galaxy_history.jpg}
\end{center}
\end{figure}

Together these lines of evidence point to a picture where star
formation at the earliest times is dominated by massive systems
in the densest regions, which rapidly exhaust their fuel, ultimately
becoming the progenitors of today's brightest cluster galaxies and cD
galaxies. At later epochs the star formation is dominated by
somewhat lower-mass systems in somewhat less dense environments,
similar to cluster outskirts, which in turn exhaust their fuel,
until at the lowest redshifts the star formation is dominated
by low-mass galaxies living in the least dense environments
(Figure~\ref{galaxy_history.jpg}). This picture is still simplistic, and omits significant
details such as how the evolving luminosity function of star forming
galaxies changes with galaxy mass and environment, and whether and how
the stellar initial mass function varies \citep[e.g.\ ][]{Gun:11}.

\subsubsection{Measuring star formation rates  from radio observations}

At redshifts between $1 \lesssim z \lesssim 3$, when galaxies were undergoing rapid evolution, the star formation activity  appears to have been dominated by dusty, heavily obscured, starbursting galaxies  \citep[e.g.][]{Chary01, Caputi07, Magnelli10, Murphy11}. Optical and near-infrared observations are therefore seriously hampered by dust extinction.
Thus, deep radio continuum surveys provide an important tool for measuring the cosmic star formation history of the Universe.   

A key result from the next generation of wide and deep radio surveys will therefore
be the measurement of the radio luminosity due to star formation in a wide
range of galaxies. Converting this radio luminosity to a star formation
rate (SFR) depends on a conversion factor which is principally based
on the infrared/radio correlation \citep{Yun01, Bell03, Seymour08}. However, this correlation has not yet been well determined at the high end
of the luminosity function, since the number of ultra-luminous IR galaxies (ULIRGs) in these papers is
very small, or at high redshifts, although evidence so far suggests the correlation remains constant \citep{Mao11a}.
    
\begin{figure}[h]
\begin{center}
\includegraphics[width=7cm, angle=0]{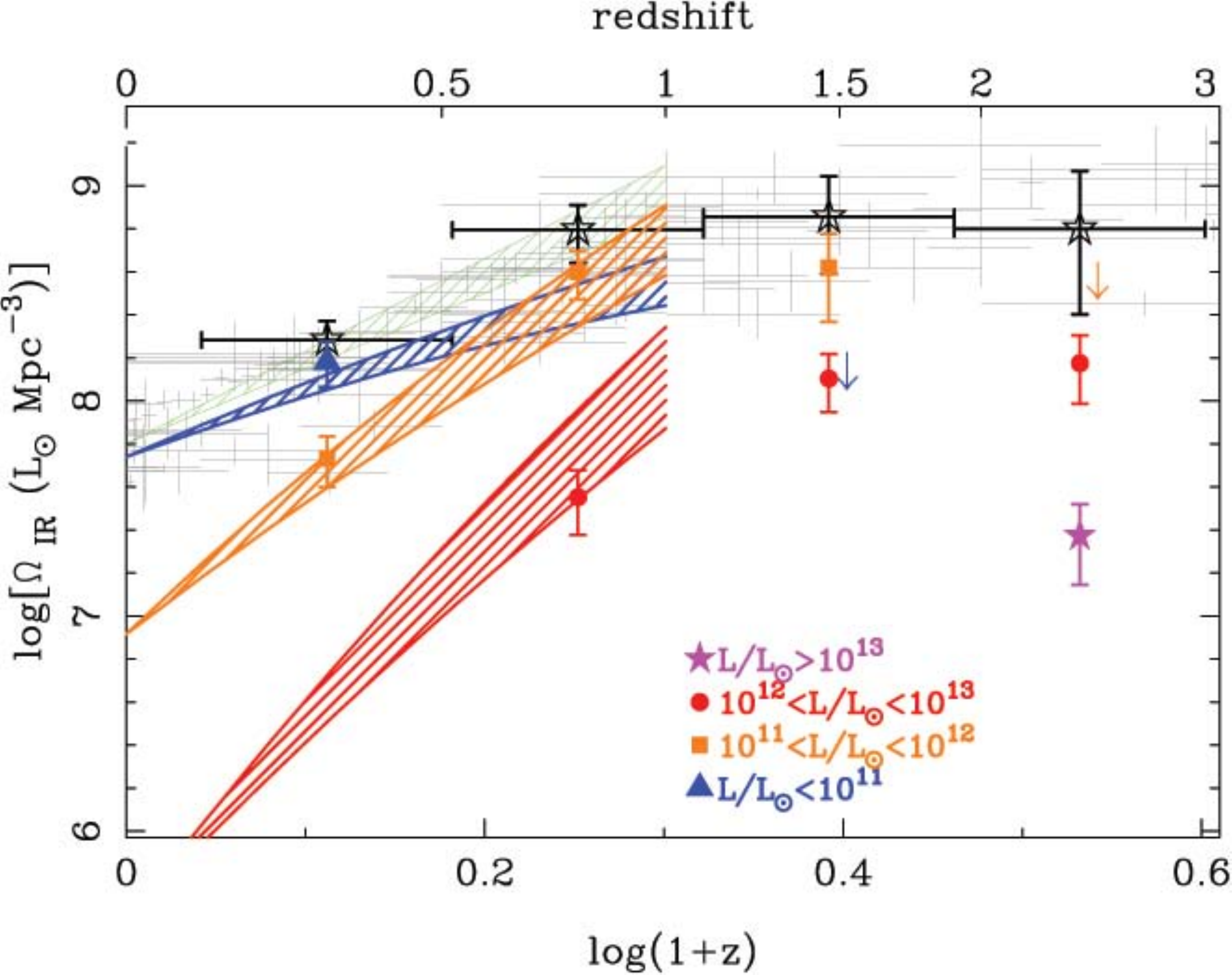}
\caption{The comoving infrared luminosity density of star forming galaxies
    as a function of redshift, separated into four infrared luminosity ranges. The shaded regions are from
    \citet{LeF:05} and the bold points are derived from the analysis of faint radio
    sources in the 13H field \citep{Seymour08} where the
    star-forming radio luminosities have been converted to infrared luminosities using the relation
    of \citet{Bell03}. The faint, grey points are from the compilation
    of \citet{Hopkins06} converted to IR luminosity density. The figure shows
    that ULIRGs make an increasing contribution to the total star formation
    budget above redshifts of unity.
}
\label{ltird.ps}
\end{center}
\end{figure}

At high redshifts, however, we know that the comoving star
formation rate density is increasingly dominated by more luminous galaxies
(with naturally higher SFRs). In Figure~\ref{ltird.ps} we show the comoving infrared
luminosity density, a proxy for the comoving SFR density, as a function of
redshift. We also separate the contribution by infrared luminosity and
find that ULIRGs represent an increasing fraction of the total star formation
budget above a redshift of unity. Hence, if we are to use deep radio surveys
to probe the star formation history of the Universe at high redshift, we
must obtain a considerably more accurate conversion of the radio luminosity
to SFR for galaxies with high SFRs. This conversion factor could also
depend upon other parameters such as stellar mass, environment,
metallicity etc.

\subsubsection{Measuring SFRs at 10\,GHz}  

Radio continuum emission from galaxies typically arises from two processes that are both tied to the massive star formation rate (SFR). At low frequencies (e.g., $\lesssim2$\,GHz), the radio continuum is dominated by non-thermal synchrotron emission arising from cosmic-ray (CR) electrons that have been accelerated by shocks from supernova remnants and are propagating through a galaxy's magnetised interstellar medium (ISM).  
This physical link to massive star formation provides the foundation for the far-infrared (FIR)--radio correlation
\citep[e.g.][]{Helou85, Condon92, Yun01, Murphy06}.  
However, this link is not at all direct given that there are a large number of physical processes affecting the propagation of CR electrons and the heating of dust (e.g. CR diffusion, magnetic field strength/structure, dust grain sizes and composition) that must conspire together to keep this relation intact \citep{Bell03}.  

High frequency ($\sim10-100$\,GHz) radio emission, on the other hand, offers a relatively clean way to quantify the current star formation activity in galaxies.  
At these frequencies, emission is generally optically thin and dominated by free-free radiation, which is directly proportional to the ionising photon rate of young, massive stars.  
While this picture could be complicated by the presence of anomalous dust emission \citep[e.g.][]{Kogut96, Oliveira97, Leitch97}, which occurs at these frequencies and is thought to arise from spinning dust grains \citep[e.g.][]{Erickson57, Draine98}, it is currently unclear whether this component contributes significantly  to globally integrated measurements \citep{Murphy10}.  
Thus, higher frequency radio observations may be particularly powerful for precisely measuring the star formation history of the Universe unbiased by dust.  

\begin{figure}[tch]
\centering
\includegraphics[width=7cm]{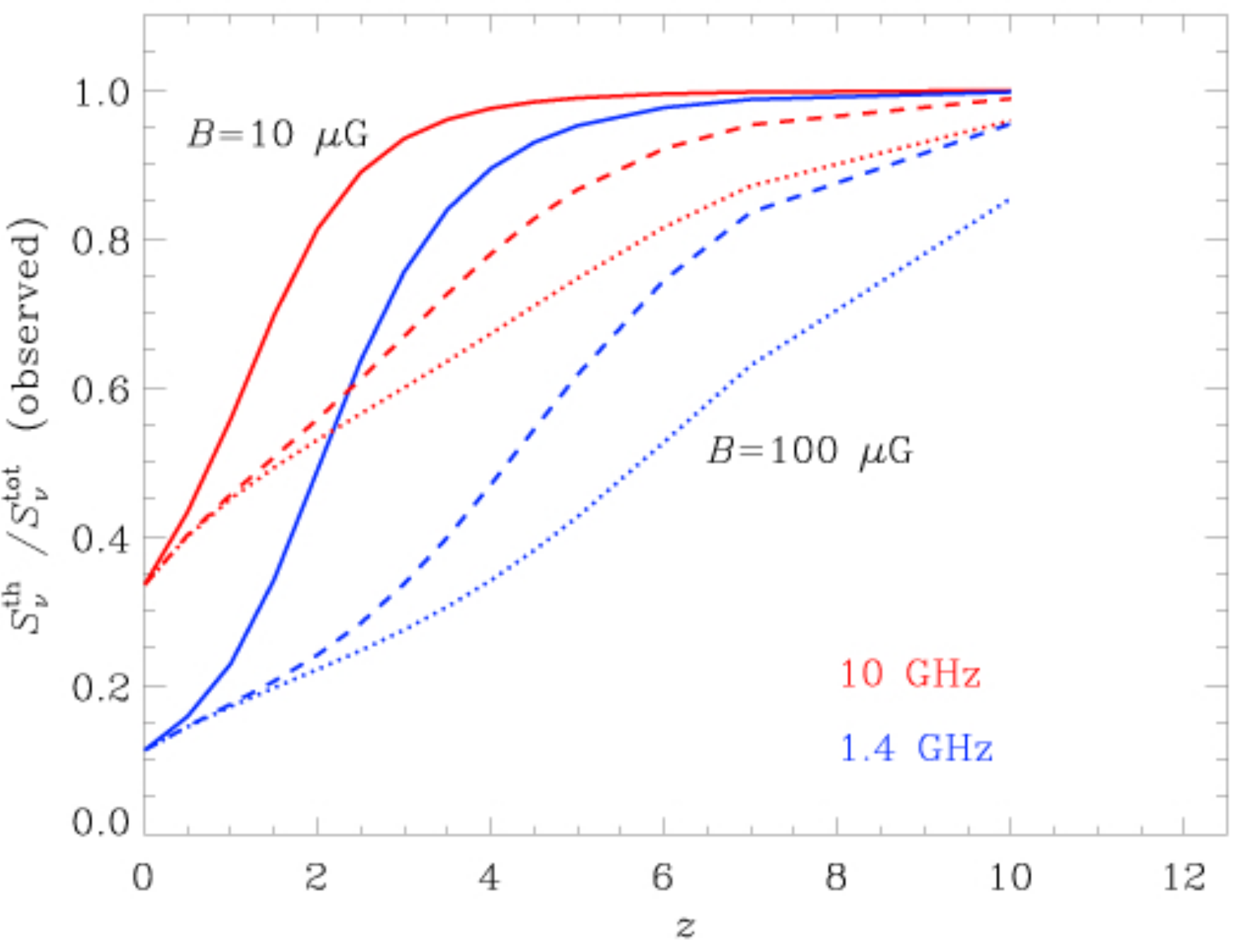}
\caption{
The fraction of radio emission of SFG due to thermal processes as a function of redshift.  
Galaxy magnetic fields of 10 (solid lines), 50 (dashed lines), and 100\,$\mu$G (dotted lines) are shown.
The radio emission from observations at 10\,GHz is dominated by free-free processes beyond a redshift of $z \gtrsim2$ even for magnetic field strengths of $\sim$100\,$\mu$G,  making it an ideal measure for the current SFR of high-$z$ galaxies \citep{Murphy09}.
}
\label{tfrac}
\end{figure}

For example, surveys at $\nu \gtrsim 10$\,GHz start to probe the rest-frame $\nu \gtrsim 30$\,GHz emission from star-forming galaxies by $z\gtrsim2$ for which $\gtrsim$50\,\% of the emission is thermal (free-free).  
This is illustrated in Figure~\ref{tfrac} where the thermal fraction of the observed 10\,GHz and 1.4\,GHz emission is shown against redshift for star-forming galaxies having intrinsic magnetic field strengths of 10, 50, and 100\,$\mu$G while still obeying the FIR-radio correlation at $z=0$.  
In addition to measuring higher rest-frame frequencies with increasing redshift, this calculation also accounts for the suppression of a galaxy's non-thermal emission due to rapid cooling of CR electrons from inverse Compton (IC) scattering off the cosmic microwave background (CMB), whose radiation field energy density scales roughly as $U_{\rm CMB} \sim (1+z)^{4}$ \citep{Murphy09}.
By $z\gtrsim4$, nearly 80\% of observed 10\,GHz radio continuum is due to free-free emission for a galaxy having a large (i.e., 50\,$\mu$G) magnetic field.

Additionally, we can compare the relative sensitivity requirements of surveys at 10\,GHz and 1.4\,GHz to detect cosmologically important galaxies at early epochs.

\begin{figure}[tch]
\centering
\includegraphics[width=7cm]{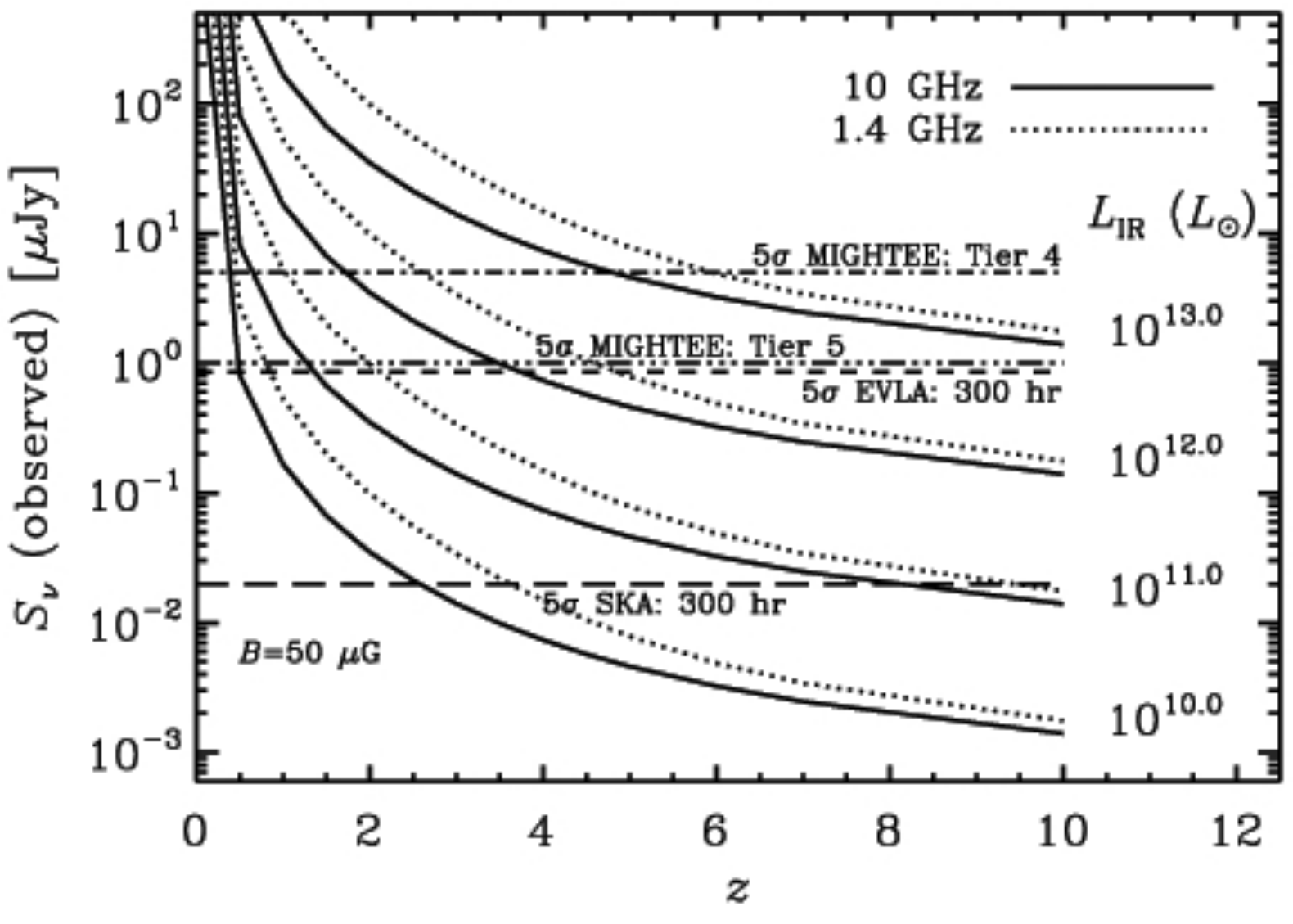}
\caption{
The expected 10\,GHz (solid line) and 1.4\,GHz (dotted line) flux densities for galaxies of different IR luminosities assuming an intrinsic magnetic field strength of 50\,$\mu$G.  
The depths of possible future surveys taken by the VLA, MeerKAT (MIGHTEE), and SKA are shown.  
Since the non-thermal emission from star-forming galaxies should be suppressed by increased IC scattering of CR electrons off of the CMB, the discrepancy between the point source sensitivity requirements of surveys at 1.4\,GHz and 10\,GHz falls below a factor of
$\sim~2$ by $z\gtrsim4$.  
}
\label{comp}
\end{figure}

This is shown in Figure~\ref{comp} where we show the expected 10\,GHz and 1.4\,GHz flux densities from galaxies having a range in IR ($8-1000$\,$\mu$m) luminosities and an intrinsic magnetic field of 50\,$\mu$G.    
This Figure essentially shows the selection function of radio surveys to star forming galaxy populations as a function of redshift.  
While higher frequency surveys need to be much more sensitive to detect the same luminosity class of galaxies at 1.4\,GHz, this becomes less of a problem at higher redshifts.  
For example, by $z\gtrsim4$, the sensitivity requirement of a 10\,GHz survey is less than a factor of $\sim~2$ deeper than a corresponding 1.4\,GHz survey to detect the same population of star forming systems.  
For an intrinsic magnetic field of 10\,$\mu$G, this small discrepancy between point source sensitivity requirements for
10\,GHz and 1.4\,GHz surveys occurs at a $z\gtrsim2$.  
Given that high-frequency observations are almost as sensitive to high-$z$ star-forming galaxies as low-frequency observations, and that higher frequency observations provide a much more robust measure of star formation activity, surveys at $\nu \gtrsim 10$\,GHz will be invaluable for accurately tracing the star formation history of the Universe.       

Figure~\ref{comp}  also shows the expected depths from potential continuum surveys using the VLA, MeerKAT, and SKA.  
The tentative Tier 4 and 5 steps of the MIGHTEE survey plan to map 0.25 and 0.01 square degrees down to an RMS of 1 and 0.2\,
$\mu$Jy at 12\,GHz, respectively. Although the precise area on the sky has not yet been defined, obvious targets are the Chandra Deep Field South and COSMOS fields.  
While the depth of a megasecond exposure is shown for the VLA, the area of such a survey would be restricted by the small primary beam at 10\,GHz ($\theta_{\rm PB} \approx 4-5 \arcmin$).  
However, VLA surveys reaching $\mu$Jy depths for entire fields such as GOODS-N at 10 GHz have already been proposed, allowing synergy with future ALMA observations.  
Clearly, the SKA will revolutionize any such high frequency continuum surveys using either the VLA or MeerKAT by being $\gtrsim$2 orders of magnitude deeper for the same integration times.

\subsection{Evolution of AGNs}

\subsubsection{The AGN component in Deep Radio Fields}

Radio source counts at  
$\mu$Jy levels are dominated by star-forming galaxies, but their contribution decreases  rapidly with increasing flux density \citep[e.g.][]{Seymour08}, and AGNs contribute significantly at radio fluxes below 1\,mJy
 \citep[e.g.][]{Gru99, Geo99, Mag00, Pra01b, Afonso06, Norris06, Mignano08, Smolcic08, Padovani09}. Recent estimates of their relative contributions are summarised in Figure~\ref{SFfraction}.

Models of the sub-mJy radio population \citep[e.g.][]{Wilman08} therefore include three main components:
\begin{itemize}
\item Star-forming galaxies,
\item The extrapolation to low flux densities of the classical radio-loud AGN population (radio galaxies and radio-QSO),
\item A radio-quiet (or low-luminosity) AGN component.
\end{itemize}

The unexpected presence of large numbers of AGN--type sources at sub-mJy levels
implies a large population of low/intermediate power AGNs, which  may have important implications for
the black-hole--accretion history of the Universe.
It will be particularly important to determine the accretion mode (see \S\ \ref{modes}) of these low-power AGNs.

Radio and optical studies of the
Australia Telescope ESO Slice Project (ATESP) sub-mJy sample \citep{Pra06, Mignano08}
have shown that at fluxes
$0.4<S<1$ mJy a prominent class of low-luminosity ($P<10^{24}$ W/Hz) AGN
associated with early-type galaxies appears. These sources have
typically flat or inverted radio spectra ($\alpha > -0.5$) and compact
sizes  ($d < 10-30$ kpc). This source class produces a flattening of the average spectral index at flux densities around 0.5-1 mJy. A similar flattening has been found also in deep
low-frequency (350 MHz) surveys \citep{Owen09,  Pra11}.
Radio spectral index studies \citep[e.g.][]{Pra10} suggest that this
class of objects is presumably associated with core-dominated
self-absorbed FRI-like radio galaxies, perhaps representing a low luminosity extension of the classical
radio-loud radio galaxy population, which dominates at flux densities
$S>>1$ mJy.
On the other hand, deeper radio fields have shown that flat/inverted radio
spectra become less frequent  at lower flux densities ($100<S<400 \mu$Jy) \citep{Owen09,  Pra09}, where steep-spectrum star-forming galaxies and radio-quiet AGN are increasingly important.

Our current understanding of the AGN component in deep radio fields can be summarised as follows:
\begin{itemize}
\item above  flux densities $S\sim 400\,\mu$Jy there is no clear evidence for a radio-quiet AGN component;
\item sub-mJy radio-loud AGN are mostly low-power, compact, (self-absorbed) jet-dominated systems, and seem to be  a low power counterpart of the classical bright radio galaxy population;
\item at lower flux densities radio-quiet AGN become increasingly important and at  $S \sim 100\,\mu$Jy they
account for about 45\% of the overall AGN component;
\item radio-quiet AGN show radio/optical/IR properties consistent with  
radio follow-up of optically
selected radio-quiet AGN \citep{Kuk98}, compact radio sizes, steep radio spectra,
luminosities of $P<10^{24}$ W Hz$^{-1}$, and Seyfert-2 optical spectra and/or mid-IR colours.
\end{itemize}

\begin{figure}
\begin{center}
\includegraphics[width=7cm]{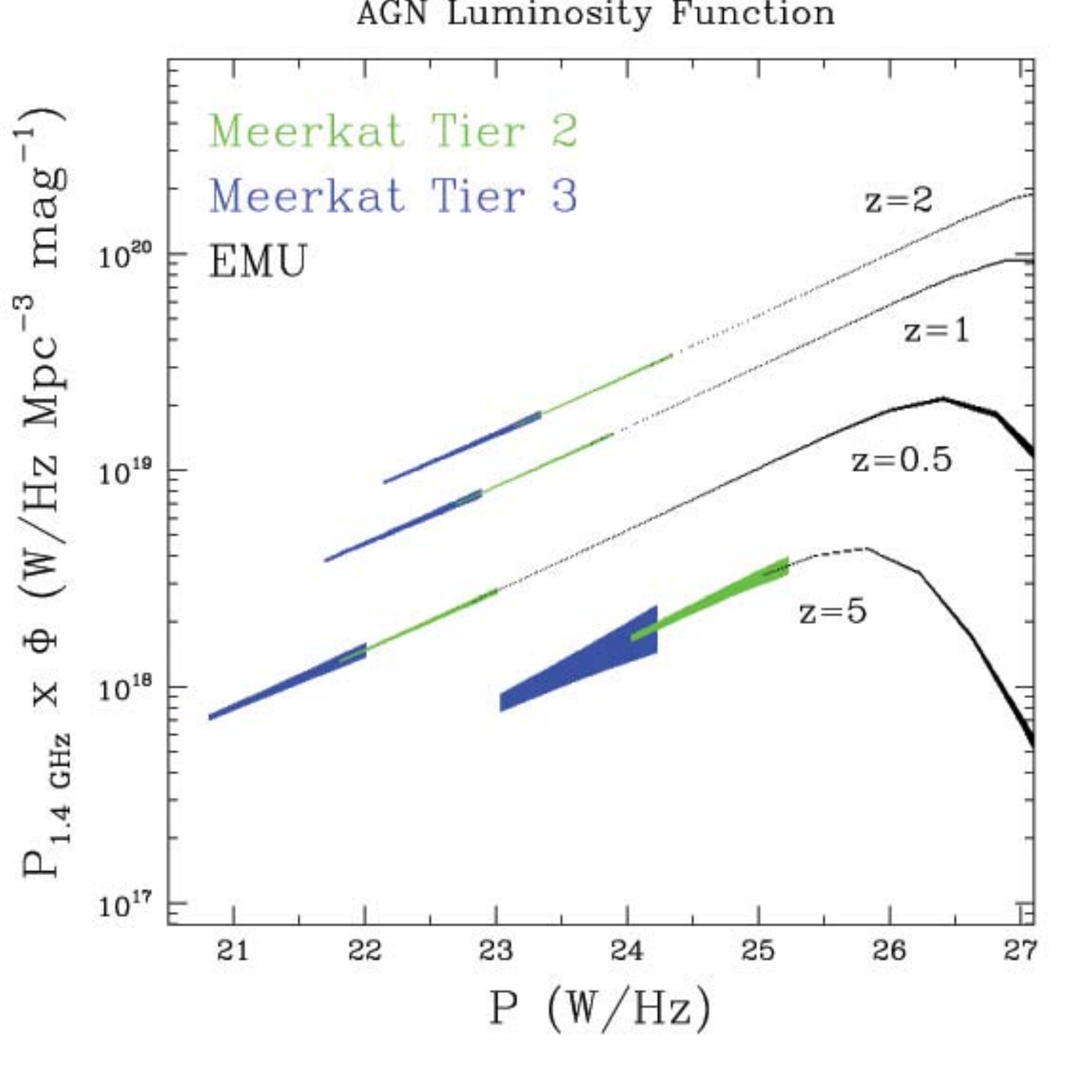}
\caption[]{AGN luminosity function at different redshifts expected from the combination of
ASKAP and Meerkat deep surveys (based on models described by
\citet{Prandoni07}.
\label{fig:flum}}
\end{center}
\end{figure}

Multiwavelength analyses of larger and deeper radio samples are needed to confirm
these results, and to obtain quantitative constraints for evolutionary modelling of faint AGN.
We can expect this field to move forward significantly in the next years, thanks to the combination of
wide-field and all-sky surveys (sampling large local volumes) and very deep fields (sampling low
powers at  high redshifts) planned with the facilities described in this paper.
An example of what can be obtained is illustrated in Figure~\ref{fig:flum}, where we show the AGN luminosity function at different redshifts that can be derived from the
combination of EMU and MIGHTEE.  

\subsubsection{Accretion Modes}
\label{modes}
AGN activity, and associated black-hole growth, occurs in at least two
different modes, each of which may have an associated feedback effect
upon the AGN host galaxy:

\begin{itemize}
\item[(i)] a fast accretion mode (often referred to as cold-mode accretion or quasar-mode accretion)
associated with quasars \citep{Silk98}; this radiatively--efficient accretion mode may be important
  in curtailing star formation at high redshifts and setting up the tight
  relationship between black hole and bulge masses observed in the nearby
  Universe \citep[e.g.][]{Magorrian98}.
\item[(ii)] a radiatively--inefficient slow accretion mode
(often referred to as hot-mode accretion or radio-mode accretion) \citep{Croton06,Bower06},
the observational manifestation of\\
  which
   is low-luminosity radio sources; this mode is
  thought to be
  responsible for maintaining elliptical galaxies at lower redshifts as
  ``old, red and dead" \citep[e.g.][]{Best06} and for preventing strong
  cooling flows in galaxy clusters \citep[e.g.][]{fabian03}.
\end{itemize}

One of the most fundamental issues in understanding the role of AGN in
galaxy formation is the need to measure accurately the cosmic evolution of
quasar activity and the accretion history of the Universe, and to compare
this with the build-up of the stellar populations of galaxies. Do black
holes and their host galaxies grow coevally, or does one precede the
other? What is the primary mode of black hole growth?

Much of the growth of black holes is believed to occur in an obscured
phase, and these ``Type-2" AGN are difficult to identify. Even the deepest
current X-ray observations do not detect the most heavily absorbed sources
which Cosmic X-ray Background synthesis models predict exist in abundance
\citep[e.g.][]{Gilli07}, implying that much high-$z$ quasar activity has
yet to be detected directly. These ``radio-quiet" quasars are not
radio-silent, and their radio luminosity distribution peaks at about
$L_{\rm 1.4GHz} \approx 10^{23}$\,W\,Hz$^{-1}$ \citep[e.g.][]{Cirasuolo03}.
Deep radio surveys with the SKA pathfinders therefore offer
an alternative route to identifying these distant AGN, in a manner
unbiased by dust and gas absorption at other wavelengths. Combined with
multiwavelength datasets to separate source populations, the relative
contribution of radio-quiet AGN to the faint radio source population can
be determined. This will enable investigation of the dependence of the
fraction of obscured AGN on luminosity and cosmic epoch, and thus the
history of radiatively-efficient accretion in the Universe to be
determined.

The deep and wide radio surveys will also allow study of the role of
low-luminosity radio sources in galaxy evolution. The low-luminosity
radio-AGN population is dominated by a population of sources in which
there is little evidence for radiative emission from an accretion disk,
and the bulk of the accretion power is channelled into the expanding radio
jets \citep[e.g.\ ][]{Merloni07,Hardcastle07,Best12}. These jets pump energy into their environments, inflating
cavities and bubbles in the surrounding inter-galactic and intra-cluster
medium, from which estimates of the mechanical energy associated with the
jets can be made \citep[e.g.\ ][]{Cavagnolo10}. Studies of radio-AGN in
the nearby Universe have shown that the fraction of galaxies that host
radio--loud AGN (with $L_{\rm 1.4GHz} > 10^{23}$\,W\,Hz$^{-1}$) is a strong
function of stellar mass  \citep[e.g.\ ][]{Best05}, and that the
time-averaged energetic output associated with recurrent radio source
activity may indeed be sufficient to control the rate of growth of massive
galaxies. However, sensitivity limits of current large-area radio surveys
mean that these low luminosity sources are only observed in the nearby
Universe. How does the relation between galaxy mass and radio-AGN fraction
(the radio source duty cycle) evolve with redshift out to the peak epoch
of galaxy formation? Out to what redshift does radio-AGN-heating continue
to balance cooling? What is the differential evolution of the radiatively
efficient and inefficient accretion modes  \citep[c.f.\ ][]{Best12}, and is
evidence for ``down-sizing" in the radio source population  \citep[e.g.\ ][]{Rigby11}
confirmed? What drives these processes? These are all
questions for the next generation radio arrays, coupled with high-quality
multiwavelength datasets. Large-area surveys are required in order to
study a sufficient volume to include the full range of galaxy environments,
given the very important role that large-scale environment can play in the
evolution of galaxies.

\subsubsection{Ultra-steep spectrum radio sources below the mJy barrier}
\label{sect:uss}
Another exciting prospect for the upcoming generation of radio continuum surveys is that of finding complete populations of high-redshift radio galaxies (HzRGs). These are among the most luminous galaxies and seem to be associated with the most massive systems \citep[e.g.][]{vanBreugel99,Jarvis01a,Willott03,Rocca04,DeBreuck05,Seymour07} often going through a phase of violent star formation at at the 1000\,M$_{\sun}$\,yr$^{-1}$ level and showing large gas and dust reservoirs
(e.g. \citealt{Dunlop94,Ivison95,Hughes97,Ivison98,Papadopoulos00,Archibald01,Klamer05,Reuland03,Reuland04,Reuland07}; but see \citealt{Rawlings04}). HzRGs are considered to be the progenitors of the brightest cluster ellipticals, and have been used as beacons to identify overdensities in the distant universe, i.e., proto-cluster environments at
${z \sim 2.5}$ \citep[e.g.][]{Stevens03,Venemans07}. Identifying and tracing the evolution of HzRGs offers a unique path to study galaxy and large-scale structure formation and evolution from the earliest epochs, and extensive studies will finally become possible over the next few years.

Until recently, searches for HzRGs have been limited to wide-area (and thus relatively shallow) surveys. This is appropriate as the luminosities of HzRGs are large enough to make them detectable to very high-redshifts (beyond $z\sim 5$) at flux densities of tens or even hundreds of mJy. For example,  TN J0924-2201, the most distant known radio galaxy at $z = 5.2$, exhibits a 1.4\,GHz flux of over 70\,mJy \citep{DeBreuck00,vanBreugel99}. But understanding a population, its origin and evolution and how it fits in the overall galaxy population, requires more than just the detection and study of the most extreme objects. The selection of more complete samples, over a wider redshift range, requires deeper surveys over wide areas. However, the large population of low-$z$ ($z\lesssim 1-2$) SF galaxies at sub-mJy  levels presents a considerable difficulty for studies of HzRGs, as efficient ways of distinguishing between AGN and SF are then necessary.

One of the most successful tracers of HzRGs relies on the relation between the  radio spectral index and redshift \citep[e.g.][]{Tielens79,Chambers96}. Although an ultra-steep (radio) spectrum (USS; $\alpha\lesssim -1$) is not a necessary condition for a high redshifts, and, in fact, the USS selection misses a possibly large fraction of HzRGs \citep[e.g.][]{Waddington99,Jarvis01b,Jarvis09, Schmidt06}, a higher fraction of high-redshift sources can be found among those with the steepest radio spectra. The USS criterion is so effective that most of the radio galaxies known at z $>$ 3.5 have been found using it \citep[see][for an overview of the selection of HzRGs]{Miley08}. Surprisingly, there is still no satisfactory explanation for this spectral-index--redshift correlation, as neither (a)~a combination of an increased spectral curvature with redshift and the redshifting of a concave radio spectrum\citep[e.g.,][]{Krolik91} to lower radio frequencies nor
(b)~radio jets expanding in dense environments, more frequently observed in distant proto-clusters, which would appear with  steeper spectral indices \citep{Klamer06,Bryant09,Bornancini10} seem to explain the observations.

With the next generation of radio continuum surveys, the search for USS HzRGs will be extended to the whole sky and to microJansky flux levels. Although such collections have been assembled before \citep[e.g.][]{Bondi07,Owen09,Afonso09,Ibar09} only very recently has it been possible to characterise them. Using deep optical and infrared imaging of the Lockman Hole, \citet{Afonso11} have started a detailed analysis of the $\mu$Jy USS radio source population, something that will be possible over a significant fraction of the sky using radio surveys such as EMU, WODAN and LOFAR.

\begin{figure}
\scalebox{0.8}{
\rotatebox{0}{
\includegraphics[width=7cm]{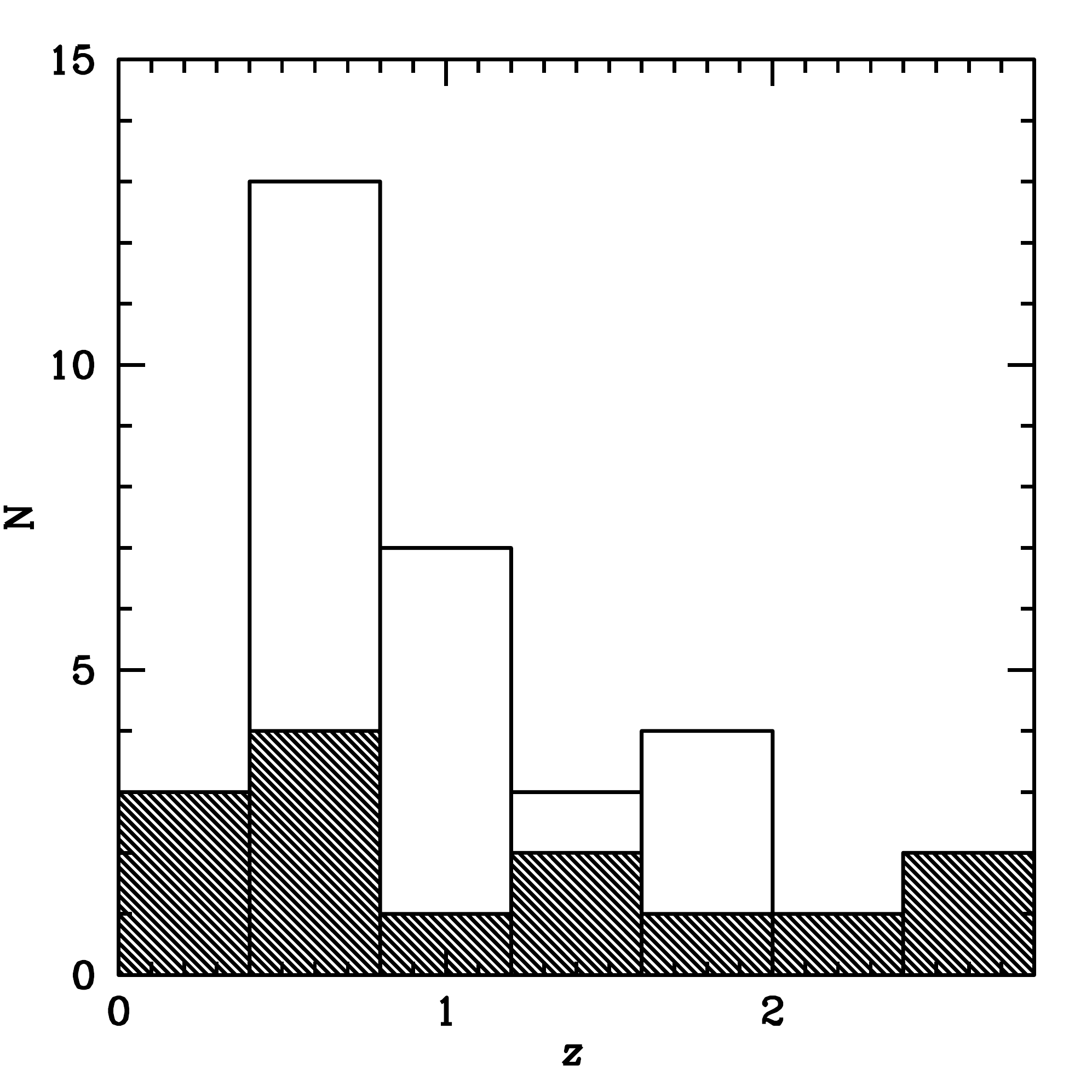}
}}
\caption{Redshift distribution for radio-faint USS sources in the Lockman Hole, from \citet{Afonso11}. Filled histogram denotes sources with a spectroscopic redshift determination, while the open region refers to photometric redshift estimates. A further 25 USS sources (43\% of the full sample) exist but with no redshift estimate, mostly at fainter $3.6\,\mu$m fluxes and likely to be found at higher redshifts.
\label{fig:z-distr}}
\end{figure}

This first detailed analysis of the USS $\mu$Jy population used the deep multiwavelength coverage of the Lockman Hole. The sample selection was made using ultra deep VLA 1.4\,GHz (reaching 6\,$\mu$Jy rms) and GMRT 610\,MHz data (15\,$\mu$Jy rms), and resulted in 58 sources with a spectral index $\alpha^{1400}_{610}\leq -1.3$. The deep IR coverage at 3.6\,$\mu$m and 4.5\,$\mu$m provided by the Spitzer Extragalactic Representative Volume Survey \citep[SERVS: ][]{Mauduit12} resulted in an extremely high identification rate, above the 80\% level. Mid-infrared colours are compatible with a significant AGN presence among the radio-faint USS population. Spectroscopic redshifts for 14 sources and photometric redshifts for a further 19 sources were used for the redshift distribution of these sources, ranging from $z\sim 0.1$ to $z\sim 2.8$ and peaking at $z\sim 0.6$ (see Figure~\ref{fig:z-distr}). Twenty-five sources have no redshift estimate, including the faintest sources at infrared wavelengths, which indicates that higher redshifts are likely in this population. In this respect they may be similar to the Infrared-Faint Radio Sources (IFRS), first identified by \citet{Norris06}, and which appear to be very high redshift dusty radio galaxies \citep{Garn08, Huynh10, Middelberg08b,Middelberg11, Norris07b, Norris11a, Cameron11, Zinn11}.

\begin{figure}
\scalebox{0.8}{
\rotatebox{0}{
\includegraphics[width=7cm]{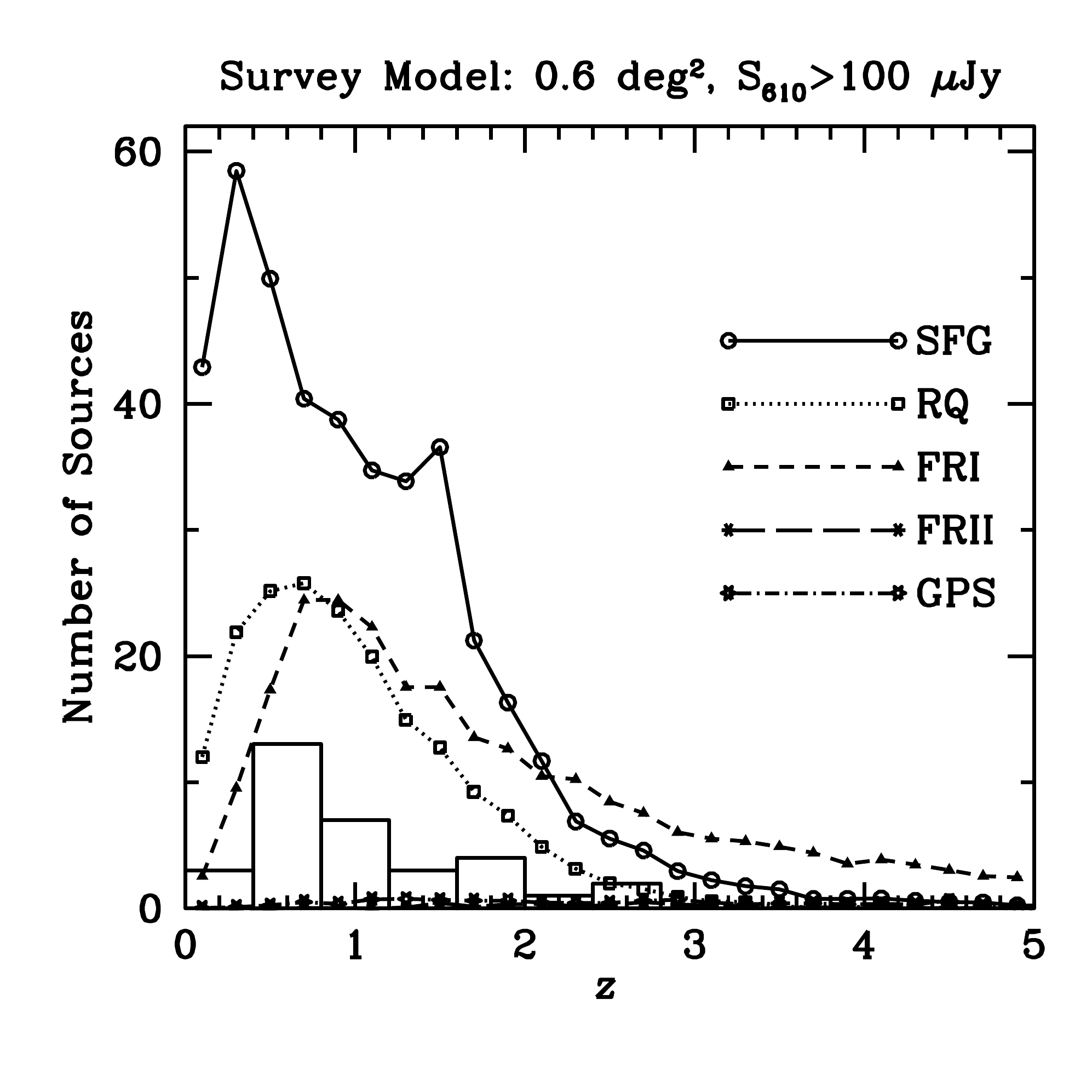}
}}
\caption{Predictions from the SKADS Simulated Skies models for the redshift distributions of radio source populations, irrespective of their radio spectral indices, for a radio survey reaching a detection sensitivity of 100\,$\mu$Jy at 610\,MHz over 0.6 square degree, similar to the Lockman Hole radio survey considered in \citet{Afonso11}. The observed redshift distribution for USS sources in that work is also displayed.
\label{fig:models}}
\end{figure}

A comparison with the SKADS Simulated Skies models (Figure~\ref{fig:models}) indicates that FRIs and RQ AGNs may constitute the bulk of the USS-population at $\mu$Jy radio flux densities (FRIIs are thought to dominate above the mJy level). According to the models, the USS technique will be as efficient for the selection of very high-$z$ radio sources at $\mu$Jy fluxes as when applied at much higher flux density levels. This raises exciting prospects for the next generation of wide-field radio surveys, as it increases the potential for detailed studies of sources at very high redshifts and understanding of the evolution of the AGN activity in the Universe.

These prospects are strengthened by the recent work of \citet{ker12} who have shown that redshift and spectral index are correlated in spectroscopically complete samples, and argue that this is due both to the dense intergalactic medium and the early evolutionary stage of the AGN, although they, together with \citet{ Zinn12a}, also show that the redshift-spectral-index correlation is weaker than other potential redshift indicators.

\subsubsection{The luminosity--dependent evolution of the radio luminosity function}
Section \ref{modes} described how AGN radio jets interact with the surrounding
intergalactic and intracluster medium to prevent
both large--scale cluster cooling flows and the continued growth of
massive ellipticals \citep[e.g.][]{fabian06,Best06,best07,Croton06,Bower06}.  Determining the evolution of the radio luminosity function (RLF) is
therefore important for understanding the timescales on which they
impose these effects.

A key early study of radio-loud AGN by \citet{Dunlop90} found increases of two to three orders of magnitude at a redshift of two, compared with the local Universe, in the comoving number density of both flat and steep spectrum radio-loud sources. They also saw indications of a high redshift ($z \sim 2.5$) decline in density, but their work, and that of subsequent studies in this area \citep[e.g.][]{shaver96,Jarvis01b,waddington01} has been limited by the inability to probe the depth and volume necessary to probe the high-redshift behaviour. This has motivated the development of the Combined EIS-NVSS Survey of Radio Sources \citep[CENSORS:][]{Best03}, which has been designed to maximise the information for high-redshift, steep spectrum sources close to the break in the RLF.  CENSORS is a 1.4\,GHz selected sample, and contains 135 sources complete to a flux density of 7.2\,mJy. It is currently 78\% spectroscopically complete \citep{ker12}, with the remaining redshifts estimated via either the $K-z$ or $I-z$ magnitude-redshift relations.  

\begin{figure}
\centering
\includegraphics[width=7cm]{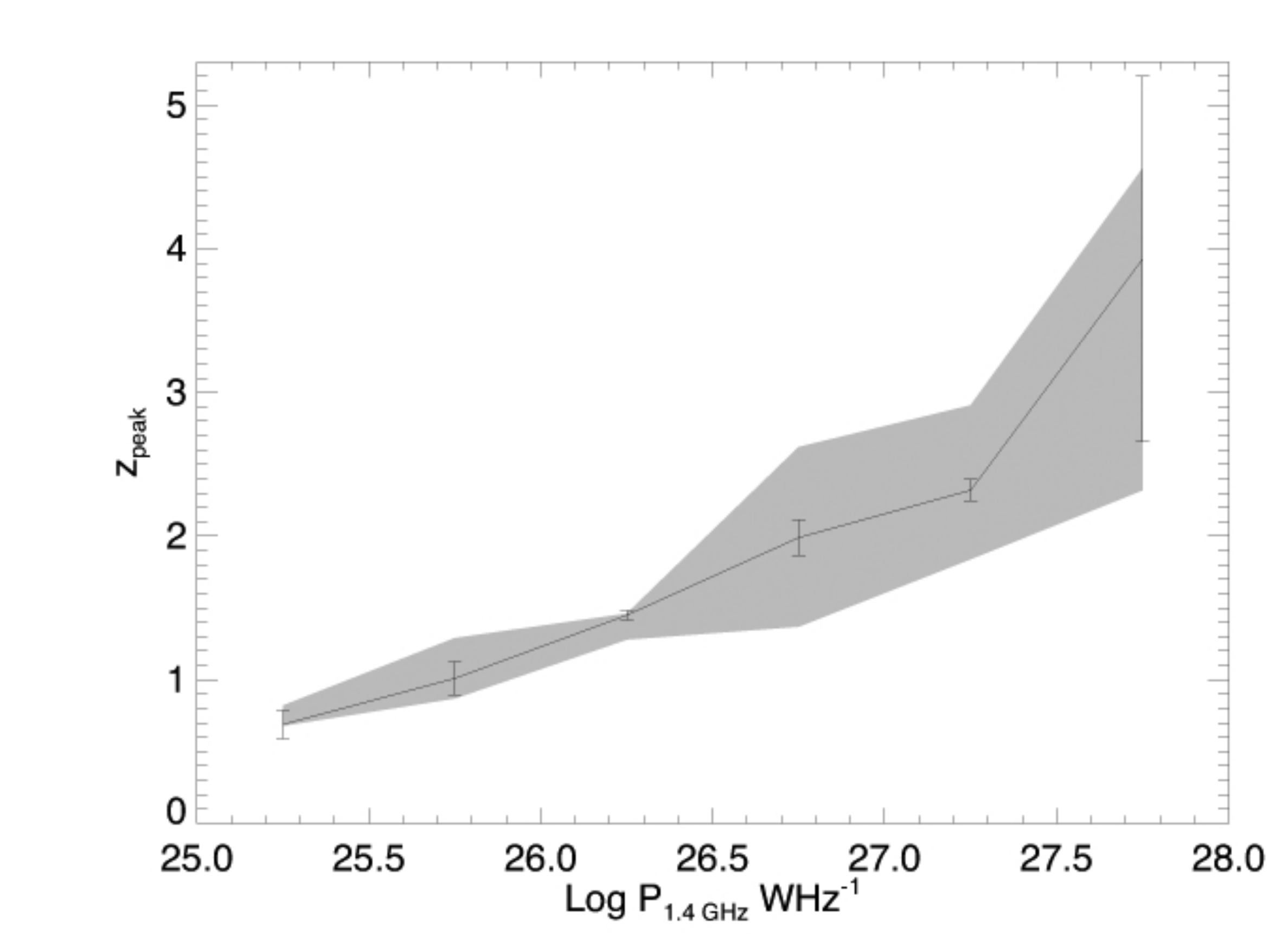}
\caption{\protect\label{zpeaks}
The dependence of the redshift of the peak in space density ($z_{\rm
  peak}$) on radio power for the best--fitting steep spectrum grid, showing that the
numbers of the most powerful radio sources peak at the highest redshifts \citep{Rigby11}.
The shaded region and error bars give two different determinations of the spread
in the measurements.
}
\end{figure}

The CENSORS sample, together with additional radio data, source counts and local RLF, is used to investigate the evolution of the steep spectrum luminosity function via a new grid-based modelling method in which no assumptions are made about the high-redshift behaviour \citep{Rigby11}.
Instead, the RLF variations are determined by allowing the space densities at various points on a grid of radio luminosities and redshifts to each be free parameters, and then simply finding the
best--fitting values to this many-dimensional problem.
The modelling finds conclusive evidence, at $>3\sigma$ significance, for a luminosity--dependent high--redshift decline in space density for the steep-spectrum sources. At low radio powers ($P_{\rm 1.4GHz} = 10^{25-26}$\,W/Hz) the space densities peak at $z \gsim 1$, but move to higher redshift for the higher powers ($z \gsim 3$ for $P_{\rm 1.4 GHz} > 10^{27}$\,W/Hz). This is illustrated in Figure~\ref{zpeaks}, and is similar to the ``cosmic downsizing" seen for other AGN populations \citep[e.g.][]{zotti09,wall08,hasinger05,richards05}. These results are robust to the estimated redshift errors and to variation in the radio spectral index with redshift.

The evolution of low-luminosity radio AGN is less well understood than that of radio-loud AGN, with some studies finding no evidence for any evolution of the RLF for low-luminosity radio AGN \citep[e.g.\ ][]{Clewley04} and others finding that low-luminosity AGN do evolve with redshift, although more slowly (less than a factor of 10 from $z=0$ to $z=1.2$) than their high-luminosity counterparts \citep{smolcic09, mcalpine11}. \citet{Best12} suggest that the luminosity dependence of the evolution of the AGN RLF may be attributed to the varying fractions of hot and cold mode sources with redshift. Recent results \citep{Mauch07,sadler07,Padovani11,Mao12} fail to resolve this discrepancy, with ambiguity added by factors such as  evolution, cosmic variance, differences in classification, differences in terminology, and the difficulty of distinguishing low-luminosity AGN (LLAGN) from SF galaxies, or the bulge from the disk of a galaxy. Next-generation radio surveys using SKA pathfinders can resolve this discrepancy provided we develop reliable techniques for distinguishing the AGN component from the SF component of a galaxy.

\subsubsection{Environment}
Studies of galaxies in groups and clusters by \citet{best07, wake08} have shown that the environment of a radio galaxy has a significant effect on the radio emission. \citet{Sabater08} and \citet{Sabater12} built on this result  by comparing (a) the radio AGN activity
in a sample of isolated
galaxies \citep[AMIGA sample; ][]{Verdes05} with that of a matched sample of galaxies
in clusters \citep{Miller01,Reddy04} and, (b) the
optical AGN activity for the AMIGA sample with that of a matched sample of
galaxies in compact groups \citep{Martinez10}.
The results of all these studies show that radio AGN
activity is strongly influenced by the environment while this influence is
less clear for optical AGN  activity \citep{Miller03}.

\subsubsection{Black Hole Spin}

A further area of study from radio observations is the study of black hole physics.
For a given accretion rate, as traced by the X-ray to mid-infrared SED, AGN show a huge range of radio luminosities. This suggests that some extra parameter, independent of the accretion rate, is controlling the jet power of the AGN. The spin of the black hole is a very attractive candidate, for both observational and theoretical reasons \citep[e.g.\ ][]{Wilson95}.
Spin is effectively a hidden parameter, which will barely show itself in the SED of the AGN. In addition,  spinning black holes can convert rotational energy into jet power, meaning that jets can be powered with nominal efficiencies close to or exceeding unity. Observations of the most powerful radio AGN suggests the jet power is comparable or larger than the rate of accretion of energy \citep[e.g.\ ][]{Fernandes11}.

Radio observations allow us to estimate the jet power, and combining these with observations at X-ray energies or optical wavelengths can allow an estimate of the accretion rate and spin. For example, \citet{Martinez11}
consider the jet powers of supermassive black holes (SMBHs) with m$_{\bullet}$$\geq$$10^{8}$~M$_{\sun}$. The authors use a suite of semi-analytic models and GR (general relativistic) MHD (magneto-hydrodynamic) simulations that predict the jet efficiencies as a function of spin, and find that these efficiencies can approximately explain the radio-loudness of optically-bright quasars.   

\citet{Martinez11} also find that they can explain the local RLF of high- and low-excitation galaxies independently (HEGs and LEGs, respectively), and that the HEGs and LEGs have different best-fitting spin distributions. The LEGs, modelled as ADAFs (advection-dominated accretion flows) or low-accretion rate SMBHs, have a bimodal spin distribution, with the SMBHs having typically low ($\hat{a}\sim0$) or high ($\hat{a}\sim1$) spins. The probability density of the high-spin peak is significant, typically $\sim$33\% of the total probability density.  On the other hand, the HEGs, modelled as SMBHs accreting at a high fraction of their Eddington-rate, are found to have typically low spins ($\hat{a}\sim0$ only), with a small or non-existent component of high spin.  
The radio LFs of HEGs and LEGs given by these spin distributions can be extrapolated to high redshift, where they correctly reproduce the total radio LF of all AGN at $z=1$. Due to the strong evolution of the high-Eddington rate SMBHs (as reflected by the evolution of the X-ray LF), this modelling also predicts that the  $z=1$ radio sources above a radio luminosity density $\sim$$10^{26}$\,WHz$^{-1}$ sr$^{-1}$ (at 151\,MHz) should show exclusively HEG spectra, while the fraction of LEGs should increase rapidly at lower radio luminosity densities.

\begin{figure}
\includegraphics[width=7cm, angle=0]{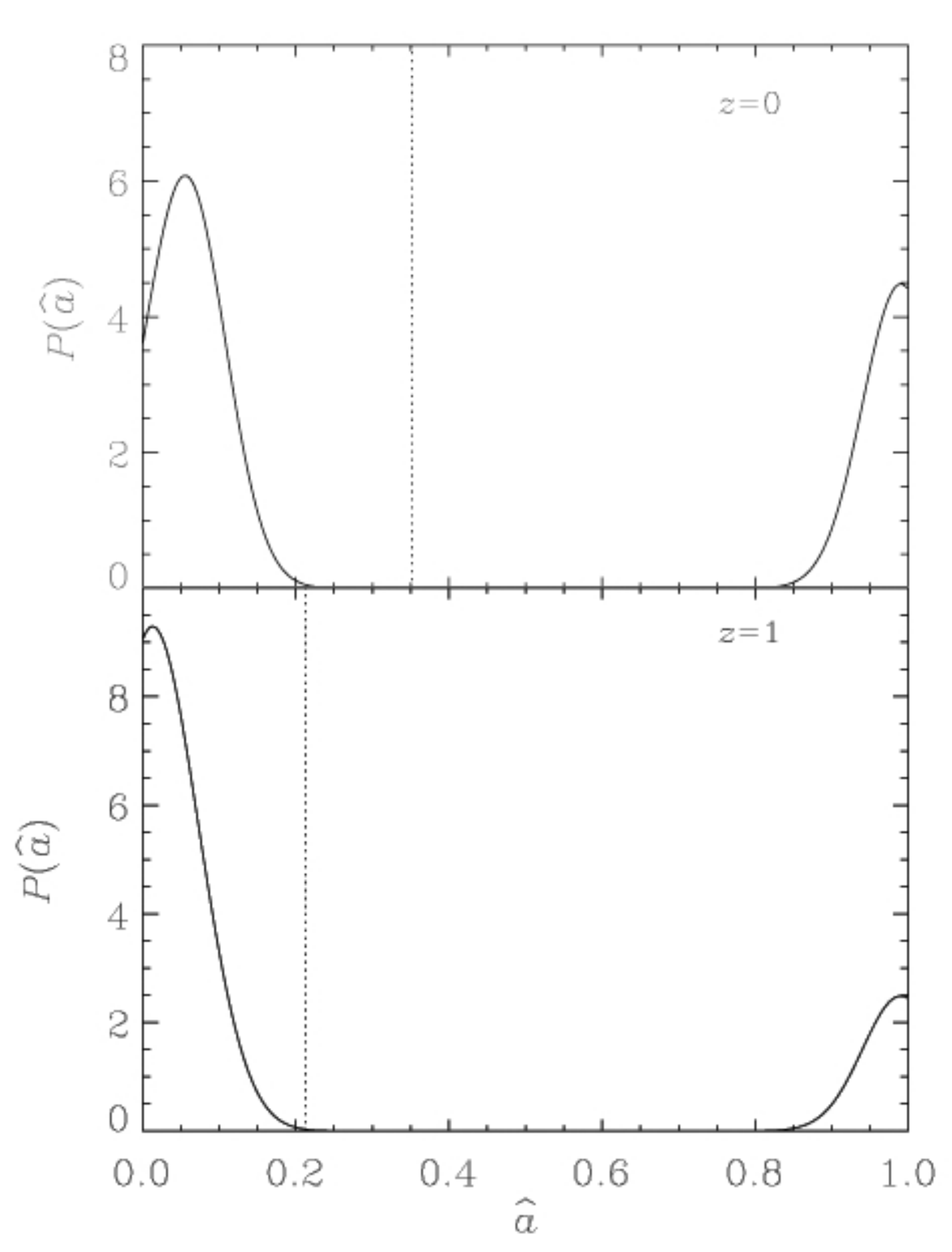}  
\caption{Inferred spin distributions for SMBHs at $z=0$ and $z=1$, from \citet{Martinez11}.
The vertical dashed lines mark the mean value at each redshift.
(Bottom) The spin distribution at $z=1$ is dominated by the low-spin population, with a minor component of high-spin SMBHs. (Top) At  $z=0$ a larger fraction of SMBHs possess a high spin, so that the  mean spin is higher. }
\label{fig:spin_dbns}
\end{figure}

Since the space density of HEGs increases rapidly, the typical spin of SMBHs will change with redshift. At $z\sim 0$, most of the SMBHs will have the bimodal distribution inferred for the LEGs, with a high fraction of SMBHs having a high spin (Figure~\ref{fig:spin_dbns},top). At $z\grtsim 1$ the space density of HEG radio sources is larger than that of LEG sources, so that the SMBHs will typically have the distribution of spins of the HEGs, dominated by a low-spin component
(Figure~\ref{fig:spin_dbns}, bottom).

\begin{figure}

\includegraphics[width=7cm, angle=0]{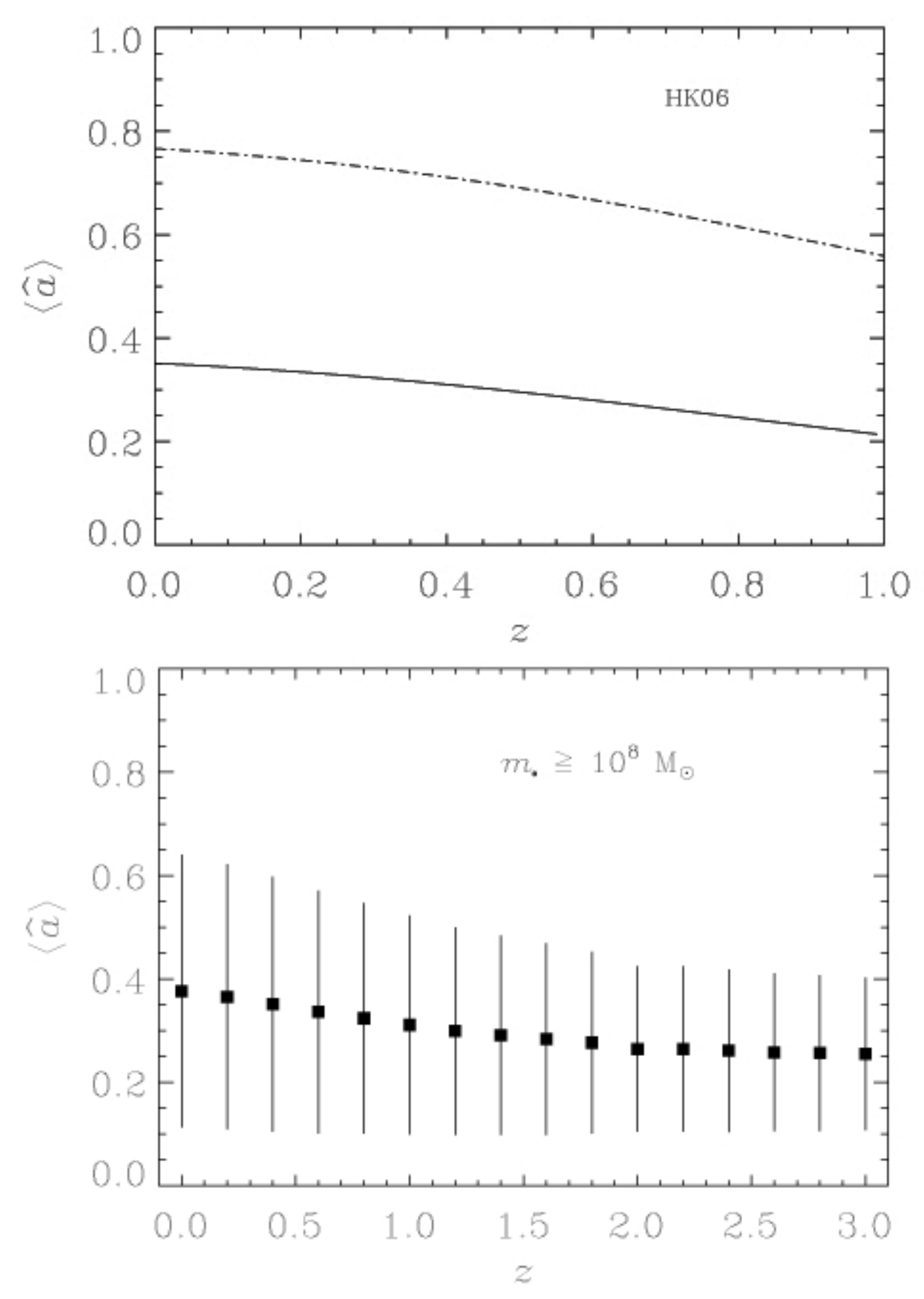}  
\caption{ (Top) Mean spin of SMBHs  with $m_{\bullet} \geq 10^{8}$\,M$_{\odot}$, from \citet{Martinez11}. There is a gradual increase of the mean spin between $z\grtsim 1$ and $z=0$, which is most likely due to a fraction of SMBHs undergoing major mergers and subsequent chaotic accretion.
(Bottom) Cosmological simulation of the mean spin from \citet{fanidakis}, which includes chaotic accretion and mergers, and which is in excellent agreement with the inference based on radio observations (top).}
\label{fig:spin_his}
\end{figure}

The evolution of the spin from SMBHs (with\\
 m$_{\bullet}$$\geq$$10^{8}$~M$_{\odot}$) can hence be described as evolving from predominantly low spins in high-Eddington rate systems at $z\grtsim 1$, to a bimodal spin distribution at $z\sim 0$, for mostly low-Eddington rate systems. At the epoch of highest accretion, the typical spin is low and at later epochs the fraction of SMBHs with high spins increases (Figure~\ref{fig:spin_his}).
This can be understood in terms of chaotic accretion \citep[e.g.\ ][]{King08} spinning the accreting SMBHs down, whereas major mergers spin the SMBHs up \citep[e.g.][]{Rezzolla08}. At earlier epochs, there is a plentiful supply of cold gas, so that accretion spins SMBHs down. As this gas disappears, however, the braking mechanism that spins SMBHs down is removed, while some SMBHs will still experience major mergers. Hence a population of high-spin SMBHs gradually appears.

\subsection{Galaxy Clusters and Large-Scale Structure}
\label{clusters}

Clusters of galaxies are the most massive systems in the present Universe.
They are not isolated regions, but evolving regions at the intersections
of filaments and sheets in the large-scale structure.


Tens of thousands of clusters are currently known, but only a few are known at $z>1$ \citep{Wilson08, Kodama07}, with the highest cluster redshift at $z = 2.07$ \citep{Gobat11}.

Radio observations of galaxy clusters are powerful probes of the physics
of the intra-cluster-medium (ICM), but they also provide a complementary
way to trace the evolution of large-scale structure of the Universe up to
very large distances. These studies also have the potential to explore the
way in which galaxy evolution (star formation, nuclear activity) depends
on its environment.

Radio observations have shown the existence of diffuse large scale
synchrotron radiation from the ICM, implying that non-thermal components,
magnetic fields and relativistic particles, are mixed with the hot ICM.
The origin of these components is still unclear and a subject of lively
debate.
Potentially these components contribute to the energy of the ICM and drive
complex physical processes that may significantly alter our
present (simplified) view of the ICM \citep{Schekochihin05, Subramanian06, Brunetti11}.

Radio emission from clusters of galaxies is generated both by the constituent galaxies and by diffuse emission in the ICM.

The radio emission from constituent galaxies often extends well beyond the
galaxy optical boundaries, and
 interacts with the ICM. This
interaction is  observed in tailed radio galaxies (see \S \ref{tails}),
and radio sources filling X-ray cavities at the centre of cool-core
clusters \citep{McNamara07, Feretti08}.

Diffuse sources are typically grouped into three classes
\citep{Feretti96, Kempner04}:
\begin{itemize}
\item Halos, which are steep-spectrum Mpc-scale diffuse sources in
the central regions
of X-ray luminous clusters \citep[e.g.][]{Ferrari08, Venturi11,Feretti12},
\item Relics, which are located in the peripheral regions of both merging
and relaxed clusters and caused by cluster-cluster shocks,
\item Mini-halos, which  are hosted in relaxed cool-core clusters, are centrally
located, and usually surround a powerful radio galaxy, such as 3C84 in the Perseus cluster.
\end{itemize}

Even amongst these classifications, there is ambiguity, with the term ``relic'' being used to describe both old galaxies and shock accelerated regions of diffuse electrons. Furthermore, some amorphous radio objects, such as the diffuse ``Phoenixes''  \citep{Kempner04} and ``roundish relics'' \citep{Feretti12}, cannot easily be so classified.

There are two ways in which these classifications could be improved. First, we could build in more physics to link observations to
 thermal and non-thermal physical
processes in galaxy clusters, but must
ensure that current
observational/theoretical expectations do not bias our classification.
Second, we could
use a comparison of radio and X-ray surface brightness distributions.

SKA pathfinder surveys will discover hundreds of diffuse radio halos in
regions of the sky where there are no known clusters \citep{Cassano12}  and even more tailed radio galaxies \citep{Mao10, Norris11c}. Radio emission in clusters is therefore likely to become an important tool both for studying clusters themselves, and for detecting large numbers of clusters to study cosmology and trace  large-scale structure formation.

\subsubsection{Radio Halos}

About 40 giant radio halos are known so far in the redshift range
$0\leq z \leq 0.55$. The GMRT Radio Halo Survey \citep{Venturi08}, in combination with the
NVSS and WENSS surveys, allows a statistical exploration of radio halos
in galaxy clusters in the redshift range $0\leq z \leq 0.4$ \citep{Cassano08}.
The cluster radio properties were found to be bimodal: only
a fraction of massive clusters host halos,
while no Mpc-scale radio emission is detected
in the majority of clusters at current sensitivity levels.
The radio luminosity of these ``radio quiet" clusters is less than one tenth
of that of the observed radio halos \citep{Brunetti07b, Brunetti09}.
X-ray follow up of these observations shows
that radio halos are found only in merging clusters while ``radio quiet" clusters are
systematically more relaxed systems \citep{Cassano10a}.

The radio properties of halos correlate with the thermal
properties of the parent clusters and the radio power
of halos increases with X-ray luminosity, suggesting a
direct connection between the thermal and non-thermal
cluster properties
\citep{Feretti99, Bacchi03, Cassano06, Brunetti07b, Brunetti09, Feretti12}.

According to a prominent scenario
for the origin of giant radio halos,
a fraction of the gravitational binding--energy
of Dark Matter halos released during cluster mergers can be channelled into turbulence. This turbulence may in  turn accelerate
relativistic particles \citep{Brunetti01, Petrosian01} explaining the  
connection observed between radio halos and cluster mergers.

\begin{figure}
\includegraphics[width=6cm]{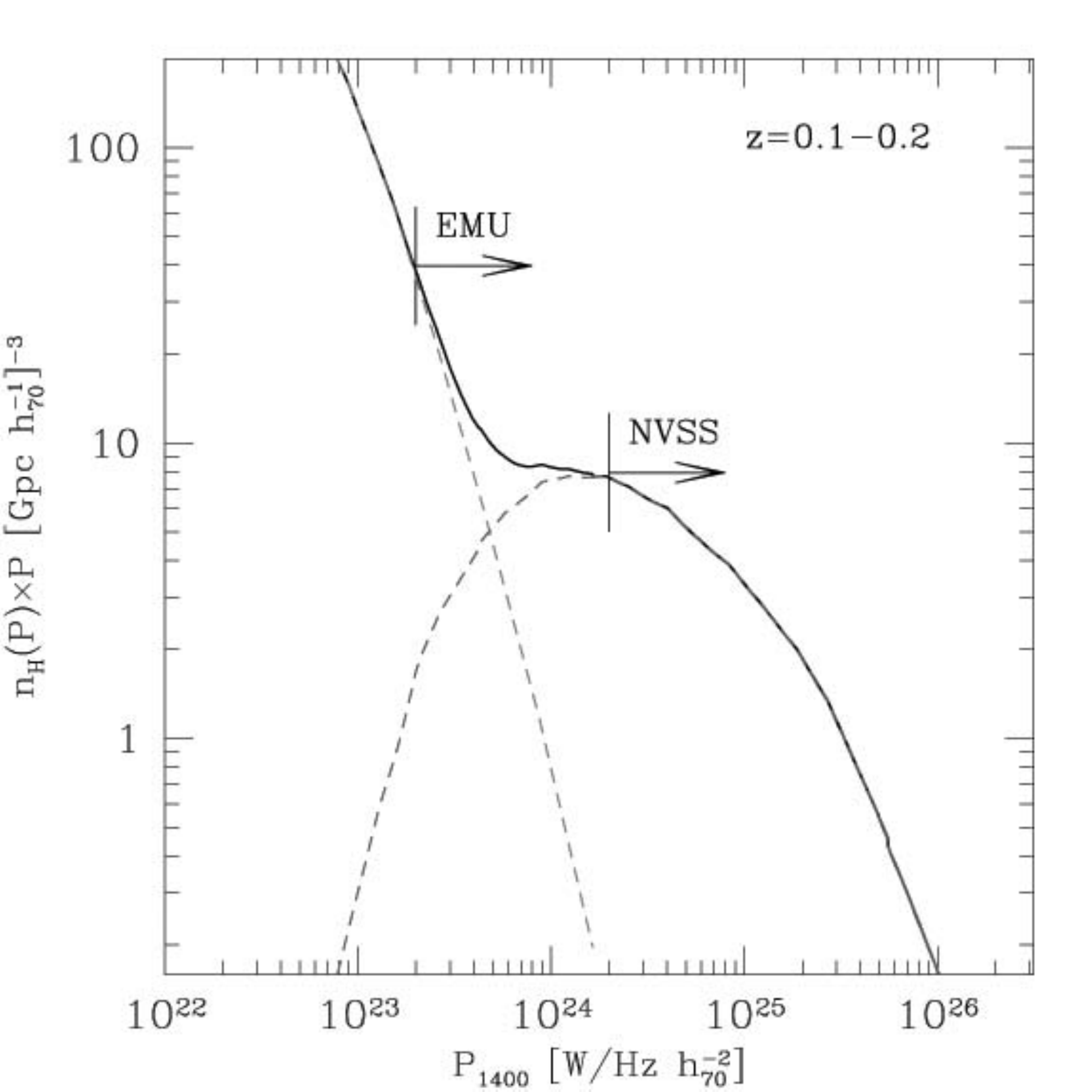}
\caption{
Luminosity functions at 1.4\,GHz of giant radio halos in the redshift range $0.1\leq z \leq 0.2$. The red dashed line marks the contribution from ``hadronic" halos in relaxed clusters, while the blue dashed line marks the contribution from halos in merging ``turbulent") clusters. The black solid line gives the total luminosity functions \citep{Cassano12}.The arrows show the EMU and NVSS sensitivities at that redshift.
}
\label{fig:RHLF_1400}
\end{figure}

The source of the synchrotron electrons in these halos is unclear, as
acceleration of electrons from the thermal pool to relativistic energies
by MHD turbulence in the ICM cannot provide sufficient energy
\citep{Petrosian08}.
Instead, there must be
a pre-existing population of relativistic (or supra-thermal) seed particles that are then reaccelerated by turbulent acceleration during mergers.
One possibility is that the seeds are secondary electrons generated
 by collisions between cosmic rays and thermal protons \citep{Brunetti05,
Brunetti11}.
These hybrid models predict that Mpc-scale synchrotron emission  
from secondary electrons (``pure hadronic halos") must also be present in relaxed clusters, with a  radio luminosity
 10-20 times smaller than that of merging
clusters \citep{Brunetti11}.
This emission should be detectable with  EMU/WODAN.
Figure~\ref{fig:RHLF_1400} shows the 1.4\,GHz Radio Luminosity Function (RLF) of radio halos in
the local Universe ($z=0.1$): at higher radio luminosities,  turbulent
cluster halos dominate, but at lower
luminosities the RLF is dominated by ``had\-ronic" halos in relaxed systems.
Additional sources of seed supra-thermal electrons include shocks, AGN, galaxies,
and magnetic reconnection, so relaxed clusters
may have an even higher radio luminosity. If EMU fails to detect radio halos in relaxed clusters, this will provide a strong constraint
on secondary particles and on the energy content of cosmic rays in the ICM.


An extrapolation of the radio power of halos at 1.4\\
GHz versus the cluster X-ray
luminosity indicates that clusters with $L_x < 10^{44}$\,erg/s should host halos of power $L_{1.4}<10^{23}$\,W/Hz. With a typical size of 1\,Mpc their surface brightness could be too low to be detected at 1.4 GHz. However less powerful radio halos should be also smaller in size \citep{Feretti12} implying that they should be easily detected by the EMU
survey.
A comparison between 1.4\,GHz data with low frequency data (LOFAR) will be crucial to to test turbulence properties.
A large number of ``hadronic halos'' may be detected at 1.4\,GHz  (see Figure~\ref{fig:RHLF_1400}), emphasising the
importance of comparing the EMU and LOFAR surveys.

If turbulence is responsible for the origin of halos,
then the radio-emitting electrons are accelerated up to energies
$m_e c^2 \gamma_m \leq$ several GeV, where the synchrotron and inverse
Compton losses quench the acceleration by MHD turbulence. Radio halos will
therefore steepen at a frequency $\nu_s \propto \gamma_m^2 b$, making it
difficult to detect halos at higher frequencies \citep{Brunetti08}.

Only the few most energetic mergers in the Universe can generate
giant radio halos with $\nu_s \geq 1$\,GHz \citep{Cassano05}.
Most mergers instead consist of major mergers
between less massive systems, or minor mergers in massive
systems \citep{Cassano06, cas10a}, which will therefore have
steeper spectra (smaller $\nu_s$).
Recent Monte Carlo calculations based on turbulent acceleration in
galaxy clusters predict that LOFAR,
reaching rms$=0.1$\,mJy/beam at 120\,MHz over the northern hemisphere,
will detect more than 350 radio halos at redshift $z\leq0.6$, increasing the number of known halos by  a factor of nearly ten  \citep{ens02b, Cassano06}.
More than half of these halos will be ``ultra-steep spectrum"
halos, with $\nu_s < 600$\,MHz \citep{cas10a}, providing a
clear test of the importance of  turbulence in particle-acceleration in merging clusters.

\subsubsection{Radio Relics}

Relics (occasionally termed radio phoenixes) are irregular, elongated diffuse sources  which are generally located
at the  periphery of the cluster, with
the linear size of 400 -- 1500\,kpc
\citep{Giovannini02}.  Most have a steep
spectrum ($\alpha < -1$),
and are linearly polarised at a level of $\sim$ 10 -- 30\%.
They show an asymmetric transversal profile and spectral index distribution.
Current models suggest they are caused by
diffusive shock acceleration of electrons in old radio-emitting regions, triggered by a cluster merger event.
Currently, $\sim$ 34 clusters are known to contain at least
one relic source. Eleven of these also have a  radio halo, and ten have a
double relic.

The existence of relic sources suggest cluster-wide magnetic
fields of about $0.1 - 1\,\mu$G and relativistic
electrons of $\sim$ GeV energies in large peripheral regions where
the density of the hot ICM is low and galaxies are scarce.

Few radio relics have been well studied and few have deep
high resolution radio maps,  so that their size
and flux density  may be under-estimated.
Most radio spectra have been obtained with a limited frequency range
and very few spectral index images have been published.
Because most relics are located at the cluster periphery,
they are often  outside, or at the edge,
of available X-ray images, making it difficult to compare radio and X-ray data.
However, a few relics have been found and studied in detail \citep[e.g.][]{JohnstonHollitt03, Weeren10, Brown11a}.

\begin {figure}[t]
\vskip 0cm
\includegraphics[width=5cm, angle=0]{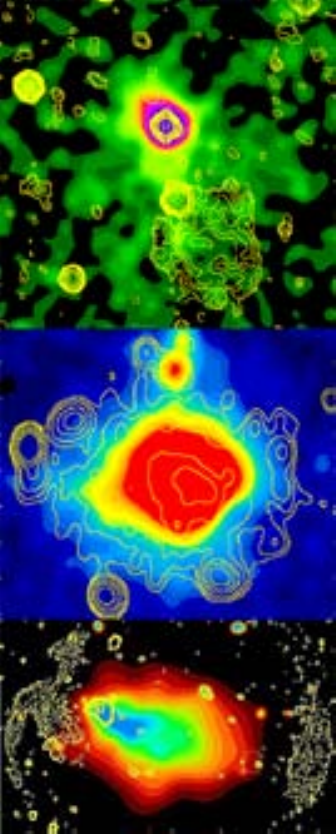}
\vskip -0.2cm
\caption{Radio image (contours) overlaid on X-ray images (colour) of three representative clusters: (top) the radio relic in the cluster A1664 \citep{Govoni01}, (centre) the giant radio halo in A2163 \citep{Feretti01}, (bottom) the double relic in A3376 \citep{Bagchi06}.
\label{cluster1}
}
\vskip -0.5cm
\end{figure}

A few relics show a curved shape radio spectrum and a diffuse, filamentary
morphology. They often have been found near the central cD galaxy suggesting
a possible connection with a past activity of this galaxy, but in A1664 (see Figure~\ref{cluster1}) and
in A548b \citep{Feretti06} the relic is far from
the cluster centre.


\begin {figure}[t]
\vskip 0cm
\includegraphics[scale=0.4, angle=0]{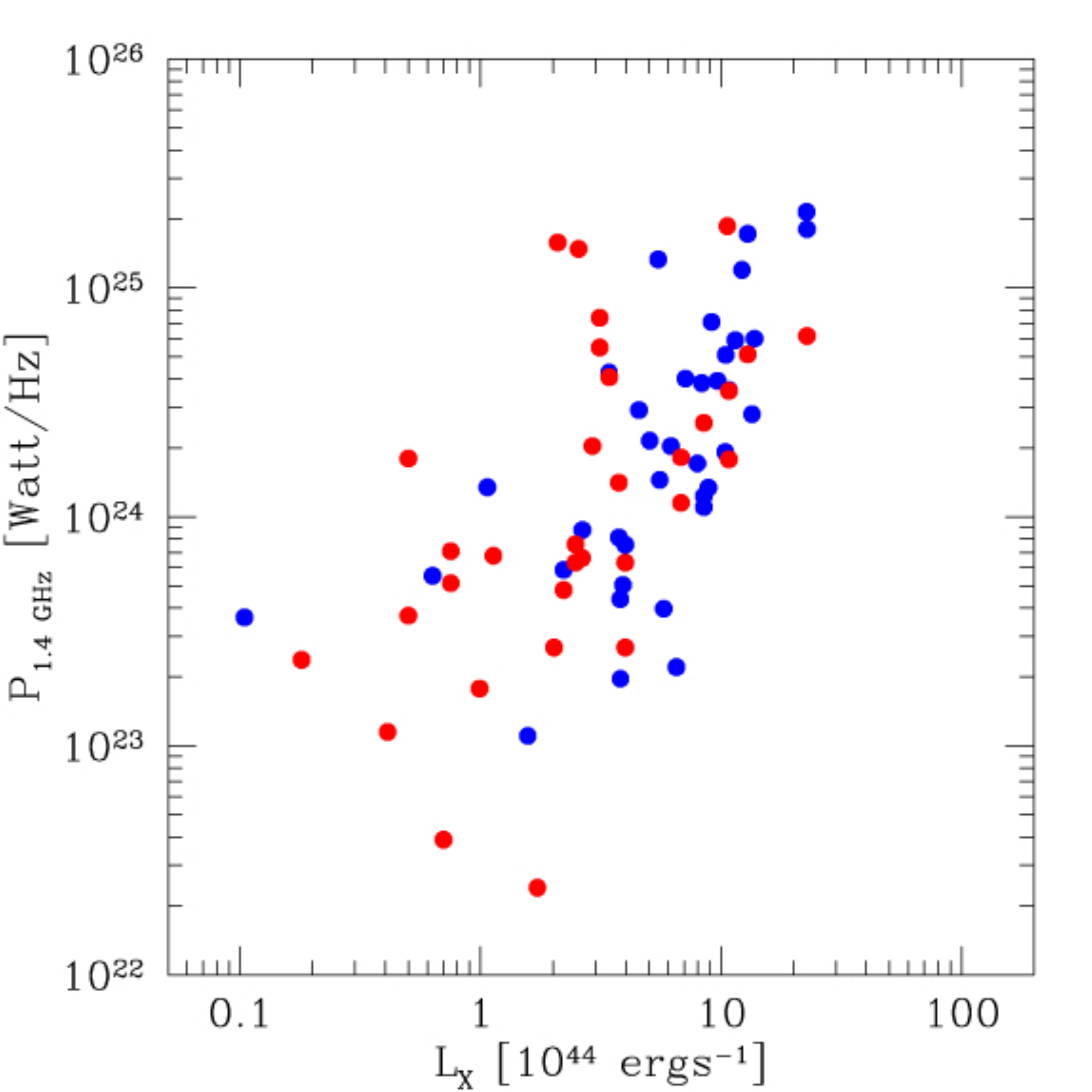}
\vskip -0.2cm
\caption{Halo (blue) and relic (red) radio power at 1.4\,GHz in units of W/Hz versus
 cluster X-ray luminosity.}
\vskip -0.5cm
\label{Xraycluster}
\end{figure}

Figure~\ref{Xraycluster} compares the radio and X-ray power of radio halos and relics, showing that
both the relic and halo power are  correlated with the parent cluster X-ray
luminosity, although this may be influenced by selection effects. The relics have a larger dispersion than the halos,
confirming that shocks are more efficient than turbulence in accelerating
relativistic electrons or amplifying magnetic fields.
This correlation confirms the link between relics and cluster
mergers and shows that the SPARCS surveys
will enable the study  of large scale magnetic fields in
low density regions at the cluster periphery. Such correlations have recently been discussed by \citet{Feretti12}.

\subsubsection {Tailed radio galaxies}
\label{tails}
Tailed radio galaxies include wide-angle tails (WATs) and narrow-angle tails (NATs). They are common in large clusters \citep{Blanton00, Blanton01, Blanton03}, and appear to represent radio-loud AGN in which the jets are distorted by the ICM \citep{Mao10}. \citet{Mao11c} have found 6 tailed galaxies in the seven square degrees of the Australia Telescope Large Area Survey (ATLAS), suggesting that $\sim$ 40,000 new clusters will be detected by EMU, using the tailed galaxies as tracers. With higher spatial resolution, more tailed galaxies become visible, and
\citet{Dehghan11},
 using high-resolution images of the same ATLAS fields, find 12 tailed galaxies in 4 \sqdeg, implying $\sim 10^5$ tailed galaxies in EMU. Importantly, such galaxies can be detected out to high redshifts of z $\sim$ 0.5 and beyond \citep{Wing11, Mao10}, providing a powerful diagnostic for finding clusters.
 
\subsection{The Magnetic Sky}
\label{magnet}

Magnetic fields are very important at all scales in the Universe, exerting their influence
 at very small scales in the interstellar medium
(ISM), at large scales in galaxies and clusters, and on the vast scale of
cosmic filaments.
Yet we still have limited understanding of the magnetic contribution to the energy balance
in these structures, and their dynamical influence and contribution to
the evolution of different classes of object. Our best tools for measuring cosmic magnetic fields are
 radio observations of synchrotron radiation and its linear
polarisation,
which are sensitive to total and ordered magnetic fields, respectively,
 and Faraday rotation, which
is sensitive to magnetic fields along the line of sight.

One of the areas which will benefit most from SKA pathfinder surveys
is the study of radio continuum polarisation. The prospects for
wide-field deep imaging, with broad wavelength coverage, and (in PAF-equipped telescopes) with excellent control over
the primary beam response,
combine to make the suite of pathfinders excellent for the study of
polarisation. The two major polarisation surveys which are
currently plan\-ned with SKA pathfinders  are (i) the
LOFAR Key Science Project on Cosmic Magnetism (see \S\,\ref{lofar}), and
(ii) ASKAP-POSSUM project
(see \S\,\ref{possum}). They are highly complementary, spanning more than a decade
in observing frequency,  and will probe
widely differing regimes of magnetised plasmas. Moreover, together they
can observe both hemispheres. There are a number of other surveys aimed at recovering high density grids of rotation measures (RM grids) towards different astrophysical regions, notably the
APER\-TIF-BEO\-WULF and
APER\-TIF-FRIGG surveys (\S\,\ref{beowulf}), the e-MERLIN Super-CLASS survey \citep[][\S\,\ref{emerlin}]{Battye2011} and the RM grid component of the MeerKAT MIGHTEE Survey (\S\,\ref{mightee}).

The primary tool for analysing the polarisation data from
the pathfinders is the ``RM Synthesis'' technique
\citep{burn66,brentjens05}. The essence of this technique is to
perform a Fourier transform of the observed complex polarisation
vector from the frequency (strictly, $\lambda^2$) domain to the
``Faraday depth'' domain. The Faraday depth refers to the
line-of-sight integral of the thermal electron density and magnetic
field strength. In simple situations it is more commonly known as the
Faraday rotation measure (RM),
while this simple relation breaks down for synchrotron emission and
Faraday rotation in the same medium, various emitting and rotating
layers along the line of sight, or more than one polarised feature
with different RMs within an angular resolution element.
The output of the RM Synthesis technique is a reconstruction of the
Faraday dispersion function, which describes the complex polarisation
vector as a function of the Faraday depth
\citep{brentjens05, heald09}. Other techniques for identifying magnetic field structures from RM synthesis observations continue to be developed \citep[e.g.\ ][]{Beck12}.

The technique
allows the detection of very faint polarised emission even when the RM
is very high (i.e.\ corresponding to the polarisation angle rotating by more than a turn across the observing band) and avoids
$n\pi$ ambiguities. Better
precision in the Faraday depth domain is obtained by observing with
wide bandwidth, and access to high values of $|\mathrm{RM}|$ is
provided by high
spectral resolution -- both of which are included in the
specifications of all of the pathfinders.

High quality polarisation observations
address diverse science goals, ranging from local  to
cosmological scales. Here, we summarise a subset of these, and discuss the impact of the pathfinder surveys on these fields.

\subsubsection{The large-scale structure of the Milky Way magnetic field}

Knowledge of the strength and orientation of the large-scale magnetic
field in the Milky Way is crucial for studies of the origin and
evolution of Galactic magnetism, but is also valuable for
extragalactic topics such as ultra-high energy cosmic rays,
cosmic microwave background polarisation, and magnetism in external
gal\-axies, galaxy clusters and intergalactic space.

Models constrained by observational RM data from pulsars,
extragalactic sources and diffuse Galactic synchrotron emission,
attempt to determine reversals in the spiral magnetic pattern of the
Milky Way, and to measure the large-scale configurations of the field
in the Milky Way disk and halo \citep[e.g.\ ][]{beck11, noutsos08, vaneck11, sun08,
jansson09, nota10, maos10},
all of which are key to understanding
the evolution of galactic magnetic fields. However, controversy still
exists
about the existence and location of reversals, and the overall 3D
configuration of the magnetic field in the Milky Way's disk and halo.

Major constraints to these large-scale magnetic
field models will come from the {\it Rotation Measure Grid} of RMs from
compact extragalactic sources, most of which probe Galactic
Faraday rotation. The SKA pathfinders ASKAP, LOFAR, APERTIF, and MeerKAT
have complementary programs to observe RMs of extragalactic point
sources  to construct an RM grid with an average
density of 100 RMs per square degree over the entire sky. These efforts will
provide a polarised source density  two orders of magnitude higher than
the recent NVSS RM catalog \citep{taylor09}, which has already
proven useful for a wide range of analyses
\citep[e.g.][]{naomi10,schnitzeler10,
stil11}.

\subsubsection{The turbulent, magnetised interstellar medium}

Magnetised turbulence in the interstellar medium (ISM) plays a major role in
many physical processes such as star
formation, total ISM pressure equilibrium, cosmic ray scattering, gas
mixing, and radio wave propagation \citep{scalo04}.

At high Galactic latitudes, in the Galactic thick disk and halo, RM
variations are small, and can be observed at long wavelengths. Spectropolarimetric observations of the
Galactic foreground emission at mid-latitudes around 350\,MHz, made
using the Westerbork Synthesis Radio Telescope (WSRT) show a wealth of
small-scale polarisation at angular resolutions down to a few arcmin
\citep{haverkorn03a, haverkorn03b, schnitzeler07,schnitzeler09}.  At even lower frequencies of 150\,MHz,
\citet{bernardi09} observed abundant structure in diffuse
polarisation at intermediate latitudes. Therefore, LOFAR RM synthesis
observations at mid- and high-latitude are expected to discover a
wealth of small-scale RM structures in the magneto-ionised gas. This
will be used to characterise the power spectrum of magnetic field
fluctuations at high latitudes, i.e.\ in the Galactic gaseous halo and
disk-halo connection.

In addition, calculation of the spatial gradient of diffuse polarisation can allow direct visualisation of turbulent structures in the ISM, from which we can infer fundamental properties such as Mach number and Alfven speed \citep{Gaensler11, Burkhart12}.

Closer to the Galactic plane, the larger RM values will cause partial
or complete depolarisation of low-frequency emission, but are well-suited to observations with ASKAP, APERTIF and MeerKAT. The
RM grid allows statistical analysis of interstellar magnetised
turbulence as a function of position \citep{haverkorn08}, while
RM synthesis of the diffuse polarised emission allows disentangling of
synchrotron emitting and Faraday rotating layers
\citep{brentjens11}. It will also be possible to map out magnetic
field strength and structure in discrete structures such as supernova
remnants, planetary nebulae, super-bubbles, H{\sc ii} regions,
Galactic chimneys, and high-velocity clouds
\citep[see, e.g.,][]{beck04}.

\subsubsection{Faraday Screens}

At 1.4\,GHz the diffuse polarised sky typically bears little resemblance to the total intensity sky \citep[e.g.\ ][]{landecker10}. A widely accepted interpretation for this characteristic is that the detected polarised features are the signature of Faraday rotation rather than structure in the Galactic synchrotron emission. The exceptions are supernova remnants
(\S\,\ref{snr}) and pulsar wind nebulae (\S\,\ref{pwn}).

Observations have shown that the synchrotron emission of external galaxies is very smooth and the WMAP results have confirmed that for our Galaxy too. However, at lower radio frequencies, surveys show a lot of small-scale structure caused by Faraday rotation. A typical radio telescope operating in the 1.4\,GHz wavelength range proves to be more sensitive to the Faraday rotation of an ionised and magnetised medium than to its bremsstrahlung signature.

The Faraday rotation structures we find in linear polarisation images are called ``Faraday Screens''. These Faraday Screens are still not well understood. Questions about these features we have to answer include:
\begin{itemize}
\item Are those features created by variation of the magnetic field or the free electron distribution
(or a combination of both)?
\item Can we correlate (or anti-correlate) these features with other tracers of the ISM?
\item How can we explain features that do not correlate with anything?
\item Can Rotation Measure Synthesis help to reveal the characteristics of these features?
\item Can we relate these features with structures seen in the rotation measure distribution
determined from extra-galactic point sources?
\end{itemize}
A large wide-field survey such as POSSUM will help to better understand these features and reveal their true nature

\subsubsection{Nearby galaxies}
Spiral galaxies have typical magnetic
field strengths of order $10\,\mu\mathrm{G}$. They typically show ordered field
patterns consistent with the spiral pattern in the plane, a higher
level of magnetic turbulence in the arms themselves, and an X-shaped
configuration in the high-latitude regions. A concise review of the
properties of magnetic fields in galaxies is provided by \citet{Beck09}.

The WSRT-SINGS Survey
has
investigated the global magnetic field properties across a small sample of
nearby galaxies with diffuse polarised emission
\citep{braun10}. This work showed that azimuthal patterns in
polarised intensity and RM \citep{heald09a} are
consistent with a magnetic field configuration consisting of
an axisymmetric spiral in the disk and a quadrupolar poloidal pattern away from the
plane, as predicted by the dynamo theory of magnetic
field amplification \citep[e.g.][]{widrow02}. A critical component of the
model is that turbulent magnetic fields in the star-forming disk depolarise
synchrotron radiation from the far side of the disk \citep[see][for a
description of this {\it Faraday dispersion} effect]{burn66}. In face-on
galaxies this effect, together with trailing spiral arms, results in the
observed patterns. At higher and lower radio frequencies, the effects of Faraday
dispersion will be reduced and enhanced, respectively, giving rise to
different patterns in the polarised synchrotron radiation.

The polarisation surveys planned with ASKAP and LOFAR will thus be complementary
in that they will trace ordered magnetic fields in significantly different
volumes within nearby galaxies. Although depolarisation by Faraday dispersion
is much stronger at LOFAR frequencies, this is mitigated by the fact that
cosmic-ray electrons (CRs) that have diffused far from their acceleration sites
(i.e.\ regions of massive star formation) will have lower energies because of
synchrotron losses. Thus synchrotron radiation from this aged CR population
will peak at LOFAR frequencies, and should trace ordered fields in the far outer
regions of galaxies. At the other end of the spectrum, high frequency observations
with MeerKAT and the VLA can ``see through'' the star forming disk, and trace
the fields in the star forming regions directly. The suite of SKA pathfinders
can thereby provide complementary polarimetric studies leading to an
``onion-peel'' polarimetric view of diffuse synchrotron emission in nearby galaxies.

\subsubsection{Magnetic field evolution in galaxies}
Beyond the nearby Universe,
SKA pathfinder telescopes have insufficient angular resolution to map the synchrotron
radiation across disks of galaxies. However,
\citet{stil09} have shown that polarised radiation with a preferred,
and wavelength-independent, orientation will be detected in unresolved
galaxies. Thus statistical studies of the magnetic fields in distant galaxies
will be possible. These same sources are also expected to be excellent
background probes in the RM grid.


Within the picture of the turbulent
dynamo, the amplitude of the regular magnetic fields in galaxies is expected to
decrease at higher redshift. The variation of galactic magnetic field
strength with epoch may be reflected in the redshift dependence of the
FIR/radio correlation \citep[see][]{Murphy09}.

\subsubsection{Galaxy clusters}
The ICM of galaxy clusters consists not only
of hot thermal gas ($T\sim 10^{7-8}$\,K) emitting in the
X-ray domain, but also of magnetic fields and relativistic particles
that may give rise to synchrotron radio emission. The most direct
evidence for the presence of magnetic fields are the so-called radio
halos and radio relics (see \S\,\ref{clusters}). Other evidence for cluster magnetic
fields comes from  the Faraday rotation of radio sources
that lie within or behind the cluster along our line of
sight \citep[e.g.][]{clarke01, jh04}.

Magnetic fields are thought to play an important role in the
development of large-scale structure in the Universe, but  their origin and evolution are
still poorly understood. Cosmological
simulations indicate that the magnetic field strength resulting from
the adiabatic compression of the gas falling into the cluster
potential well cannot account for more than 1\% of the magnetic field
strength that is inferred from observations, so that other
amplification mechanisms are required \citep{Dolag08}. Cosmological
simulations also predict that processes related to the cluster
formation, such as merger events and shear flows, can in principle
amplify the magnetic field up to the observed values.
Hence, measurements of the
ICM magnetic field, such as the power spectrum and
radial profile, are necessary to test these predictions, and to
measure the amplification of the magnetic field  during
cluster formation. Magnetic fields may also affect thermal
conduction in the ICM, and so measuring the magnetic field
strength is crucial to understand  the origin of the relativistic
particles responsible for the radio halo and relic emission.

\subsubsection{Faraday Rotation Measures and ICM magnetic fields}

A powerful method to constrain the magnetic field properties in the
ICM is to study the Faraday rotation of sources that are within or
behind the cluster along our line of sight
\citep[e.g.\ ][]{krause09}. Recently
\citet{Bonafede10} have constrained the magnetic field in
the Coma cluster by observing 7 sources at different projected
distances from the cluster centre. The resulting RM images have been
compared to the mock RM images obtained with the FARADAY code
\citep{Murgia04} for different magnetic field configurations. The
magnetic field that best reproduces the Coma observations has a central value $B_0=4.7\,\mu$G, and declines with the
gas density profile $n$ according to $B(r)=n(r)^{0.5}$.
\citet{Murgia04}, \citet{Govoni06}, and \citet{Vogt05} have achieved similar
results on smaller samples. These studies require deep multi-frequency
observations of several sources located at different projected
distances from the cluster centre, so only
a small number of clusters have been studied.
The sensitivity of current radio telescopes is insufficient to study a large number of galaxy clusters with
many RM probes per cluster, but the next
generation of radio telescopes will make this possible.

\subsubsection{Fractional polarisation of radio sources and magnetic field in the ICM}

When synchrotron emission from a cluster or background source
crosses the ICM, regions with similar intrinsic direction of the
polarisation plane, $\Psi_{int}$, going through different paths, will
be subject to differential Faraday rotation. If the magnetic field in
the foreground screen is tangled on scales smaller than the
observing beam, radiation with similar $\Psi_{int}$ but opposite
orientation will be averaged out, and the observed degree of
polarisation will be reduced (beam depolarisation). In the central
region of a cluster, $B$ and $n$ are higher, resulting in a
higher  RM, and lower fractional polarisation $F_P$. Sources
at larger radii experience a lower RM, and so suffer from less depolarisation.

 \citet{Bonafede11}
selected a sample of massive galaxy clusters and used the
NVSS data \citep{Condon98} to analyse the polarisation
of radio-sources as a function of the projected distance
from the cluster centre. They find that, statistically, the fractional polarisation decreases with
the cluster projected distance. By
comparing this trend with that predicted by magnetic
field models, they estimate the magnetic field in the cluster centre to be
$\sim 5\,\mu$G. The advantage of this
approach is that it does not need multi-frequency
observations. This effect can be investigated with radio surveys at a
single frequency in full polarisation mode. Although such
studies do not provide detailed information on magnetic fields in
specific clusters, they allow us to understand the average properties
of magnetic fields in the ICM, and potentially even on larger scales.

\subsubsection{The origin of cosmic magnetism}

Galaxies and the ICM can only account for about one third of
the baryon density in the local Universe expected from a concordance
cosmology. The majority of the missing matter is likely to reside in
a warm-hot intergalactic medium (WHIM), which in turn is expected to
reside in the cosmic web of the large-scale structure. Detection of
the magnetic field found in this cosmic web, or placing stringent
upper limits on it, will provide powerful observational constraints on
the origin of cosmic magnetism.

Various mechanisms have been suggested for the origin of magnetic
fields in our Universe. One possibility is that the fields
are truly primordial, and that a seed field formed prior to
recombination in the very early Universe \citep{Banerjee03}.
Alternatively, fields could have been produced
via the Weibel instability, a small-scale plasma instability formed
at structure formation shocks \citep{medvedev04}. Both of these
mechanisms occurred early in cosmic history, prior to galaxy formation,
and are therefore referred to as \emph{early-type} mechanisms.
In contrast, possible \emph{late-type} mechanisms are processes
where the field was injected into the WHIM via the action of
jets from AGN and other outflows such as galactic winds
\citep[e.g.\ ][]{kronberg04}.

The SKA pathfinder polarisation surveys will together generate a
rich dataset from which a wide, high density, broad bandwidth
RM Grid may be built, and weak magnetic fields in
intergalactic filaments may be detected for the first time. A lower limit for the
intergalactic magnetic field has been determined
\citep[$10^{-15}\,\mathrm{G}$;][]{Dolag11}.

In the filamentary
structures of the cosmic web, field strengths are expected to be
far more sensitive to the \emph{origin} of the magnetic
fields than in larger gravitationally bound structures where turbulent
amplification will have led to saturation of the field. Specifically
the strength and penetration of magnetic fields in filaments is expected
to be much lower if the field arises from a ``late''
astrophysical source (AGN, galactic winds etc), compared to that arising from
a primordial one, and simulations have demonstrated that the
strength of these turbulently amplified early-type seed fields
range typically between 0.1 and 0.01\,$\mu$G \citep{Ryu08}.
Consequently observations of the filamentary cosmic web are invaluable for inferring the origin of cosmic
magnetism. Although the predicted field strengths are extremely low,
such RMs are observationally accessible with the SKA pathfinder
telescopes.

In addition, the rotation of emission from background
radio sources through the large-scale structure filaments within
e.g. super-clusters of galaxies can also be used to infer the origin
of cosmic magnetism.
A statistical analysis of these RMs to measure the power-spectrum of
the magnetic field of the cosmic web \citep{kolatt98}, using
cross-correlations with other large-scale structure indicators, will
also provide stronger constraints than individual RMs. These
measurements will also probe the distribution and density of thermal
gas at the interface between clusters and the cosmic web, which has
important consequences for a number of related areas in astrophysics
and will be highly complementary to X-ray and Sunyaev--Zel'dovich
studies of these structures. However, because of the low temperatures of these regions,
RM measurements may provide the only way to probe them.

\subsubsection{High redshift rotation measures}

Studies of high-redshift objects \citep[e.g.\ ][who examined the redshift
dependence of the spread of RMs]{Kronberg08} suggest
that the amplification of magnetic fields occurred more rapidly than the
expectations of mean-field dynamo theory
The conclusions of this type of
work are so far limited by small samples, together with uncertainties in the
removal of the Milky Way foreground RM structure. SKA pathfinders will provide
RM grids that will enable this kind of study to be done with significantly
lower uncertainty. A better comparison with the predictions of dynamo theory
\citep[see e.g.\ ][]{Arshakian09} will thus become possible.

\subsection{Cosmology}
\label{cosmology}
\label{sec:intro}

The SKA pathfinder surveys (here referred to as SPARCS surveys) will not only be an important preparation for SKA surveys, but will also be able to deliver important cosmological results themselves. Earlier surveys were able to provide valuable constraints on cosmology and cosmic evolution \citep[e.g.][]{Blake02, Overzier03}. Here we show that SPARCS surveys will be able to constrain cosmological models and parameters via several probes: the Integrated Sachs-Wolfe effect \citep{sachs67}, the lensing effect known as cosmic magnification, and source count correlations.

The source count correlation function (or if measured in Fourier space, the power spectrum) is a well-known statistic measuring the excess probability that sources are clustered together on different scales. In the radio surveys considered here, we assume that individual redshifts will not be available, so the power spectrum will be measured on the 2-D celestial sphere, by measuring the spatial auto-correlation function of the radio sources in the SPARCS surveys.

Cosmic magnification is a gravitational lensing effect, where the number density of background sources is affected by foreground structures in two competing ways. First, lensing dilates the background patch of sky, reducing the background number density. Second, the lensing expands the size of background sources while maintaining their surface brightness, resulting in an increase of apparent luminosity of sources, increasing the number density above any given detection threshold. Either of these effects can dominate depending on the luminosity function, so there may be either a correlation or anti-correlation of background number density with foreground structures. The effect is sensitive to the geometry of the Universe and the growth of structures in the Universe. In our case, we correlate background SPARCS radio sources with foreground sources from shallower optical surveys, such as SDSS \citep{Abazajian09} for northern surveys, Skymapper \citep{Keller07} or DES \citep{DES2005} for southern surveys.

The ISW effect is a gravitational redshift, due to the evolution of the gravitational potential while photons pass through under- or over-densities in their path from the last scattering surface to us. In an Einstein-de Sitter universe, the blueshift of a photon falling into a well is cancelled by the redshift as it climbs out.
But in a cosmology where there are evolving gravitational potentials, photons will experience a net blue- or red-shift, leading to a net change in photon temperature, which accumulates along the photon path. This translates into CMB temperature anisotropies proportional to the variation of the gravitational potentials:
\begin{equation}
\frac{\delta T}{T} \propto \int \left[ \dot{\Psi}(z) +\dot{\Phi}(z) \right] dz \, .
\end{equation}
Here we have introduced the gravitational potentials $\Psi$ and $\Phi$, which are the time-part and space-part perturbations of the FRW (Friedmann--Robertson--Walker) background in the conformal Newtonian gauge, $ds^2= -a^2 ((1+2\Psi)d\tau^2 - (1-2\Phi)dx^2)$. In GR, in the absence of anisotropic stress, we expect $\Psi$ and $\Phi$ to be identical.

The ISW effect can contribute significantly to the CMB temperature fluctuations on large angular scales, but it enhances only the low $\ell$ multipoles, and is smaller than other CMB anisotropies.
For this reason, in order to detect the effect, we have to cross-correlate CMB temperature maps with tracers of the potential wells \citep{Crittenden96} such as galaxies; if the potential evolves, then we should observe a correlation between CMB temperature anisotropies and the galaxy distribution. For the tests considered here, we cross-correlate the SPARCS radio sources with the CMB data from WMAP or Planck.

\subsubsection{Calculating probe predictions}
Here we consider the measurements of the three probes discussed above (source count   correlations, cosmic magnification and ISW), and calculate the resulting constraints  obtained with the LOFAR, WODAN and EMU surveys (see \citealt{Raccanelli11} for details about the surveys and the assumed distributions and bias).

To predict the redshift distribution for these surveys, we use the latest empirical simulations developed for the Square Kilometre Array continuum survey \citep{Wilman08}. This simulation provides different prescriptions for the redshift evolution of the various populations which dominate the radio source counts at five different radio frequencies: 150, 610, 1400, 4860 and 18000\,MHz.
Catalogues are generated from the $S^{3}$ database\footnote{http://s-cubed.physics.ox.ac.uk}, using the radio flux-density limits of different surveys, and we perform our analysis by extracting a catalogue from the database with cuts at the $10\,\sigma$ detection levels of the forthcoming surveys.

The $S^3$ simulation provides us with a source catalogue in which sources are identified by type (e.g.\ SB, SFG, FRI, FRII, RQQ). The bias is computed separately for each galaxy population using the formalism of \citet{mo96}, in which each population is assigned a dark matter halo mass chosen to reflect the observed large-scale clustering. With this framework, the increasing bias $b(z)$ with redshift would lead to excessively strong clustering at high redshift, so the bias for each population is held constant above a certain cut-off redshift \citep[see ][for details]{Wilman08}.

We calculate the angular power spectra for each of the probes, i.e.\ source-source, radio background -- optical foreground, and source-CMB. Each of these is a projection of the density power spectrum; see \citet{Raccanelli11} for details. For instance, the ISW cross-power spectrum is given by

\begin{equation}\label{eq:ClgT}
C_{\ell}^{gT} = \langle a_{{\ell}m}^g a_{{\ell}m}^{T*} \rangle = 4 \pi
\int \frac{dk}{k} \Delta^2(k) W_{\ell}^g(k)
W_{\ell}^T(k),
\end{equation}
where $W_{\ell}^g$ and $W_{\ell}^T$ are the galaxy and CMB window functions respectively,
and  $\Delta^2(k)$ is the logarithmic matter power spectrum today.

The galaxy window function, which enters all three power spectra, is written as:
\begin{equation}\label{eq:flg}
W_{\ell}^g(k) = \int \frac{dN}{dz} b(z) D(z) j_{\ell}[k\eta(z)] dz,
\end{equation}
where  $(dN/dz)dz$ is the mean number of sources per steradian with redshift
$z$ within $dz$, brighter than the flux limit, $b(z)$ is the bias factor relating
the source to the mass over-density, $D(z)$ is the linear growth factor of mass
fluctuations, $j_{\ell}(x)$ is the spherical Bessel function of order $\ell$,
and $\eta(z)$ is the conformal look-back time. Similar window functions exist for the CMB and foreground elements for the different power spectra.

We can write corresponding correlation functions as a function of the angular separation $\theta$:
\begin{equation}
C(\theta) = \sum_{\ell} \frac{2\ell+1}{4 \pi}
C_{\ell} L_{\ell}(\cos \theta),
\end{equation}
where $L_{\ell}$ are the Legendre polynomials of order $\ell$.

\subsubsection{Predictions for the pathfinders}
\label{cosmoresults}

\begin{figure}
\begin{center}
\includegraphics[width=8cm]{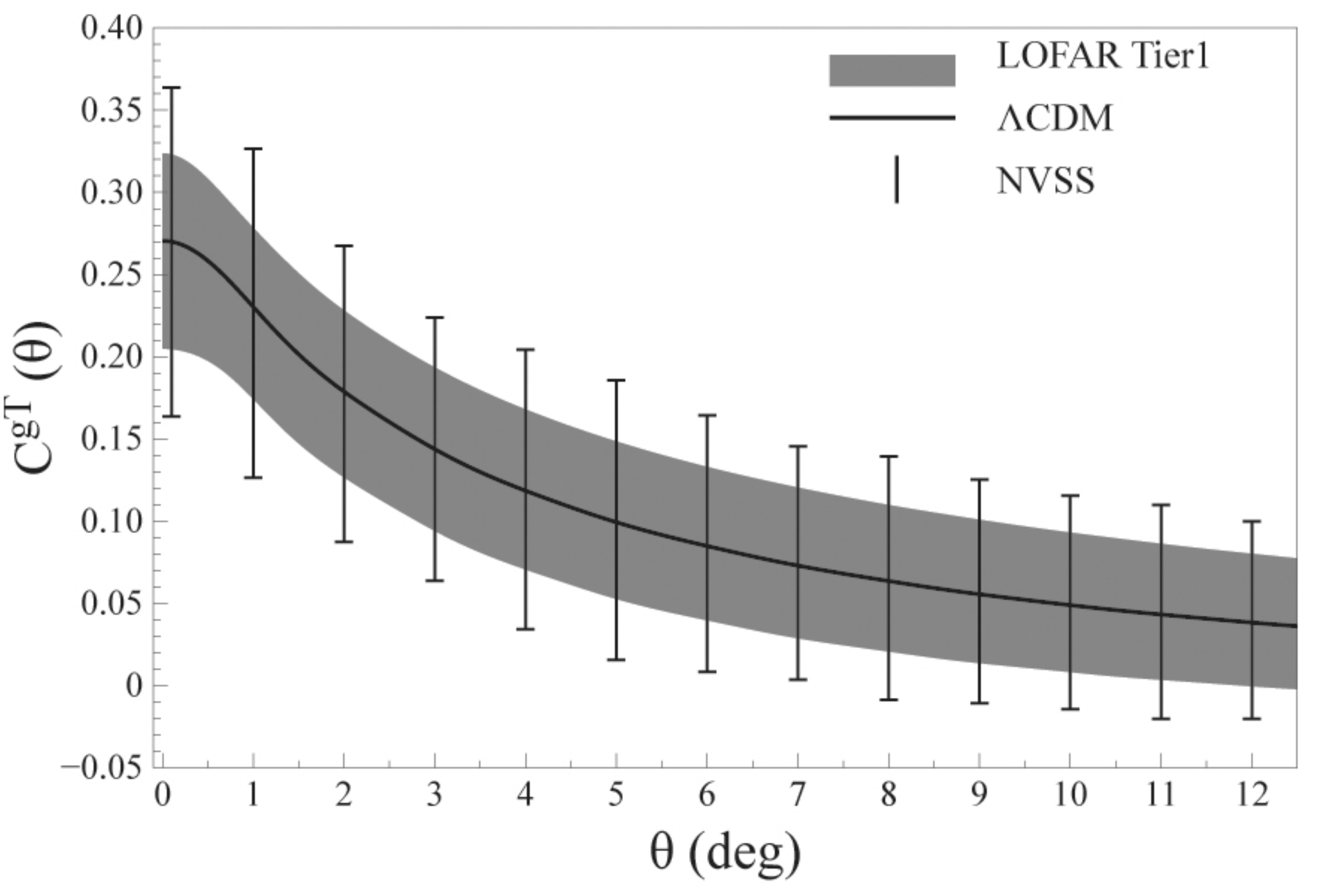}
\caption{Predicted correlations of the CMB with the LOFAR Tier 1 survey data. As a comparison we also show error bars from current NVSS measurements. Note the improvement we obtain with one of the pathfinders over current constraints on all scales.}
\label{fig:isw_results}
\end{center}
\end{figure}

We have calculated correlation functions or power spectra for the ISW, cosmic magnification and source counts for the forthcoming radio surveys. For example, in Figure~\ref{fig:isw_results} we show our result for the CMB-LSS cross-correlation using the predicted redshift distributions and bias for the LOFAR Tier 1 survey, where we assume the standard $\Lambda$CDM+GR model and the shaded areas are cosmic-variance errors, calculated via:
\begin{equation}
\label{eq:err-clgt}
\sigma_{C_{\ell}^{gT}} = \sqrt{\frac{\left(C_{\ell}^{gT}\right)^2 + C_{\ell}^{gg}C_{\ell}^{TT}}{(2\ell+1)f_{\rm sky}}},
\end{equation}
where $f_{\rm sky}$ is the sky coverage of the survey; $C_{\ell}^{gg}$ and $C_{\ell}^{TT}$ are the auto-correlations density-density and temp\-erature-temperature, and contain shot noise and detector noise; the errors are then converted to real space as:
\begin{equation}
\label{eq:errorW}
\sigma^2 (\theta)  = \sum_\ell \left(\frac{2\ell+1}{4\pi}\right)^2 ~P_{\ell}^2(cos{\theta}) ~ \sigma_{C_{\ell}^{gT}}^{2} \, .
\end{equation}

\noindent

Now we wish to predict the combined cosmological measurements that the SKA pathfinders should allow. In order to do this, we perform a joint analysis including CMB+SNIa priors, ISW, cosmic magnification and the auto-correlation of radio sources.
We test deviations from the standard $\Lambda$CDM+GR model for two cases: we first parameterise modifications of the theory of gravity; and then alternatively, we assume GR to be the correct model to describe gravity, and consider a dynamical dark energy model.

\subsubsection{Modified Gravity Constraints}
We want to test deviations from GR using the $\{ \eta, \mu \}$ parameters, that are defined via:
\begin{align}
& \frac{\Phi}{\Psi} = \eta (a,k), \\
& k^2 \Psi = -4 \pi G a^2 \mu (a,k) \rho \Delta,
\end{align}
where $\Delta$ is the gauge-invariant comoving density contrast:
\begin{equation}
\Delta = \delta + 3 \frac{aH}{k}v;
\end{equation}
in General Relativity $\eta (a,k) = \mu (a,k) = 1$, while in alternative models they deviate from this value, and can also be functions of time and scale \citep[e.g.\ ][]{Zhao10}.

\begin{figure}
\begin{center}
\includegraphics[width=8cm]{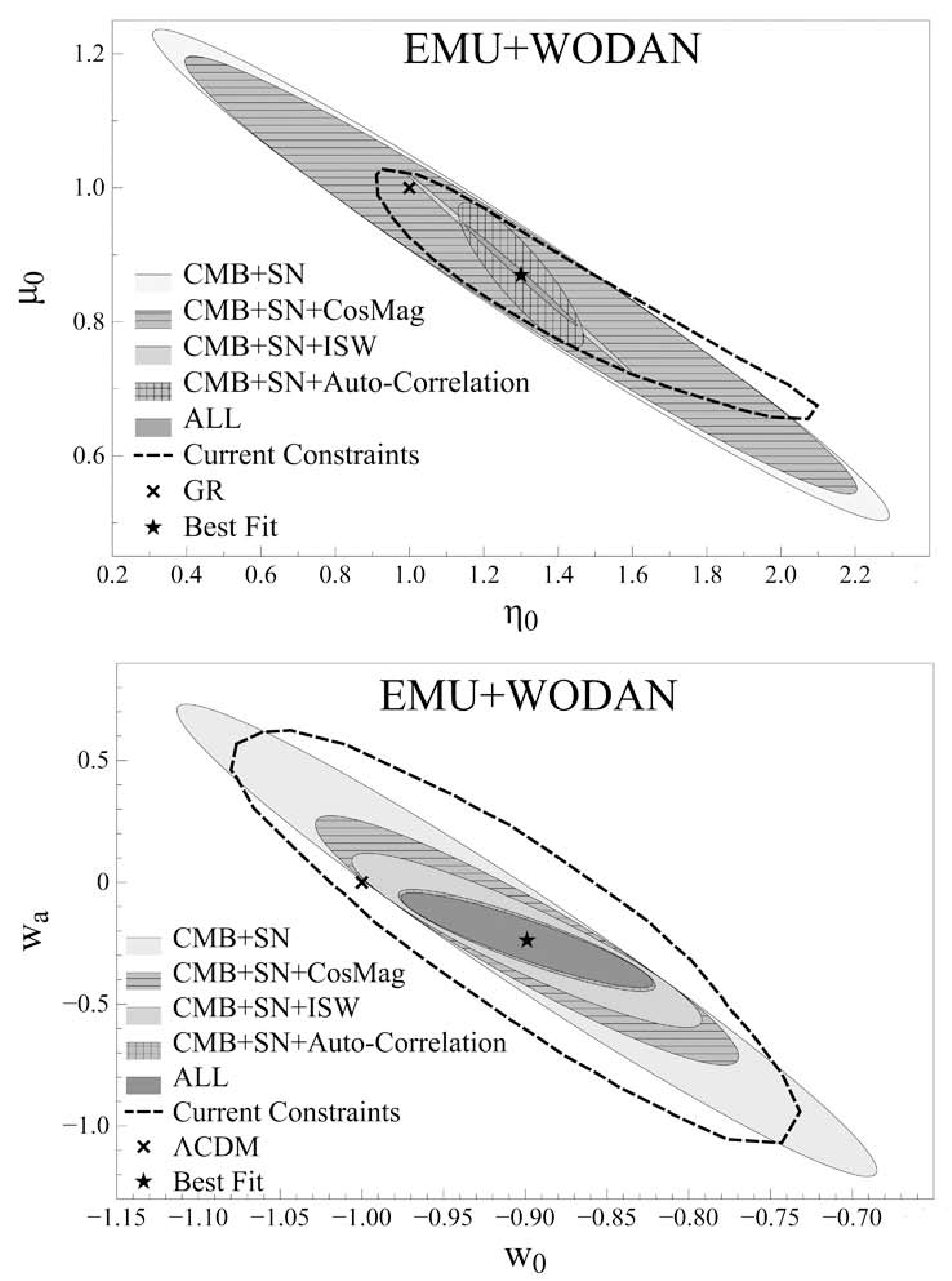}
\caption{Predicted constraints on parameters using EMU+WODAN. The outer dotted ellipse shows the current constraints, and the innermost grey ellipse shows the constraints available using EMU+WODAN with the three probes described. Bottom panel: dark energy parameters, showing current and predicted 68\%CL constraints. Top panel: modified gravity parameter constraints.}
\label{fig:forecast-mg}
\end{center}
\end{figure}

We model the time evolution of $\mu$ and $\eta$ as (see Zhao et al. 2010 for details):
\begin{align}
\eta (z) = \frac{1-\eta_0}{2} \left( 1+\tanh \frac{z - z_s}{\Delta z} \right) + \mu_0, \\
\mu (z) = \frac{1-\mu_0}{2} \left( 1+\tanh \frac{z - z_s}{\Delta z} \right) + \mu_0.
\end{align}
\noindent
In Figure~\ref{fig:forecast-mg} (top) we show predictions for how well we can constrain the modified gravity parameters using the EMU survey. Also displayed are the 68\% confidence level for current surveys \citep[SNe, WMAP and SDSS; ][]{Raccanelli11}. We use the current best-fit model as our fiducial model; note that if this current best fit remains the best fit, the standard GR+LCDM model would be excluded with forthcoming radio surveys at high significance. In any case, the surveys impressively reduce the permitted parameter space.

\subsubsection{Constraints on Dark Energy evolution}
To study how well the SKA pathfinders will constrain the evolution of dark energy we use the $\{ w_0, w_a \}$ parameterisation \citep{linder03} of the dark energy equation of state:
\begin{align}
w(a) = w_0 + w_a (1-a).
\end{align}
In Figure~\ref{fig:forecast-mg} (bottom panel) we show predictions for the constraints on the parameters $\{ w_0, w_a \}$, using the EMU survey. Again, the constraints are substantially improved over the currently permitted region of parameter space.

\subsubsection{Bias}

To obtain constraints on cosmological parameters from continuum surveys, the redshift distribution and bias of radio sources needs to be known. Here we suggest a method for measuring the bias of radio sources at high redshift. We cross-match FIRST radio sources \citep{Becker95} with SDSS-DR7 data \citep{Abazajian09} and use photometric redshifts in the DR7 catalog to obtain an estimate of the redshift distribution of the optically identified sources. To probe the high-$z$ radio continuum population, we assume that the $z$-distribution of all FIRST sources is given by the SKADS simulation \citep{Wilman08} and we subtract off the distribution of the optically identified sources to estimate the $z$-distribution of the unidentified sources. By measuring the angular correlation function of the unidentified sources and comparing this with predictions for dark matter clustering, we will then be able to estimate the bias at $z\sim 0.7$. Some initial results using this technique are presented in Passmoor et al (in preparation).

If SKA pathfinder surveys are to be used to measure cosmological parameters, it will be important to continue and extend this work to obtain a better estimate of bias for different classes of radio source, and also to understand the sensitivity of the derived cosmological parameters to uncertainty in the value of the bias parameter.

\subsubsection{The effect of redshifts on cosmological tests}

The cosmology discussion to this point makes the conservative assumption that no redshifts are available for individual radio sources, although we do know the redshift distribution of the different classes of source, and the already impressive results shown in Figure~\ref{fig:forecast-mg} are calculated using this assumption. While less than about 1\% of radio sources in SPARCS surveys are likely to have spectroscopic redshifts at the time of the data releases, photometric and statistical redshift information is likely to be available for $\sim$50\% of sources, as described in \S\,\ref{redshifts}.

\citet{Camera12} have explored the impact of photometric redshifts on the cosmological measurements from EMU and WODAN. Using only the galaxy clustering angular power spectrum and dark energy constraints, they find that even approximate photometric redshifts obtained from SDSS and SkyMapper yield a significant improvement to the constraints provided by the radio data alone.

An alternative approach has been suggested by \cite{Norris11b}, using radio data alone. Sources with detected polarisation in the POSSUM survey (see \S\,\ref{possum}) are virtually all AGN, with an expected median redshift of $<z>\sim$1.8, while the remaining unpolarised EMU sources will be  a mixture of AGN and SFG, with an expected median redshift of $<z>\sim$1.1. The unpolarised sources can therefore be treated as a separate population from the polarised sources for the purposes of the cosmological tests described above, giving two radio source samples at different redshift ranges, thus potentially yielding a measurement of the time-dependent component of the dark energy and modified gravity parameters. However, modelling is required to establish whether this technique yields a significant improvement over other techniques.

\subsubsection{Weak Lensing}

Weak gravitational lensing by large-scale dark matter
in the Universe distorts the images
of faint background galaxies. Measurements of this ``cosmic shear'' effect as a
function of redshift can provide powerful constraints on  dark energy and its evolution over
cosmic time \citep[e.g.\ ][]{albrecht06,peacock06}. The
measurements require a large number of well-resolved galaxies
to reduce the ``shape noise'' associated with the intrinsic
variety of galaxy shapes. Most
cosmic shear studies to date have therefore been at optical
wavelengths. Radio measurements have typically been less effective because of
smaller number densities and poorer spatial resolution. \citet{chang04}, however, made a statistical
detection of cosmic shear in the VLA FIRST survey, and  \cite{patel10} used  VLA
and MERLIN data to explore radio weak lensing techniques.

Measuring weak lensing at radio wavelengths potentially offers a number
of advantages over optical surveys in terms of minimising
instrumental and astrophysical systematic effects. For example, radio
telescopes have highly stable and well-understood point spread functions which
makes them particularly suited for measuring the small distortions
caused by weak lensing. In addition, a future weak lensing survey
conducted with the SKA could yield redshifts for a significant
fraction of the lensed galaxies through the detection of their H{\sc i}
emission lines \citep[e.g.][]{blake07}. Uncertainties and biases
associated with photometric redshift errors would consequently be
greatly reduced with an SKA lensing survey.

A major astrophysical cause of systematic error in weak lensing
measurements is the intrinsic alignment in galaxy shapes, caused by
tidal torquing during the galaxy formation process. Cleanly separating
these from the extrinsic alignments caused by weak lensing will be
crucial for interpreting precision weak lensing measurements in the
future. \cite{brownM11a} showed that  polarisation, which is routinely measured at radio wavelengths
can,  in principle, discriminate between the lensing distortions  and intrinsic
alignments.

\cite{blake07} have explored the potential of the SKA to perform a precision weak lensing survey. The potential
advantages described above would make such a survey highly
complementary to future large-scale optical lensing surveys.
The SKA pathfinders  offer
excellent opportunities to develop the field of radio weak lensing and
to demonstrate the above techniques and advantages on real data. For
example, large-area surveys with  arcsec resolution with LOFAR
can test weak lensing techniques on large radio
datasets. Moreover, the e-MERLIN telescope offers an exciting
opportunity to perform deep sub-arcsec radio imaging over a large
field of view which will be an ideal test-bed for investigating radio
lensing techniques. This potential has recently been recognised with
the award of $\sim$ 800 hours of e-MERLIN legacy survey time for the
Supercluster Assisted Shear Survey (SuperCLASS), which will image a 1.75\,\sqdeg\ region of sky containing a supercluster  to a depth of 4\,\ujybm\ with 0.2\,arcsec resolution, including polarisation information.
This study
 is expected to detect  $\sim$1--2 galaxies per
sq. arcmin, which will be used to map   the
dark matter distribution in the vicinity of the supercluster \citep{brownM11b}.

\subsection{Variability and Transients}
\label{transients}

Transient and variability studies
are key science drivers for SKA pathfinder instruments, and potentially for the SKA as well.  
Here we briefly discuss some of the key issues that need to be considered and addressed on the road to the SKA, potentially via pathfinder instruments. We also highlight some of the outcomes that can potentially be achieved through future surveys.
 
Blind and dedicated radio transient surveys can potentially
\begin{itemize}
\item discover new classes of energetic objects\\ (`unknown-unknowns'),
\item monitor known transient and variable radio sources ('known-knowns'),
\item  search for previously undetected transient and variable sources (`known-unknowns'),
\item explore the intervening medium via extreme scattering events, and
\item search for a SETI signal.
\end{itemize}

Table \ref{var_table} summarises our current knowledge of variable radio sources
(see also \cite{Con79} and references therein),  showing observations  at a variety of cadences, frequencies and Galactic latitudes. A complex variety of variability is reported; in some cases short time-scale variations at low frequencies are attributed to interstellar scintillation, while long-term variations at GHz frequencies are in some cases attributed to AGN activity. A systematic experiment to map the surface density of variable radio sources as a function of flux, frequency and cadence would be an invaluable step on the road to the SKA.

Source variability \citep[including bright calibrators, see ][]{Bryan} may affect the calibration and clean strategies required for future surveys and will ultimately impact on the final dynamic range and image fidelity (e.g. see \citealt{Stewart_2011}). Therefore characterising the variability and abundance of known radio sources is a necessity for planning future surveys and calibration strategies. Many of the compact nJy sources to be studied by the SKA will be affected by scintillation \citep{Den_Con, Rickett}. Furthermore, at these sensitivities highly variable sources in the range 1$\mu$Jy Ð 1mJy will yield a dynamic range of 1000:1 (and greater). \cite{Carilli} report a surface density of highly variable sources of up to 18 deg$^{-2}$ (see Table \ref{var_table}) at a detection threshold of $\sim$0.1 mJy: such a high abundance could seriously impact SKA surveys.

Possible transient sources range from extrasolar planets at frequencies below 50\,MHz \citep{JM} to brown dwarf flares at 8\,GHz \citep{Berger02}. Early blind searches for transient sources suggest that a bright ($\geq 0.1\,$Jy), frequent, population of GHz transients does not exist \citep[see e.g.][]{Croft,Bower2010,BowerCAL,ATATS2,Bell}. The benchmark survey of \citep{Bower_2007} which predicted a substantial number of radio transients in the sub-mJy range has recently been revised by \cite{Frail_2012},
 lowering the predicted surface density of sources.  An all-sky survey could however still potentially detect several thousand new transients, though a much lower yield would also be consistent with the current uncertainties  \citep[see e.g.][]{Keith, Murphy12}.

So far many surveys have been conducted in the GHz regime, at low frequencies, synchrotron sources may be optically thick with lower peak fluxes. Coherent processes therefore might dominate transient detections below 1\,GHz, and rate estimates should differentiate between the two processes. Exploring the transient rates at low frequencies is an especially important goal for pathfinders such as LOFAR and the LWDA \citep{Fender_2008, Lazio2011} and ultimately SKA-Low.

A goal for the transient key science projects is to sample a given field on a logarithmic range of time-scales, perhaps commensally with a project revisiting a given field at a variety of time-scales to build up the necessary sensitivity. A further goal of the variability studies will be to follow up on objects (such as pulsars or radio stars) which may be detected in the large continuum surveys, and may be manifested in the image as having spoke-like features, as discussed in \S \ref{pwn}.
When the EMU and WODAN surveys start, transient and variability studies through projects such as VAST \citep{Vast10, Murphy12} can be performed commensally via comparison with previous catalogues such as SUMSS \citep{SUMSS} and NVSS  \citep{Condon98}. Through commensal and targeted observations a more complete census of the abundance of transients and variables can be achieved prior to SKA operations.

\begin{table*}
\centering
\caption{A selection of variable source statistics taken from the literature. $\rho$ gives the snapshot rate of sources (deg$^{-2}$). $t_{char}$ gives the characteristic timescale on which the variability was sampled. $\frac{\Delta S}{S}$ gives the fractional change in flux (please refer to individual publications for further details).}
\label{var_table}
\begin{tabular}{|l|c|c|c|c|c|c|c|}
\hline  Study & Flux (mJy)  & $\rho$(deg$^{-2}$)& $t_{char}$ & $\nu$ (GHz) & $\frac{\Delta S}{S}$ \\
\hline
\cite{Keith} & $>$14 & 0.268 & days - years &  0.843 & $\geq$50\% \\
\cite{Carilli} & $>$0.1 & $<$18 & 19days \& 17mths & 1.4 & $\geq$50\%\\
\cite{Becker} & $>$0.1 & 1.6 & $\sim$ 15 years & 4.8 & $\geq$50\% (d)\\
\cite{Frail_CAT} & $>$0.25 &  5.8 & $\sim$ 1 day & 5 and 8.5 & $\geq$50\% \\
\cite{Bryan} & $>$ 2500 & - & days - years & 0.843 & $<$20\% (d)\\
\cite{Condon_1975} & $\sim$ 1000--25000  & - & days & 2.695 \& 8.085 & 0.5\% \& 0.98\% (a) \\
\cite{Den_1981} & $\sim$ 500--33000  & - & 5 to 10 years & 0.318 & 8--100\%\\
\cite{Simonetti} & 400--12000 & - & days & 0.820 \& 1.41 & 4.1\% \& 3.5\% (b) \\
\cite{Ryle} & 200--2000 & - & months & 2.7 to 15.4 & 10\%--50\% (c) \& (d)\\
\cite{Taylor} & 18--1200 & - & days--months & 5 & 10\%--400\% (c) \& (d)\\
\hline
\end{tabular}
\begin{flushleft}
(a) \textit{0.5\% at 2.695 GHz and 0.98\% at 8.085 GHz. Values are derived from the average (daily) fractional change in 16 sources.}\\
(b) \textit{4.1\% at 0.820 GHz and 3.5\% at 1.41 GHz. Values are derived from the average modulation index (rms/mean) in 13 flat spectrum sources.}\\
(c) \textit{also see \cite{Rickett}}\\
(d) \textit{Observations in the direction of Galactic plane.}\\
\end{flushleft}
\end{table*}    

\subsection{Galactic science}

\label{galaxy}

The proposed wide-field surveys such as EMU and POSSUM include  the
Galactic plane, giving the opportunity to create a sensitive atlas of
discrete Galactic radio continuum sources. Most of these sources represent the
interaction of stars at various stages of evolution with their
environment. The high sensitivity of the surveys allows access to all stages
of HII regions from the optically thick hyper-compact kind presumably related
to the earliest stages of a ``turned-on'' star to large HII regions
related to massive mature or giant stars producing strong stellar winds.
Due to the high resolution of these surveys we can detect the youngest, most
compact supernova remnants and pulsar wind nebulae. We can detect and study
planetary nebulae giving us insight in the late stages of the evolution of
low mass stars. And we can detect radio stars and pulsars. High sensitivity
polarisation surveys such as POSSUM will also open the door to the
identification of currently
unexplained polarisation features known as Faraday screens.

At 1.4\,GHz the main mechanisms for radio emission from
discrete Galactic objects are thermal free-free emission from HII regions and
planetary nebulae and non-thermal synchrotron emission from supernova remnants
and pulsar wind nebulae. Resolved Galactic sources can be separated by their
spectral
and linear polarisation properties. Young HII regions such as compact,
ultra-compact,
and hyper-compact sources are optically thick at frequencies
around 1.4\,GHz, and thus easy to separate from not only Galactic but also
extragalactic sources even if they are not resolved. Large evolved HII regions
have flat optically thin spectra compared to supernova remnants that may show
similar structures but have steeper spectra, and are typically linearly
polarised. Pulsar wind nebulae also have flat radio spectra, but are
linearly polarised. Comparisons with surveys at other wavelength such as
infrared or X-ray will further help to discriminate individual sources of interest for follow-up studies.

\subsubsection{HII regions}

The high sensitivity of the planned wide-field 1.4\,GHz surveys such as EMU and WODAN allows access to all stages of the evolution of HII regions, even though free-free emission near 1.4\,GHz will be optically thick for the very dense ultra- and hyper-compact HII regions (UCHII and HCHII, respectively), which are too small to be resolved by a 10\,arcsec
beam, unless they are very nearby. UCHII and HCHII represent a very young phase in the development of an HII region \citep{churchwell2002}.

\begin{figure}[htb]
\centerline{
\includegraphics[width=7cm]{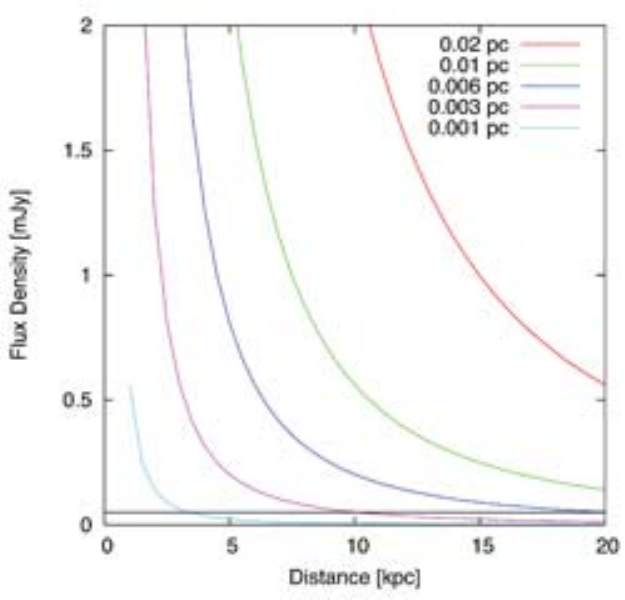}}
\caption{\label{fig:hii} Predicted peak radio flux density of unresolved
UCHIIs and HCHIIs at 1.4\,GHz as a function of distance from the sun. UCHIIs and
HCHIIs with different diameters are displayed. We assumed a constant
electron temperature of 8000\,K, a homogenous
distribution of material, and high optical depth ($\tau \gg 1.0$).}
\end{figure}

Expected flux densities at 1.4\,GHz for the most compact HII regions are displayed in Figure~\ref{fig:hii}. The separation between ultra-compact and hyper-compact HII regions is at a diameter of about 0.02\,pc. Hence, EMU will be able to detect {\em all\/} ultra-compact HII regions and also many hyper-compact objects. There is still confusion about what a hyper-compact HII region actually is. Are they just extremely compact ultra-compact HII regions or a new class of objects? There is evidence that HCHIIs have somewhat different spectral characteristics than the UCHIIs, that might be explained by a density gradient within the ionised gas \citep{kurtz2005}. So far only a small number of HCHIIs are known. A more complete census of these objects with a sensitive wide field survey is required to get a more complete picture of the evolutionary process of HII regions. Follow-up observations at higher radio frequencies or a comparison with high radio frequency surveys is then required to determine physical characteristics which will help to unravel the true nature of these objects.

We can also probe the magnetic field inside HII regions by either studying the rotation measure signature of polarised point sources ``shining through'' the object \citep[e.g.\ ][]{harvey11} or by studying overlapping or even related diffuse polarised emission \citep[e.g.\ ][]{foster06}.

\subsubsection{Planetary nebulae}

Planetary nebulae (PNe) are the most abundant compact Galactic sources in the NRAO VLA Sky Survey \citep[NVSS,][]{Condon98}. \citet{Condon98b} identified 680 NVSS radio sources brighter than 2.5\,mJy with planetary nebulae. Assuming a detection limit of $50\,\mu$Jy, EMU will be able to detect unresolved PNe up to seven times farther away than NVSS, potentially leading
to a significant increase of the number of PNe detected at radio frequencies.  PNe can be very useful
tools for measuring extinction and probing star-formation rates of stars not massive enough to produce supernovae or HII regions \citep{Condon98b,Condon99}. Identification of these PNe, however, will
be difficult.

The ionised low density circumstellar material of nearby planetary nebulae, released during the AGB wind phase of the central stellar object, is illuminated by background linear polarised emission through Faraday rotation. Free thermal electrons embedded in a magnetised medium rotate
background linear polarisation emission to create a rotation measure signature observable at 1.4\,GHz. A typical radio telescope operating in this wavelength range proves to be more sensitive to the Faraday rotation of a plasma region than to its bremsstrahlung signature.
Linear polarisation studies of nearby planetary nebulae can reveal important information about the mass loss history of old AGB stars or white dwarfs that cannot be obtained easily any other way
\citep[e.g.\ ][]{ransom08,ransom10}.

\subsubsection{Supernova Remnants}
\label{snr}

Only 274 supernova remnants (SNRs) have been discovered so far in our Galaxy \citep{Green09}\footnote{http://www.mrao.cam.ac.uk/surveys/snrs/snrs.info.html}. However, the estimated population of SNRs is much higher  
\citep[maybe 500--1000; ][]{Helfand06}. This discrepancy is mainly caused by a strong bias towards bright extended objects. These are the mature SNRs expanding into a medium to high density environment. There are two groups of missing SNRs. The compact, presumably very young, objects were missed by previous wide-field surveys, because they are either
confused by other nearby objects or mistaken for extragalactic sources if they are not resolved. The second group of missing SNRs are low surface brightness SNRs. A population of more than 1000 SNRs for our Galaxy can also be extrapolated using the Canadian Galactic Plane Survey \citep[CGPS,][]{taylor2003} and its supernova remnant catalogue \citep{kothes06}. The CGPS is a radio continuum and HI survey of the Outer Galaxy. Since the distance to the edge of the Galaxy, and hence to every object in the survey, is very short, the $\sim$ 1 arcmin resolution at 1420\,MHz radio continuum ($\le 1$\,pc) is sufficient to resolve any SNR - even those which are very young.

EMU and POSSUM will have a very similar spatial resolution than the CGPS but much higher sensitivity. The 10 arcsec beam translates to about the same spatial resolution at the most distant edge of the Galaxy through the inner part of the Milky Way ($\approx 20$\,kpc) than the CGPS beam does towards the Outer Galaxy. Hence, we would expect to detect a similar number density of SNRs. This indicates that EMU in conjunction with POSSUM could discover more than 500 new SNRs. Resolved SNRs can be easily distinguished from extragalactic sources and Galactic HII regions by an analysis of structure and spectral and polarisation signatures.

Radio polarisation observations of SNRs can also be a powerful tool to probe the magnetic field into which these objects are expanding. SNRs act as magnifying glasses of ambient magnetic fields, since they freeze and compress it into the expanding shell of the swept up material. Radio polarisation and rotation measure studies can reconstruct the 3-dimensional ambient magnetic field, as demonstrated in the studies by \citet{uyaniker02, kothes09, harvey10}. Radio polarisation studies of SNRs with EMU and POSSUM can probe the large-scale magnetic field of the Inner Galaxy, detect potential field reversals, and probe the transition from the Galactic plane to the halo.

\subsubsection{Pulsars and their wind nebulae}
\label{pwn}

Pulsars generate magnetised relativistic particle winds, inflating an expanding bubble called a pulsar wind nebula (PWN), which is confined by the expanding supernova ejecta. Electrons and positrons are accelerated at the termination shock some 0.1\,pc distant from the pulsar and interact with the magnetic field to produce synchrotron emission across the entire electromagnetic spectrum. At GHz frequencies the flat spectrum PWNe stand out from steep spectrum SNRs and can be distinguished from HII regions by their linear polarisation signal.

The high resolution and sensitivity of the proposed wide field 1.4\,GHz surveys will enable us to built a more complete census of Galactic pulsar wind nebulae down to unprecedented levels.

The number of pulsars at high Galactic latitudes is poorly constrained since most surveys concentrate on a narrow band along the Galactic plane. Out of the 1433 pulsars with measured 1400\,MHz fluxes listed in the ATNF Pulsar Catalogue
\citep{manch05},
1410 have integrated fluxes above $50\,\mu$Jy. Only 310 of those have Galactic Latitudes $|b| > 5 \deg$. Sensitive high resolution radio continuum surveys will enable us to get a more complete picture of the pulsar distribution in the Milky Way Galaxy and their rotation measures will help to constrain the configuration and strength of the large-scale Galactic magnetic field.

Pulsars should be distinguishable from other point-like radio sources because of their steep radio continuum spectra, their polarisation, and
artefacts caused by short term variability. These observational signatures will be complemented by the variability and transient surveys discussed in \S \ref{transients}.

\subsubsection{Radio stars}

Radio stars emit a tiny fraction of their total luminosity in the radio band.
For example, the quiet Sun has a radio  luminosity  of $10^{14}$\,W, which is only
$\sim 10^{-12}$  of its bolometric luminosity.
Nevertheless, in many cases, radio observations of stars and stellar systems have revealed astrophysical phenomena,
not detectable by other means, that play a fundamental role in our understanding of stellar evolution and of physical processes
that operates in stellar atmospheres.

Generally, the brightest stellar radio emission appears to be associated with magnetically-induced phenomena, such as stellar flares,
related to the presence of a strong and/or variable stellar magnetic field (high brightness temperature) or with enhanced mass-loss\\
(large emitting surface).

Among the brightest radio sources are active stars and binary systems,
including flare stars, RS~CVns, and Algol binary systems, characterised
by strong magnetic activity which drives high energy processes in their  atmospheres.
Their radio flux density is highly variable and it usually shows two
different regimes: quiescent periods, during which a  basal flux density of few mJy is observed, and
active periods, characterised by a continuous
strong flaring which can last for several days \citep{Umana95}.

Both quiescent and flaring radio emission show
spectral and polarisation characteristics consistent with non-thermal
radio emission, probably driven by the magnetic activity manifested at other wavelengths.
In this scenario, the radio flux is due to gyrosynchrotron emission \citep{Gudel09}, caused by  the interaction
of the stellar magnetic field with mildly relativistic
particles. A similar mechanism causes the radio emission
from pre-main sequence (PMS) stars and X-ray binaries. Non-thermal radio emission is also seen from shocks of colliding winds in massive binaries.

Stellar radio flares can occur also as
narrow band, rapid, intense and highly polarised (up to 100\%) radio bursts, that are
particularly common at low frequency ($< 1.5$\,GHz). They are believed to be the result of coherent emission mechanisms, requiring  a strong magnetic field (which may be variable) and a source of energetic particles.
Coherent burst emission has been observed in RS~CVns, flare stars, Brown dwarfs (BD), and chemically peculiar stars (CPs) \citep{Slee08, Berger02,Trigilio00}. Coherent emission has been detected in just a few tens of stars,
 because of the limited sensitivity of the available instruments.
 
Stellar radio emission often exhibits significant circular polarisation \citep{Trigilio11,Ravi10}.While some extragalactic sources do exhibit circular polarisation \citep[e.g.][]{Rayner00}, the degree of polarisation is small (typically $\sim$ 0.01\%), compared to the levels up to 100\% seen in radio stars. It is therefore important, for the purposes of distinguishing stars from extragalactic objects, that surveys record circular as well as linear polarisation.

Thermal emission (bremsstrahlung emission) is expected from winds associated with WR and OB stars, shells surrounding PNe, and Novae
and jets from symbiotic stars  and class 0 PMS stars \citep{Gudel02}. In all these cases, radio continuum  emission is an important diagnostic to understand the underlying physics. For example, radio continuum  observations are the most precise way yet to determine the mass-loss of stellar winds. This is particularly important  when other
 diagnostics cannot be used, such as in the case of dust enshrouded objects \citep{Umana05}.

\begin{figure}
\includegraphics[width=8cm]{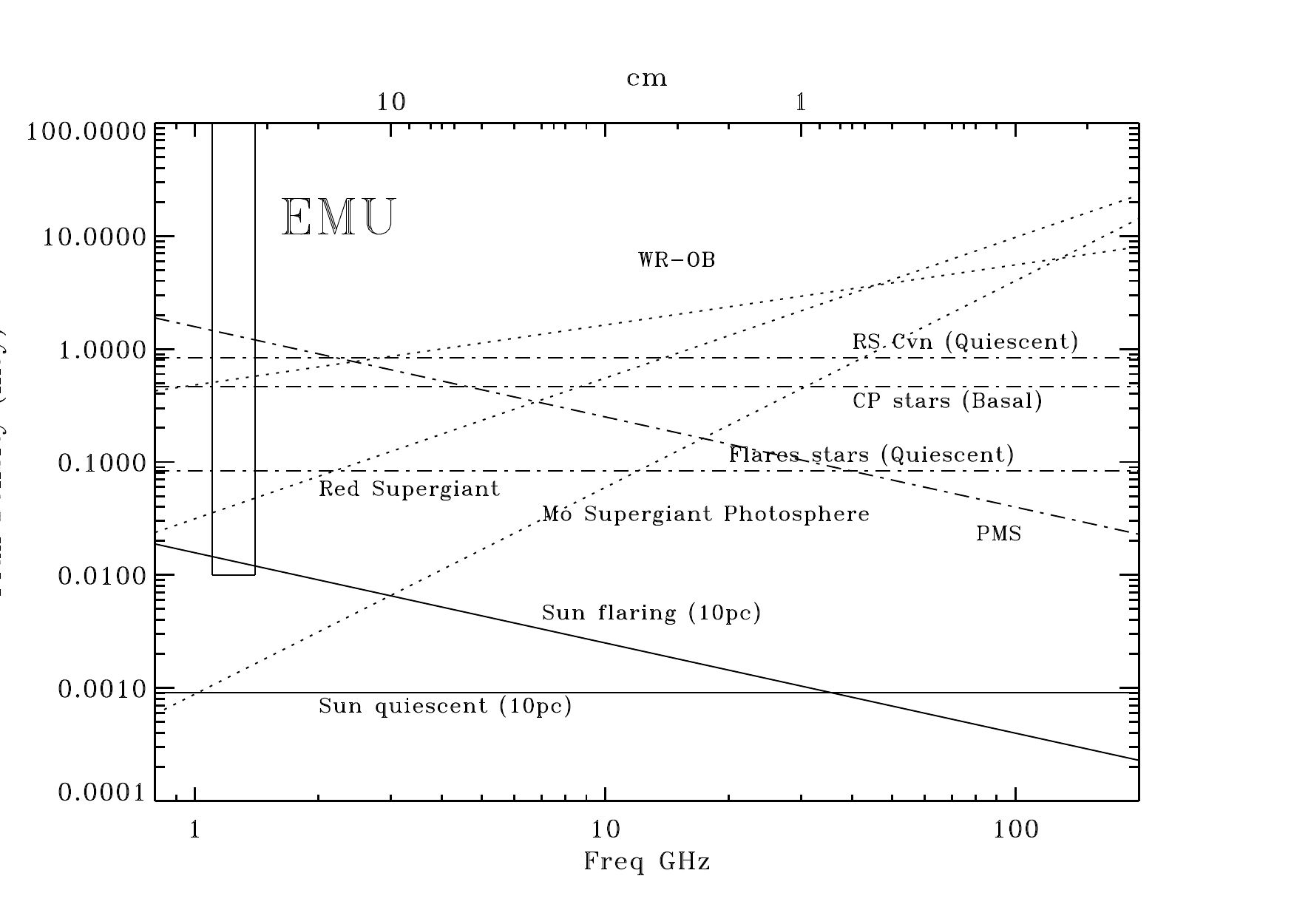}
\caption{Typical radio spectrum of several classes of radio emitting stars. Fluxes have been derived from the radio luminosity
\citep{Seaquist93, Gudel02, Umana98, Trigilio08, Berger06}  assuming an appropriate distance
for each type of radio star (10\,pc for flares stars, including late M-L, 100\,pc for active binary systems, 1\,kpc for supergiants, OB and WR, 500\,pc for CP stars).
The bandwidth and sensitivity of EMU have been also indicated.}
\label{fig:radiostars}
\end{figure}

Since no sensitive survey for radio stars has yet been conducted, we have little statistical information on stellar radio emission is available. Forecasts of the number of sources detectable with SPARCS surveys  rely on radio observations of small samples of stellar sources,  usually selected on the basis of observed peculiarities at other wavelengths.
Assuming typical radio luminosities for each type of radio star, we conclude that  stellar winds and non-thermal radio emission from many active binaries, flares stars, PMS will be easily detected  with an rms sensitivity of 10\,\ujybm\ (EMU),  while SKA  (1\,\ujybm\ rms) will be able to detect a quiescent Sun at 10\,pc (see Figure~\ref{fig:radiostars}).
 
Until now, targeted  observations of well known radio stars have constituted the best approach to  investigate  their radio emission mechanisms (e.g. flare development, spectral and polarisation evolution, emission mechanism, etc.).
An all-sky  survey  would  significantly enlarge the known stellar radio emitting database, free from  selection effects.
Results from  such a survey would provide new insights into the physics of active stellar systems
and plasma  processes.

Two areas of stellar radio emission    will particularly benefit from SKA and SPARCS surveys.
One is the study of gyrosynchrotron stellar flares, where  multi-epoch, multi-frequency  observations will enable the detection of serendipitous flaring activity,  allowing the derivation of a typical behaviour
(occurrence rate, variation amplitude, etc.) from a statistical study of a larger source population. Moreover, detailed studies of a large number of stellar coronae will be possible, allowing us to understand
the nature of energy release in the upper atmospheres of stars of different mass and age and to investigate the correlation between radio and X-ray emission and thus on the Neupert effect
 \citep{Gudel09}.
The expected sensitivity will also  test the presence of
super-flares on solar-type stars, and search for  correlations with an orbiting hot Jupiter  \citep{Maehara12}.

The other important area is related to the study of coherent emission.
The EMU plus WODAN surveys will offer the best opportunity  yet to determine  how common coherent radio emission  is from stellar and sub-stellar systems.
The detection  of coherent emission from a large fraction of surveyed stars will have immense implications for our understanding of both stellar magnetic activity
and the dynamo mechanism generating magnetic fields in fully convective stars
and brown dwarfs  \citep{Hallinan08,Ravi11}.   
Coherent emission observed in active and BD stars\\  
shares several characteristics with that observed in CP stars
\citep{Trigilio00}, since both require a large-scale magnetosphere, and are similar to the low frequency coherent
radio emission observed from the magnetised planets in our solar system \citep{Trigilio11}.
The well-known topology of CPs magnetic fields, independently derived from optical
observations, makes this kind of object an excellent laboratory for stellar magnetosphere studies. Matching of
predicted emission with that observed yields parameters of the radio source, such as the surface magnetic field,
the number density of the emitting electrons, and the energy spectrum of the electrons.  In turn, this provides
clues to the acceleration process, the size of the magnetosphere (Alfven radius) and the inclination of the
rotational axis.

If coherent emission is  present in many radio active stars, with the same characteristics,  it will
constitute an excellent diagnostic for star magnetospheres,
and  a powerful probe of magnetic field topology.
In one CP star (CU~Vir) the coherent emission is stable on a time-scale of
years, and has been used to time the rotation of the star, revealing a likely change of its period
\citep{Trigilio08, Ravi10}.
The
discovery of other similar radio lighthouses  will enable high precision studies of
the rotation period, and  thus angular momentum evolution, in different classes of star.

\subsection{Unexpected discoveries}
\label{WTF}

Experience has shown that many great discoveries in astronomy have been made, not by testing a hypothesis, but by observing the sky in an innovative way. The necessary conditions for this to take place are (a)~a telescope observing an unexplored part of the observational phase space (frequency, resolution, time-domain, area of sky, etc), (b)~an intelligent observer who understands the instrument sufficiently well to distinguish between artefact and discovery, (c)~a prepared and enthusiastic mind ready to accommodate and interpret a new discovery.

SKA pathfinder surveys  will easily satisfy (a), if only in terms of the numbers of objects surveyed. However, their petabytes of data, and arms-length access, may prevent an observer from satisfying (b) and (c). Although we may hope that someone will eventually stumble across any unexpected phenomena in the data, the impenetrable size of the database implies dark nooks that may never be fully explored. If we rely on serendipity, discoveries may remain undiscovered, for ever.

An alternative is to harness data-mining techniques to help the intelligent observer search for the unexpected.
For example, the WTF (Widefield ouTlier Finder) \citep[e.g.\ ][]{Norris11b} project will systematically mine the EMU database, searching for the unexpected by 
discarding objects that already fit known classes of object, using a variety of approaches including:
\begin{itemize}
\item decision tree analysis
\item cluster analysis
\item k-furthest neighbour (kFN) by analogy with the kNN technique\citep{Zinn12}
\item Bayesian analysis
\end{itemize}

Identified objects/regions will be either processing artefacts (which are themselves important for data quality control),
statistical outliers of known classes of object, or, in a few cases, genuinely
new classes of object.

\newpage

\section{Technical Challenges}
\subsection{Survey design \& quality}

The SKA pathfinder radio continuum surveys may be characterised by the following parameters:
\begin{itemize}
\item point-source detection limit,
\item angular resolution,
\item brightness sensitivity,
\item dynamic range,
\item frequency coverage,
\item polarisation,
\item accuracy:
  \begin{itemize}
  \item flux densities,
  \item positions,
  \item polarisation,
  \item image fidelity,
  \item uniformity in sensitivity.
  \end{itemize}
\end{itemize}

All of these need to be addressed for these surveys. Ironically, the more ambitious a survey, the more susceptible it will be to small systematic errors. The ambition of the surveys therefore needs to be matched by a corresponding level of effort to understand subtle sources of error and uncertainty. For example, many radio-astronomers have been content to tolerate 10\% uncertainty in their measurements of flux density, while we know that our instruments are, in principle, capable of measuring flux densities to 1\%. Reasons for this failure include deconvolution errors, errors in primary beam shape and pointing, failing to correct for missing extended flux, resolution bias, bandwidth smearing, time smearing, CLEAN bias, Eddington bias, etc. To overcome these effects (and probably find new effects which we have not yet met)  will doubtless require a significant amount of effort to understand the instrument and its errors, but that effort is justified by the scientific return from these enormous surveys.

It will also be important to present the data to users in ways that are clear and unambiguous, with uniform coverage and easily understood survey parameters. For example, most surveys produce catalogues chosen at  a $\sim 5\,\sigma$ level of confidence. For some purposes, a less reliable catalogue at a $3\,\sigma$ level (for greater numbers of sources) may be useful, while for other studies a threshold of $8\,\sigma$ or $10\,\sigma$ level (for greater reliability) may be more appropriate \citep[e.g.\ ][assumes $10\,\sigma$]{Raccanelli11}. While an $8\,\sigma$ subset can always be chosen from a
$5\,\sigma$ sample, it may cause less confusion amongst users for the survey to generate these different levels of catalogue.

It is also important to remember that these surveys will be used for purposes that  haven't yet been thought of. For example,
one of the most exciting results from the NVSS was the measurement of the
RM sky (and hence the Galaxy's magnetic field) \citep{taylor09}. That was not planned  at all during the survey, but was subsequently possible because the survey had excellent   continuum calibration.

Management of large
surveys is likely to be an issue, with previous large radio surveys
having had relatively few people with
the responsibility for conducting the survey in contrast to the
emerging model in which large teams are responsible.

\subsection{The SPARCS Reference Fields}

The uniformity required from the SPARCS surveys cannot be achieved by simply ensuring that the calibrators are on the same flux scale, although this is an essential first step. For example, the fluxes of strong sources in the first ATLAS data release \citep{Norris06, Middelberg08a}  agreed with those measured by the VLA, while weaker ATCA sources were systematically weaker than weak VLA sources, an effect which turned out to be caused by incorrect assumptions about bandwidth smearing in mosaiced ATCA data \citep{Hales12a}. Other subtle flux-dependent effects doubtless await  us, and consistency between our calibrator sources will not be sufficient to detect or fix them. There have even been suggestions of dec\-lination-dependent calibration effects \citep{parra10}.

During the commissioning of the SPARCS telescopes, particularly those using the relatively untried PAF technologies,  it will be important to repeatedly survey well-studied  fields containing a grid of sources, pushing the sensitivity and dynamic range until all the effects are well-understood and can be corrected for.

The measured flux densities of sources in radio surveys are subject to a large number of subtle corrections and bias effects, and even expert radio-astronomers do not always agree on these. For example, there is a continuing debate on the measured scatter in the radio source counts at low flux densities (see Fig. \ref{srccnt}), with some proponents arguing that it is caused by  cosmic variance, and others arguing that it is due to processing and calibration differences. In some cases, reprocessing  or reobserving  \citep[e.g.][]{Condon12}  has shown that some variations are caused by different processing paths, although cosmic variance may be sufficient in other cases.  
The problem may be even worse for EMU and WO\-DAN, as both rely on new  phased array feed (PAF) technology. To extract the maximum science output from these major projects, relying on new PAF technology that has yet to be sufficiently battle tested,
requires careful characterisation of individual survey systematics.

To achieve this, we propose three reference fields at  declination $\sim +30\deg, 0\deg$, and $-30\deg$), which can be observed by all existing and new radio telescopes. Where possible, they have been chosen  to overlap with a field that is well-studied at other wavelengths, to maximise the  science to be obtained from these observations. The two Southern fields are in the CDFS-ATLAS field \citep{Norris06} and the COSMOS field \citep{Schinnerer07}. A Northern field (15:30, +29:00) has been chosen within the overlap range of ATCA and Westerbork, within the SLOAN survey area, and at least 2.5$\deg$ away from any strong ($>$0.5 Jy) sources.  
It is planned to observe all three fields, as far and deeply as possible, with all existing survey telescopes (VLA, ATCA, Westerbork, LOFAR, GMRT, etc.) as well as the new
 SPARCS surveys.

\subsection{Calibration}

\label{calibration}

Calibration and imaging of SKA pathfinder surveys is likely to present significantly new challenges compared to earlier surveys. For example:
\begin{itemize}

\item Direction-dependent calibration is likely to be important not only for low
frequencies, such as \hbox{LOFAR}, but also for intermediate
frequencies, such as  EMU and probably even for
 \hbox{VLA} and \hbox{MeerKAT}.
 
\item Because all
data are acquired in a spectral-line mode,  flagging the data for RFI affects
 the weighting of the data.
 

\item The time scale of gain variations, such as those resulting from
pointing errors, is important.  For example, wind gusts have a
particularly dramatic effect on pointing errors, in part because the
errors scale as the square of the wind velocity.  The result is quite rapid
pointing variations, which may be difficult to track and correct.

\item Experience from multi-epoch observations with the WSRT has
illustrated the importance of monitoring the calibration information
over time.  In particular, there can be time variability of the
calibration information (e.g., a variable source in the field of view
of the calibrator), which could affect the calibration or result in
apparent variables.

\item The scientific drivers for high dynamic range imaging need to be
identified and kept clear.  There may not be a need for uniform and
high dynamic range across the entire sky.
\end{itemize}

\subsection{Imaging}

At the excellent sensitivities offered by SKA pathfinder telescopes, imaging
and calibration algorithms need to deal simultaneously with issues of
\begin{itemize}
\item wide-bandwidth (e,g, primary beam corrections and polarisation leakage terms change with frequency),
\item wide-field imaging,
\item advanced multi-scale deconvolution techniques for imaging extended emission,
\item directionally-dependent effects such as instrumental and  ionospheric errors,
\item the need for high dynamic range and high polarisation purity.
\end{itemize}

 Furthermore,
 the inherent data volumes are large (typically petabytes) so that
computational efficiency and the number of data traversals are
important parameters to consider in designing post-processing
strategies.

\subsubsection{Existing algorithms}

Two distinct approaches are being pursued in the community to correct
for wide-field and direction-dependent effects, which can be classified as (a) projection
algorithms and (b) faceting algorithms \citep{SB1}.  

Projection algorithms
are based on physical modelling of the various direction dependent (DD)
terms of the measurement equation and incorporating them as part of
the forward and reverse transforms.  These algorithms therefore
do not require assumptions about the source
brightness distribution. This allows them to take advantage of the FFT
algorithm for Fourier transforms, as well as integrate well
with advanced deconvolution techniques for scale-sensitive wide-band
imaging.

Faceting algorithms on the other hand are based on data partitioning in
such a way that standard direction {\it independent} techniques can be
applied to the partitions \citep{SB2a, SB2b, SB2c}.  This approach is phenomenological
by design requiring no physical understanding or modelling of the DD
term in the measurement equation.  They can however suffer from
non-optimal use of the available signal-to-noise ratio (due to data
partitioning), curse of dimensionality (too many degrees of freedom)
and computational load.  To alleviate some of these problems, in
addition to the problem of degrees of freedom, assumptions
about the structure of the brightness distribution typically have to be made.

The W-Projection algorithm \citep{SB3} incorporates the effects of non
co-planar baselines and corrects for its effects during imaging. The
A-Projection algorithm \citep{SB4} similarly accounts for the time,
frequency and polarisation dependence of the antenna primary beams
modelling the antenna aperture illumination patterns (or the
measured illumination patterns).  The combined AW-Projection algorithm
therefore can be used at low frequencies where the effects of the
W-Term as well as that of antenna primary beams limit the imaging
performance.  Antenna pointing errors are thought to limit the imaging
performance at high frequencies and in mosaic imaging at all
frequencies.  Using A-Projection for forward and reverse transforms,
the Pointing SelfCal algorithm \citep{SB5} solves for the time dependent
antenna pointing errors.  The pointing solutions are then incorporated
as part of the A-Projection to correct for the effects of antenna
pointing errors.

The Multi-Scale Clean \citep{Cornwell08} and Asp-Clean \citep{SB7} algorithms are
advanced image deconvolution techniques for imaging fields with
extended emission.  The MS-MFS algorithm \citep{SB8} for wide-band
multi-frequency synthesis not only deals with the deconvolution of
extended emission, but also accounts for spatially resolved spectral
index variation across the field of view.  The combination of
AW-Projection and MS-MFS algorithms can therefore be used to account {\it
simultaneously} for all the dominant wide-field wide-band
effects which are expected to limit the imaging performance with
SKA-pathfinder telescopes.

The algorithmic design of the combined AW-Proj\-ection and MS-MFS
algorithm is such that it falls in the category of embarrassingly
parallel algorithm.  To mitigate the fundamentally higher computing
load, work is in progress to deploy these combined advanced algorithms
on High Performance Computing (HPC) platforms, typically consisting
of a cluster of computers with multi-core CPUs and high bandwidth
interconnect \citep{SB9}.  

Since these algorithms also need to iterate over a large volume of
data, computing clusters connected to a parallel file system
(e.g.\ the Lustre file system) and using multi-threaded I/O techniques
are currently being used.  Tests with up to a few Terabytes of data show close
to linear scaling in computing with number of computing nodes, although this is subject to the achievable compute-to-I/O ratio.
Work is in progress to address the problem of improving this ratio
while keeping the resulting memory footprint in reasonable limits.
For SKA-sized data sets, iterating over the entire data set might not
be feasible. Work is in progress to devise computing
schemes and algorithms that may eliminate or at least reduce the
number of iterations involving the entire data.

\subsubsection{Development of new deconvolution algorithms}
\label{CS}

In the last few years,
the   theory of Compressed
Sensing
\citep[CS:][]{Donoho06,Candes06} has been developed by the signal/image
processing community, and may be applicable to radioastronomical
image deconvolution.

The theory of CS mostly addresses random sampling,
sparse signals, and coarse-grained dictionaries.  In radio interferometry, sampling is usually
structured, astronomical images are not sparse, and
dictionaries  must be  fine-grained to adequately represent the images. Theoretical CS results therefore do not usually improve
estimates of astrophysical sources from interferometric data.  However, a key ingredient of the CS framework, sparse representations \citep{Mallat08}, has demonstrated significant improvements in
radio-astronomical deconvolution, and has been used  as a tool for deconvolution of extended and diffuse sources in the image plane \citep{Wiaux09a, Li11, Dabbech12}, as well as in Faraday depth \citep{Li11b}.



Sparse representation theory  \citep{Fornassier10}
may be seen as a generalization of ideas exploited in
CLEAN \citep[e.g.][]{Hogbom74, Wakker88,Cornwell08}.  CLEAN performs an iterative deconvolution
by using a dictionary of
shifted point spread functions (PSFs) to construct a sparse dictionary of point
sources.
Sparse representation theory aims to construct a dictionary of more complex geometrical
features than point sources, addressing three fundamental issues of astronomical image restoration. First, the sparsity of the dictionary
reduces the indeterminacy and instability caused by the
zeros of the transfer function of an interferometer.
Second,
the theory allows flexible and sophisticated models that can
cope with complex astrophysical sources. Third, it offers
computationally efficient optimisation techniques to solve the
resulting deconvolution problem.  
%

Sparse representations can take either of two approaches: synthesis and
analysis \citep{Elad05}.

In the synthesis approach, the unknown intensity distribution $\bf x$
(of size $(N,1)$, say) is assumed to be sparsely synthesisable by a
few atoms of a given full rank dictionary $\bf S$ of size
$(N,L)$. Hence, we write ${\bf x}$ as ${\bf x=\bf S}\pmb{\gamma}$,
where $\pmb{\gamma}$ (the synthesis coefficients vector) is sparse.
In contrast, the analysis approach assumes that $\bf x$
is {\it{not correlated}} with some atoms of an analysis dictionary $
{\bf A}$ of size $ (N,L)$: $ {\bf A}^T {\bf x}$ is sparse.

Since real images can often be approximated by a linear combination
of a few elementary geometrical features, the synthesis approach is
more intuitive and has been the focus of
more research. Its design simplicity has also made it popular in image processing
applications such as compression, de-noising, and in-painting. However, because a synthesis-based deconvolved image is
restricted to a subspace of the synthesis dictionary, the resulting astrophysical images
may be too rough to be realistic.  
On the other hand, since the signal is not built from a small
number of atoms, the analysis approach may be more
robust to ``false detections'' .  However, the number of unknowns in the synthesis approach
(the number of atoms in the dictionary) may be much larger than in the
analysis approach.
Thus, while
analysis-based optimisation strategies may be computationally and
qualitatively more efficient for large dictionaries
\citep{Starck10}, the question of which approach is best remains
open \citep[][and references therein]{Gribonval09}.

Regardless of the approach chosen, a sparsity-based
deconvolution method still requires a dictionary, chosen from a class of
    images \citep{Mallat08}.  Astronomical wavelet dictionaries
are widely used, but sometimes fail to represent
asymmetric structures adequately. In such cases, it may be necessary to use  other transforms
designed for specific classes
of objects, such as curvelets, which are optimum for curved, elongated
patterns such as planetary rings or galaxy arms, and
shapelets, which are often optimum for  galaxy morphologies.  
Modelling complex images may require several dictionaries to be concatenated into a larger dictionary
(\cite{Chen98,Gribonval03}), although computational issues limit
the size of dictionaries, especially for radio synthesis images
containing hundreds of thousands
of Fourier samples.

Several deconvolution methods that combine these approaches have recently been suggested
\citep[e.g.][]{Suksmono09, Wiaux09a, Wiaux09b, Vannier10,Wenger10, McEwen11}),
including successful simulations of SKA pathfinder
observations. \citet{Li11} present a classical synthesis approach with an
IUWT (Isotropic Undecimated Wavelet Transform) synthesis dictionary,
\citet{Carillo12} used an analysis approach using a concatenation of wavelet bases,
and
\citet{Dabbech12} defined a hybrid analysis-by-synthesis approach.


While the theoretical results of CS have not yet generated tools which can be used routinely for synthesis imaging,
they have brought
new perspectives from the domain of
sparse representations: improved sparsity models, and improved
optimization algorithms.  The increased research effort on fundamental
models for sparse representations (analysis / synthesis / hybrid
models) leads to improved models and thus to improved
reconstruction.
%
Research in sparse representations is active and growing,  and promises valuable future
developments in radio astronomical deconvolution algorithms.

\subsection{Source Extraction and Measurement}

There are many approaches to source extraction, each with different strengths and
weaknesses, so it is important to mount a joint effort to determine how an optimum
source finder would work, and what algorithms to implement. In such a joint effort,
a reference data set should be subjected to the various algorithms, for optimum comparability.

Building a source finder is likely to be subject to compromise,
and current source finders fall well short of what may be theoretically achievable. For example, no source finder currently accounts for variations
in the point spread function across the image. Development of an ``optimum" source
finder may well require collaboration with computer scientists
with no astronomy background to develop signal extraction algorithms.
It is likely that no single algorithm will be optimum for all scenarios.

\subsubsection{Compact Source Extraction}

 Two main areas of recent focus have been investigations into the impact of background
and noise estimation on the detection of compact sources, and a rigorous comparison of existing
source finding software tools. The first of these focuses on source detection, where the background and noise levels in an image are characterised \citep{Huynh12}.

The second focuses on source characterisation, where islands of pixels are described or fit in
some well defined manner \citep{Hancock11}. The most common way to describe a source in
radio astronomy is as a group of Gaussian components.

The process of background characterisation has been explored by \citet{Huynh12}, who tested the steps of background and noise measurement, and of the choice of a
threshold high enough to reject false sources yet not so high that the catalogues are significantly
incomplete. This analysis explored the results from testing the background and thresholding
algorithms as implemented in the SExtractor, Duchamp/\-Selavy, and SFIND tools on simulated data.
The result of this analysis shows clearly that it will be crucial to first develop and then implement an
automated algorithm for establishing the appropriate scale size on which to estimate the ``local"
background and rms noise values for each image, and that this may not be common to all
images. This analysis also suggests that the False Discovery Rate approach \citep{Hop:02}
to thresholding may be the most robust, optimising for both completeness and reliability.

The source characterisation stage has been found to have its own distinct issues \citep{Hancock11}.
For a selection of commonly used source finding tools (SFind, Sextractor, IMSAD, and Selavy)
there are significant differences in the way that islands of sources are divided into components
and subsequently fit. This is exacerbated with examples of sources which are blen\-ded or
adjacent, especially when there is a large difference in the relative flux densities (even
though the fainter component(s) may in their own right be very high signal-to-noise
objects). Such behaviour has particular implications for variable objects. If one such source
were variable or transient, this behaviour could be missed if a source finding program isn't able to
correctly decompose the island. If a source finder's ability to decompose an
island into components is related to the clipping limit, or background noise, then a source of
constant flux could be interpreted as transient or variable due to irregularities in the source
characterisation process.

These issues have triggered the development of new algorithms \citep[e.g.][]{Hancock11, Hales12a} that aim to
determine the correct number of components within an island, and to generate an initial parameter
set for the fitting of multiple gaussian components. \citet{Hancock11} use a Laplace transform,
to identify regions of negative curvature in the pixel intensity profiles, combined with the map
defined from the islands of pixels above a threshold, to automate the process of identifying
the numbers of components in each island, and their initial position and peak flux density
estimates.

Another clear requirement that has developed from these analyses is the need to quantify
the false detection rate of sources as a function of signal-to-noise level, and perhaps
other image or telescope parameters, in order to assess the likelihood that any detected
source is actually real. The impact of even a 0.1\% false detection rate when catalogues of
many tens or hundreds of millions of sources are being anticipated, is clearly large.

These initial steps have demonstrated that while source-finding has been ubiquitous in
radio astronomy for many decades, there is still scope for optimising and quantifying the
process, and that careful attention will need to be paid to these details in the lead up
to the new generation of radio telescopes. There will be many lessons to be learned
in order to ensure that source measurement using the SKA, ultimately, can be made as
robust as possible.

\subsubsection{Extended Source Extraction}

Many of the feature recognition algorithms widely applied in fields such as optical character recognition, machine vision or medical imaging are absent from the repertoire of techniques routinely employed by astrophysicists.  
However, the increasing complexity and scale of astronomical data are driving an interest in more advanced feature extraction techniques for astronomical contexts. In particular, radio surveys on next generation telescopes will generate image data containing tens of millions of astronomical sources, many of which will be extended.

\begin{figure}[h]
\begin{center}
\includegraphics[width=8cm, angle=0]{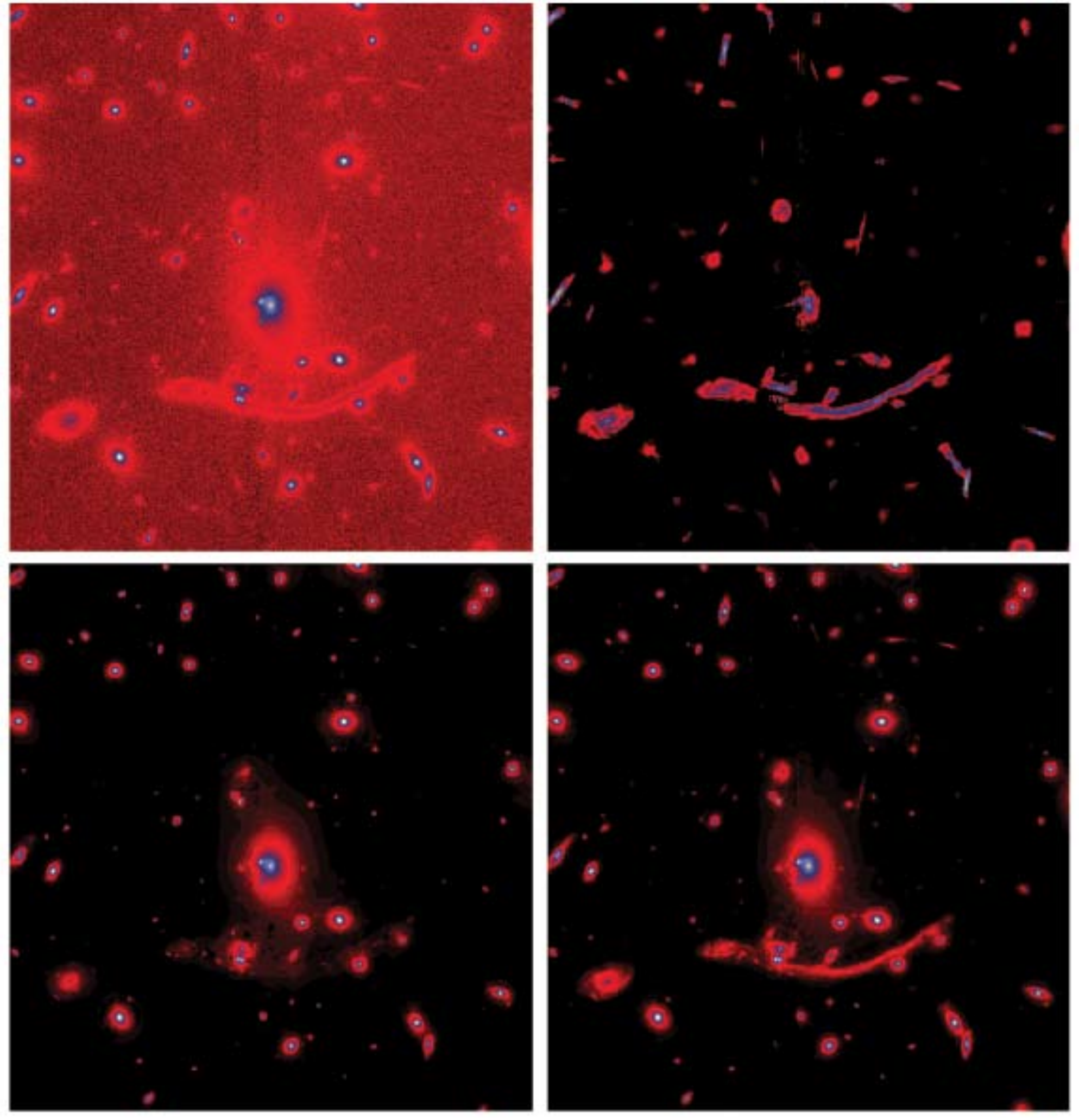}
\caption{Feature extraction in the HST image of Abell 370 from \citet{Starck03}. The top left panel shows the original HST image, top right shows the co-added image from the ridgelet and the curvelet transforms which extract extended features, bottom left shows the reconstruction from the \`a trous algorithm which responds to point sources, and bottom right shows the addition of the results of all three transforms.}\label{fig:A370}
\end{center}
\end{figure}

Localised kernel transforms, such as the Wavelet transform, are often used for astronomical purposes, including diffuse source detection \citep[e.g. detection of diffuse X-ray emission in the COSMOS field:][]{Fino06}. They are created either by modifying a conventional kernel transforms, or by using exotic basis functions, and are supplemented by a localisation algorithm to allow the target to be located. The success of these methods relies on ``matching" the basis function to the problem at hand. Given the variety of source morphologies in astronomical images, no single transform will adequately detect all sources.
Localised kernel transforms work best in cases where a single class of objects can be matched to a suitable set of basis functions, and can also be used to separate diffuse from point source components using curvelet, ridgelet or \'a trous transforms (e.g. Fig. \ref{fig:A370}).

Compressive processes rely on the fact that pixels in an image are not independent, so
 it is possible to encode the patterns in the image rather than the pixel-by-pixel information. This provides a more efficient  ``compressed" representation of the information by providing a `model' of the image. Object detection is then applied to this `model' rather than to the real data, greatly reducing the complexity of the problem. Compression attempts to preserve the information inherent in the original image but does not attempt to abstract geometric or semantic information. Compressive techniques therefore preserve large-scale noise features in addition to diffuse sources. The use of compressive  sampling for deconvolution has been discussed in \S \ref{CS}.

Template matching is a process where an image is correlated with a target object of known form, and is similar to ``matched filtering".
This technique has been used in radio astronomy for pulsar timing prediction \citep{Straten06, Oslowski11}, transient source detection \citep{Trott12}, and to remove radio point source contaminants from diffuse source backgrounds \citep{Pindor11}.
Two disadvantages of template matching are (a) it is hard to control the required number of filters, resulting in a greater number than for localised kernel transforms, and (b) each template  must fit a range of  orientation and scaling. Even when the computational complexity of template matching is reduced by combining it with a scale invariant transform such as the Mellin transform, orientation must still be calculated making this technique very computationally expensive. On the other hand, it has the advantages that (a) the templates do not have to be precise, and cartoon-like models suffice, (b) the method is robust against artefacts, (c) it can be applied in either the image or Fourier domains, and (d) it returns the size, orientation, location  type of detected sources.

\begin{figure}[h]
\begin{center}
\hbox{\includegraphics[width=8cm, angle=0]{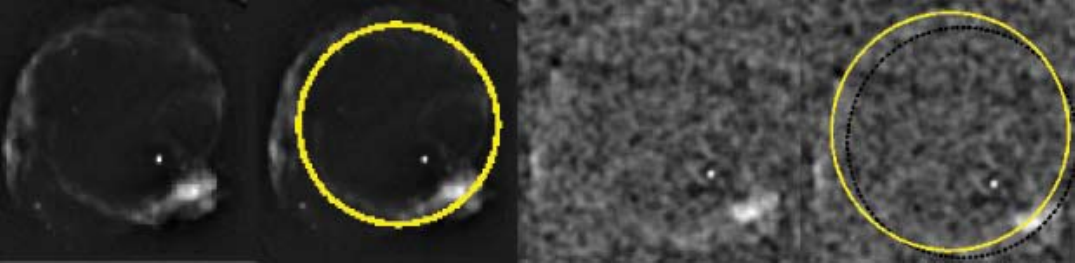}}
\caption{Application of the circle Hough transform to detect radio supernova remnants in the Molonglo Galactic Plane Survey. Left to right: the original image of G315.4-2.3, the location and size of the remnant as found by the CHT, the original image with ten times the Gaussian noise and the response of the CHT in the noise case (yellow) compared to the original (black) which demonstrates the robustness of CHTs to noise.  \citep{Hollitt12b}.}
\label{fig:Hollitt}
\end{center}
\end{figure}

Hough transforms can find and characterise geometrical objects such as lines, circles and ellipses. In general, the Hough transform is equivalent to template matching for some geometric shape, although it inherently incorporates  scaling and orientation constraints. Circle or elliptical Hough transforms are particularly relevant to astronomical source detection to determine questions such as the size and location of a circle/ellipse, and have the advantage that the geometrical abstraction limits the number of filters required (e.\ g.\ different types of objects contain circular and arc-like features).
The Circle Hough Transform (CHT) can find partial circular objects and is robust to noise. The CHT inputs an image  and outputs a three dimensional array in which two dimensions represent the possible locations for the centres of circles and the third dimension spans the set of possible circle radii. Peaks in this so-called ``Hough space" therefore correspond to circles in the input image. In its simplest form the circle Hough transform is computationally challenging, with both computational effort and memory consumption scaling as $O(n^3)$ when transforming an n $\times$ n pixel image. Recently the transform has been recast via a convolution approach which reduces the complexity to $O(n^2)\log n$ \citep{Hollitt09, Hollitt12a}. Although this is not the fastest approach for astronomical source detection, it has been applied as a source detection method for a variety of extended radio sources with arc-like features including  supernova remnants, tailed radio galaxies and radio relics \citep{Hollitt09b, Hollitt12b}.  For example, Figure \ref{fig:Hollitt} shows the detection of a  supernova remnant.
Hough transforms return semantically useful information such as position, size, orientation and eccentricity, but will produce somewhat less information than a full template matching process. For example, it cannot distinguish between a supernova remnant and a tailed radio galaxy.
 
Segmentation methods assume nothing about a source but rather seek to group pixels together based on differences with the background.
Three main classes of segmentation methods are used in astrophysics: region growing, edge detection, and waterfalling.
The simplest version of a segmentation method is to set a threshold and assume everything above that threshold is a target. In practice adjoining pixels above the threshold are grouped together as a single target. This approach is usually too simplistic and requires an extension via one of the three subclasses above.
Region growing, which includes floodfilling techniques, compares adjacent pixels at the edge of a target and groups them together based on some tolerance in pixel value \citep[e.g. Duchamp and BLOBCAT][] {Whiting12, Hales12}). This technique alone is poor at dealing with disjointed areas, although it may be heuristically augmented to
define nearby areas as one object. Another method, edge detection, uses thresholds to find the borders between different ``types" of pixel. This is rarely useful in astronomy because sources have ``soft' edges . Waterfalling
involves considering the image as a topographic surface and calculating the intensity gradient, pixels with the lowest gradient denoting watershed lines which are used to bound regions.
An advantage of segmentation methods is their speed, making them good for large area surveys. One of the downfalls of segmentation methods is that no semantic information is returned. An additional classifier step is required to be applied to the output of segmentation algorithms to determine semantic properties.

While algorithms for the detection of point sources are well-established, automatic detection and characterisation of diffuse sources presents a significant challenge to large radio surveys. While methods such as local kernel transforms, template matching and Hough transforms have been demonstrated to detect and characterise diffuse sources, they are too computationally expensive to apply to large images. The favoured approach at present is therefore a combination of segmentation algorithms with heuristics to extract diffuse source catalogues. This combined process is not likely to be sufficient for future survey science goals and some combination of techniques, possibly applied on different resolution scales, will be required.

\subsection{Classification and Cross-Identification}
\label{crossid}
The large area surveys proposed by the SKA pathfinder projects  will return many tens to hundreds of millions of sources, requiring automated techniques both to group components into sources and to identify  the counterparts at other wavelengths.

Table \ref{surveys} lists the current, or planned, large area multiwavelength surveys that have significant  overlap with next generation radio surveys. Based on analysis of current deep multiwavelength fields such as
COSMOS and GOODS-N  \citep{Scoville07,Schinnerer07,Morrison2010,Giavalisco04}, we estimate that, assuming a  limit of
 rms $\sim$ 10\,\ujybm\ at 1.4\,GHz, $\sim50$\% will have counterparts in areas co-incident with deep optical surveys (i.e.\ DES, Pan-starrs), while $\sim70$\% will have near-IR counterparts in the overlap regions with surveys such as VISTA VHS and VIKING. However, because most of the deeper surveys only cover small areas, about 50\% of radio sources will have no counterparts at other wavelengths, with this fraction falling to at most 30\% in the
$\sim 1000-2000$\,\sqdeg\ overlap with the deepest large area optical-near IR surveys.

\begin{figure}
\centering
\includegraphics[width=7.0cm]{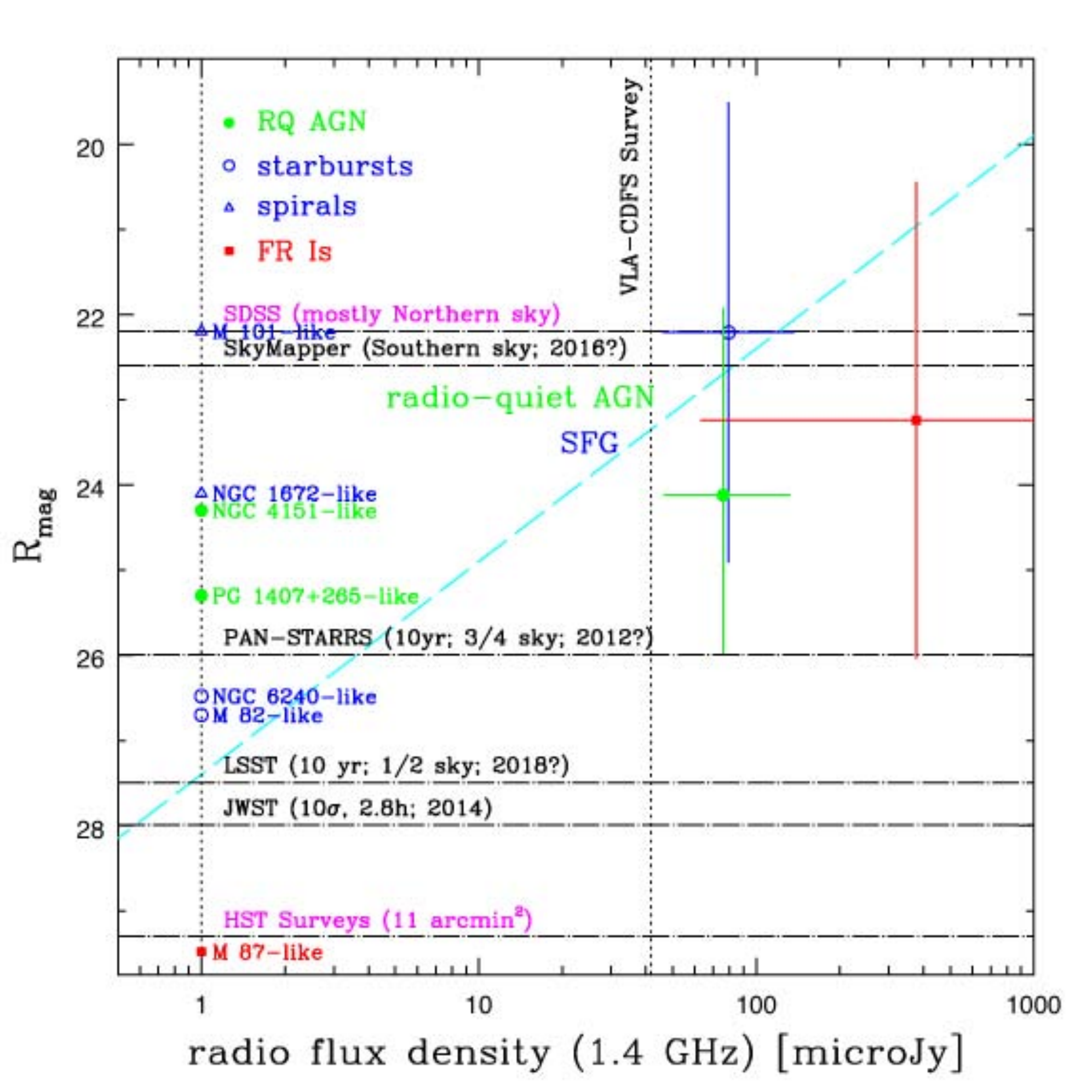}
\caption{$R_{\rm mag}$ versus the 1.4\,GHz radio flux density for faint radio
  sources. The diagonal dashed line indicates the maximum value for SFG
  and the approximate dividing line between radio-loud and radio-quiet AGN,
  with SFG and radio-quiet AGN expected to populate the top left part of
  the diagram. The typical R magnitudes of
  the three classes at $S_{1.4} = 1\,\mu$Jy are also shown, with SFG
  split into starbursts and spirals. Finally, the mean radio and $R_{\rm
    mag}$ values for sources from the VLA-CDFS sample, with error
  bars indicating the standard deviation, are also marked. The horizontal
  dot-dashed lines indicate the approximate point-source limits of planned and existing surveys.}
\label{Rfr}
\end{figure}

 \cite{pad11}
has estimated  cross-identification rates for the SKA down to nJy
radio flux densities.
Figure~\ref{Rfr} plots $R_{\rm mag}$ versus 1.4\,GHz radio flux density
and shows the expected $R_{\rm mag}$ for various classes of sources
that will be found in the radio surveys.

Figure~\ref{Rfr} shows that, down to $50\,\mu$Jy, $\approx 50$\% of SFGs should
be detected by the SDSS (northern hemisphere) and SkyMapper (southern
hemisphere), while most AGN will be fainter. PAN-STARSS and the LSST should
detect most of the radio sources down to this flux density but on longer
timescales (e.g.\ by $\sim $2028 in the case of the LSST, assuming a start of
operations in 2018). Surveys reaching $1\,\mu$Jy (e.g.\ those from the VLA
and MIGHTEE) will obviously have much fainter counterparts and might
require JWST and the Extremely Large Telescopes (ELTs) for the optical
identification of most of their sources, which however will be covering a
relatively small field of view (up to a few arcmin$^2$).
X-ray surveys are intrinsically much less sensitive than optical/IR surveys,
and will detect only a very small fraction of the AGNs detected by radio surveys  \citep{pad11}.

The requirements for spectroscopic redshifts are obviously even more difficult. For example, long exposures
($\sim 10$h) with 8/10\,m telescopes
can secure redshifts {\it in the case of strong emission lines} down to
$R_{\rm mag} \approx 26$.  Large-area spectroscopic surveys, however, cannot reach these magnitudes. For example, current magnitude limits are $r \le 17.8$ (galaxies)
and $i \le 19.1$ (quasars) for the SDSS and $b_J \le 20.85$ for the 2dF QSO
redshift survey.

The source density of potential matches will probably be $>1$ per radio beam, so simply taking the nearest neighbour to the radio source in other catalogues is not a feasible approach. Furthermore, radio sources will typically have more than one component per source, so a simple nearest-neighbour algorithm will attempt to match sources to the lobes of an FRII galaxy as well as to the core. Clearly, more sophisticated matching algorithms must be considered

A significant number of techniques exist to facilitate the matching of catalogues. Techniques such as the $p$-statistic \citep{Downes86} and the likelihood ratio \citep{Sutherland92} make use of the number density of background (and true counterpart) sources to ``weigh'' potential counterparts within a given search radius, but cannot take into account the likelihood that several radio components may correspond to one optical component. More recently \citet{Budavari08}  outlined a Bayesian method to determine the most likely match. This method, while similar to the likelihood ratio, is distinct in that it does not consider the source density, and hence is less biased towards faint counterparts (although potentially less reliable for very high source densities). A Bayesian approach \citep{Budavari11} also allows a number of other constraints to be built into the process for selecting the most likely candidate. This can in principle even extend to including the knowledge that the core of an FRII  is associated with a host galaxy while the lobes are not. This balance between reliability of matches, and completeness as a function of flux, will be key in deciding which algorithm is appropriate for the scientific exploitation of radio surveys.

Finally, it is likely that any automated algorithm will fail on complex radio sources, which may include those which are most scientifically interesting. For these cases,  a ``Citizen Science'' approach is being developed to enlist the help of thousands of enthusiastic amateurs, in collaboration with the Galaxy Zoo group  \citep{Lintott08} to examine each of the millions of complex sources by eye. The ``Radio Zoo'' project expects to release its first prototype in mid-2012.

\begin{table*}[h]
\begin{center}
\caption{Key multiwavelength surveys with which EMU/WODAN data will be cross-identified (restricted to surveys larger than 1000\,\sqdeg) adapted from \cite{Norris11b}. All magnitudes are in AB. The ``detectable'' column is the fraction of 1.4\,GHz EMU/WODAN sources that are in principle detectable by the multiwavelength survey to its $5\,\sigma$ limit. The ``matched'' column is the fraction of 1.4\,GHz sources  which are both detectable and in the area of sky covered by the multiwavelength survey.  The sensitivity shown for the WISE survey is for the $3.5\,\mu$m band.
}
\vspace{0.5cm}
\begin{tabular}{lllllll}
\hline
Survey  & Area & Wavelength & Limiting& Detectable & Matched & Data\\
Name & (\sqdeg) & Bands & Mag. &  &   & Release\\
&&& flux$^{\rm a}$&(\%)&(\%)&Date\\
\hline
WISE$^{\rm 1}$ & 40000 & 3.4, 4.6, 12, 22 $\mu$m & $80\,\mu$Jy & 23 & 23 & 2012\\
Pan-Starrs$^{\rm 2}$ & 30000 & $g$, $r$, $i$, $z$, $y$ & $r<24.0$ & 54 & 41 & 2020\\
Wallaby$^{\rm 3,b}$ & 30000 & $20\,$cm (HI) & $1.6\,$mJy$^{\rm c}$ & 1 & 1 & 2013\\
LSST$^{\rm 4}$ & 20000 &$u$, $g$, $r$, $i$, $z$, $y$ & $r<27.5$ & 96 & 48 & 2020\\
Skymapper$^{\rm 5}$ & 20000 & $u$, $v$, $g$, $r$, $i$, $z$ & $r<22.6$& 31 & 16 & 2015\\
VHS$^{\rm 6}$ & 20000 & Y, J, H, K & K$<20.5$ & 49 & 25 &2012\\
SDSS$^{\rm 7}$ & 12000 & $u$, $g$, $r$, $i$, $z$ & $r<22.2$ & 28 & 8 & DR8 2011 \\
DES$^{\rm 8}$ & 5000 & $g$, $r$, $i$, $z$, $y$ & $r<25$ & 71 & 9 & 2017\\
VST-ATLAS$^{\rm 9}$ & 4500 & $u$, $g$, $r$, $i$, $z$ & $r<22.3$& 30 & 4 & 2012?\\
Viking$^{\rm 10}$ & 1500 & Y, J, H, K & K$<21.5$ & 68 & 3 & 2012\\
Pan-Starrs Deep$^{\rm 2}$ & 1200 & $0.5-0.8$, $g$, $r$, $i$, $z$, $y$ & $g<27.0$ & 57 & 2 & 2020\\
\hline
\end{tabular}
\medskip\\
\label{surveys}
\end{center}
$^{\rm a}$Denotes $5\,\sigma$ point source detection.  However, in many cases, \emph{a priori} positional information will enable  $3\,\sigma$ data to be used, resulting in a higher detection rate.\\
$^{\rm b}$Being an HI survey, WALLABY will measure redshifts for all detected galaxies out to $z=0.26$.\\
$^{\rm c}$per $4\,$km\,s$^{-1}$ channel achieved in $8\,$hours integration.\\
$^{\rm 1}$\citet{Wright2010}\\
$^{\rm 2}$\citet{Kaiser2010}\\
$^{\rm 3}$\citet{Koribalski11}\\
$^{\rm 4}$\citet{Ivezic2008}\\
$^{\rm 5}$\citet{Keller07}\\
$^{\rm 6}$http://www.ast.cam.ac.uk/research/instrumentation.surveys.and.projects/vista\\
$^{\rm 7}$\citet{Abazajian09}\\
$^{\rm 8}$\citet{DES2005}\\
$^{\rm 9}$\citet{Shanks05}\\
$^{\rm 10}$http://www.eso.org/sci/observing/policies/PublicSurveys/sciencePublicSurveys.html\\
\end{table*}

\subsubsection{The Likelihood Ratio technique for radio\\ source cross-identification}

The Likelihood Ratio (LR) technique \citep[e.g.\ ][]{Richter75,Sutherland92,Ciliegi03}
can be used to identify optical/near--infrared
counterparts with sources selected from low resolution survey data. \citet{Smith10}
discuss a recent application of this Bayesian technique to identify 2423 optical counterparts to
6621 sources from the {\it Herschel}-Astrophysical TeraHertz Large
Area Survey \citep[{\it H}-ATLAS; ][]{Eales10}, selected at 250\,$\mu$m.

The LR technique relies on the brightness and astrometric properties
of the input catalogues in general, and each individual source in
particular, to determine the ratio of the probability that two
objects are related, to the probability that they are unrelated. After
using an empirical method to account for the fraction of sources which
are not detected in the optical/near-infrared catalogue, we can
determine the reliability of each association, with values ranging
from 0 (not related) to 1 (a match).

The likely positional uncertainties, coupled
with the probable optical/near--infrared catalogues that will be
available for identifying counterparts to the new catalogues that
these surveys will generate, suggest that, apart from the problem of
radio sources with multiple  components,  the problem will be very
similar to that addressed using the LR technique in the
{\it Herschel}-ATLAS.

\citet{Mcalpine12}  have begun detailed investigations based on existing deep
radio survey data and new near-infrared data from the VISTA VIDEO
survey to quantify how effective the LR technique will be for these
purposes, and how it will be affected by using data of different
sensitivity and spatial resolution.

\subsubsection{Bayesian approaches}

\citet{Budavari08} propose Bayesian hypothesis testing for cross-identification. Given a set of detections one can ask directly whether the measured directions are consistent with a common single object. This hypothesis is tested against its complement using the Bayes factor, which is the likelihood ratio
\begin{equation}
B = \frac{L_{\rm{}same}}{L_{\rm{}not}}
\end{equation}
where the likelihoods are calculated as a sum over all possible model parameters, e.g., the true position of the source. Assuming the astrometric uncertainty is well approximated with a normal distribution, the derivation is analytic. For example, in the case of two detections of point sources with high-precision circular uncertainties and an all-sky prior on the true position
\begin{equation}
B = \frac{2}{\sigma_1^2 + \sigma_2^2}\,    \exp \left\{
    -\frac{\psi^2}{2(\sigma_1^2 + \sigma_2^2)} \right\}
\end{equation}
where $\psi$ is the angular separation of the detections and $\sigma_i$ are the uncertainties; all in radians. When this dimensionless quantity is larger than 1, the data support the association. The higher the value, the greater the evidence.
The probability is analytically given in terms of $B$ and the prior of the hypothesis, which is a function of the source densities of the input data sets.
The approach is applicable to different kinds of data, such as fluxes or polarisation. The corresponding Bayes factors are computed independently and simply multiplied together.
The method's firm statistical foundation along with its explicit dependence on the geometry and physical modelling provides a clean framework for catalog associations at all wavelengths and straightforward extensions. Such examples include the association of stars with unknown proper motions \citep{Kerekes10} or transient events \citep{Budavari12}.

\subsection{Measuring Redshifts}
\label{redshifts}

To obtain the full value of a radio survey, it is essential to have multiwavelength identification, and redshifts. However, only
about 1\% of radio sources from these surveys will have spectroscopic redshifts
(\S\,\ref{crossid}), and so it is necessary to rely on photometric redshifts, or even statistical redshifts.

The most popular way to compute photometric redshifts is via fitting of templates of galaxies to the measured multi-band photometry of target galaxies.
Clearly, the accuracy of the result depends not only on the library of templates, but also on the number of photometric points available and their associated errors. Other techniques, such as k-nearest neighbour (kNN) implementations, Support Vector Regression (SVR) \\
models, Self-Organised maps (SOMs),  Gaussian processing (GP), neural-networks (ANNz) and rainbow forest  (RF) do not make any assumptions  about the spectral energy distribution (SED) of the galaxies, or the type and amount of extinction. Instead, they build on a  simple colour-redshift relation using a training sample of sources with reliable spectroscopic redshifts. These methods are very efficient and fast whenever the training sample is representative of the entire population, and become less reliable outside the redshift range for which they were trained.

About 20\% of EMU sources will have photometric data from the 6 optical bands of the  Skymapper survey,  the 2
near-infrared bands of the Vista  Hemisphere Survey (VHS)    and  the 4 mid-infrared bands  of WISE  (see Table~\ref{surveys}). Typically, with these available datasets, an error in redshift of $\sim 0.03$ can be reached for normal galaxies up to redshift $z\sim 1.5$ using either standard  (SED template  fitting) or empirical (machine learning)  techniques.

About 20\% of identified sources will be classified as AGN, based on one or more of the following criteria: (a)~radio morphology, (b)~radio polarisation, (c)~X-ray detection, (d)~a radio spectral index very different from $\alpha \sim -0.7$,
(e)~a radio/IR ratio very different from the canonical star-forming value.  About 5\% of the sources will be QSOs,  detectable by the next X-ray all sky  survey, eROSITA \citep{Cappelluti11, Merloni12},
and for these sources a special treatment must be adopted \citep{Salvato09,Salvato11}.

About 80\% of the EMU sources will have an optical counterpart that is fainter than $r=22$\,mag$_{AB}$ and  for these sources the ancillary multiwavelength  data set  will be coarse and non homogenous.  Finally, the training sample with spectroscopic redshift available  at high redshift may not be as complete as at $z<1$, hampering the empirical photometric redshift techniques.

With the final goal of providing the most reliable photometric redshift  for the larger number of sources possible, the EMU redshift working group is testing {\it all\/} the methods  mentioned above, trying to understand the advantages and disadvantages of each of them,  and thus applying  the most suitable, or a combination, of the methods, depending on
(a)~the available depth of the multiwavelength data  and  (b)~on the redshift range available for the training sample.  
We  started  experimenting  on COSMOS field \cite{Scoville07}  representative of a pencil beam survey ( 2\,\sqdeg) with a very deep and homogenous multiwavelength data set, but for which the training spectroscopic sample is limited to low redshift radio sources, and with a strong component of X-ray detected sources. Other fields which we will be testing are   the larger area ATLAS \citep{Norris06, Middelberg08a} which has deep but non homogenous ancillary data,  and GAMA ( 310\,\sqdeg) for which the training spectroscopic sample is rich  but representing only the nearby universe ($r<20$\,mag) and with limited availability of ancillary data.

In addition to the optical, near, and mid-infrared data, we will make use of the radio properties of the sources.
While for the empirical methods the radio flux will be used as an additional band, for the SED fitting technique, we will use the radio properties as a {\it prior\/} for limiting the redshift range or the templates used for  the possible solutions.

The ``statistical redshift'' concept, introduced by \cite{Norris11c}, recognises that a probabilistic estimate of redshift for a large sample can be useful, even if many of them are incorrect, provided that the level of incorrectness and incompleteness can be accurately calibrated.
Even a non-detection can carry useful information, and radio data themselves can add significantly to the choice of SED template. For example,
\begin{itemize}
\item high redshift radio galaxies can be identified from their strong radio emission coupled with a NIR non-detection \citep{Willott03, Norris11a}.
\item a steep radio spectral index increases the probability of a high redshift \citep{Breuck02},
\item the angular size of a particular galaxy class can be loosely correlated with redshift \citep{Wardle74},
\item polarised radio sources are nearly always AGN \citep{Hales12}, while unpolarised sources are mainly star-forming galaxies. Therefore, for the EMU/WODAN surveys,  polarised sources have a mean $z \sim 1.8$, while unpolarised sources have a mean $z \sim 1.1$.
\end{itemize}

\subsection{Data Issues}

\subsubsection{Stacking and Data-Intensive Research}

Stacking is used to explore the properties of a class of objects which are below the detection limit of a  survey. For example, the radio flux of RQQs can be measured by averaging the measured radio flux at the position of each RQQ in  a radio survey, even though each individual RQQ has a flux well below the detection limit.

More generally, the process of stacking involves combining (typically by taking a censored mean or median) the data at the position where such objects are expected in the survey. The noise tends to cancel, while any low level of flux density in the sources adds, resulting in a detection threshold very much lower than that of the unstacked survey. Stacking at radio wavelengths has been used very successfully  \citep[e.g.][]{Boyle07, White07, Ivison07, Ivison10, Dunne09, Messias10, Bourne11} on high-resolution data for the purpose of studying faint populations which are below the detection threshold of the radio image, and has proven to be a powerful tool for studying star formation rates, AGN activity, radio-infrared correlation, and measuring the fraction of the extragalactic background contributed by various source populations. Jack-knifing techniques are currently being explored to provide more information about the flux density distribution of stacked sources  than is available from a simple mean or median (Rees et al., in preparation).

The unprecedented area-depth product of SPARCS surveys makes them very suitable for stacking. For example, stacking a sample of a million optically selected galaxies in the EMU data will result in a noise level of $\sim10$\,nJy. Because of the wide area of the EMU, WODAN and LOFAR surveys, even rare classes of source can be stacked, and it should also be possible to create stacked images of extended sources, such as clusters. However, the extent to which such deep stacking will be successful  will depend on the extent to which imaging artefacts are cancelled by stacking.
Stacking is also not without its hazards, and can be biased by a number of effects such as failure to account for the point-spread function, particularly in the presence of confusion and in highly clustered fields  \citep{Bourne11, Greve10, Penner11, Chary10, Beelen10}.

Other examples of data-intensive research include:
\begin{itemize}
\item Identification of sources which do not fit into known categories of radio source, and so are likely to be artefacts or exciting new classes of source, as discussed in \S\,\ref{WTF}.
\item Extraction of low surface brightness radio emission to detect the WHIM synchrotron emission from cosmic filaments or sheets, by cross-correlating with optical galaxies selected from other surveys.
\item Cross-correlation of low-redshift galaxies with high-redshift galaxies, or the CMB, to test cosmology and fundamental physics as discussed in \S\,\ref{cosmology}.
\end{itemize}

\subsubsection{Survey versus Pointed data}

Since surveys compete for observing time with conventional ``pointed'' observations, it is useful to consider their role compared to projects doing detailed science on
    particular sources or fields
Clearly both modes are valuable, and complementary. Surveys provide source lists for follow-up observations of particular sources, as well as
 finding and characterising calibrators for other radio
observations.
However, the huge volume of data  from the new radio surveys will trigger a shift in the way astronomers ``observe''. The ingenuity currently applied to devising a key experiment, and crafting a successful proposal, will be applied to devising novel ways of mining the data. To some extent, this is already happening, and about three times as many citations are delivered by results from the HST/STScI archive than from the papers written by PIs.  
The surveys discussed here will also have an enormous legacy value, since they will be hard to improve on until the technology makes a further major step forward.
    
\section{Conclusion}

The SKA Pathfinder Radio Continuum Survey working group (SPARCS) has been established to coordinate the science and technical developments associated with continuum surveys being planned or undertaken with the SKA pathfinder telescopes (APERTIF,  ASKAP, VLA, e-MERLIN, e-EVN, LOFAR, Meerkat).
This review paper was triggered by the first SPARCS workshop, and is an attempt not only to share the knowledge residing in the individual pathfinder projects, but also to establish a baseline of knowledge from which all projects can move on, and which can be used to optimise the development of the SKA.  A primary outcome of the first SPARCS workshop was to rethink aspects of the way we are constructing the SKA pathfinder telescopes and designing the surveys. For example, it became clear that the impact of these surveys on cosmology and fundamental physics may dwarf the key science goals we have been working on for the last few years. That places demands on source-measurement accuracy and uniformity which in turn impacts on calibration and imaging accuracy.

Key demonstrable outcomes of the workshop include this review paper, and
 the establishment of a series of reference fields in both hemispheres which are now starting to be observed with major existing telescopes (VLA, ATCA, Westerbork) and will be observed with the new pathfinder telescopes (LOFAR, Aperitif, ASKAP, Meerkat) to ensure consistency and uniformity across the surveys.

In this review paper, we have explored the current state of science to be tackled with the SPARCS surveys, and identified the most pressing challenges. We conclude that:
\begin{itemize}

\item The SKA pathfinder telescopes and upgrades, and their planned continuum surveys, have complementary capabilities which will lead to ground breaking science. Particularly exciting are (a) the ability of EMU and WODAN to cover the entire sky with  high sensitivity and resolution, detecting about 100 million galaxies, (b) the ability of LOFAR to make complementary observations of much of that region at low frequency,(c) the ability of POSSUM and BEOWULF to obtain detailed information about the magnetic sky over the same region, and (d) the ability of e-MERLIN, VLA, e-EVN, and MIGHTEE to study selected areas and objects with very high sensitivity and resolution.

\item Together, these surveys are likely to make major inroads on some of the pressing questions about the origin and evolution of galaxies, including distinguishing AGN and SF components of galaxies, tracing the evolution of the luminosity function for both, and exploring the causes and signatures of hot-mode and cold-mode accretion.

\item Amongst the sources detected by the surveys will be very high redshift galaxies, possibly the highest discovered, but identifying them, and measuring their redshift will be challenging.

\item The SPARCS surveys are likely to detect tens to hundreds of thousands of new clusters, and will also map the extended halo and relic emission from a fraction of these. Comparison with X-ray observations is likely to yield new insights into the physics of clusters and the growth of structure in the Universe.

\item The measurement of polarisation and rotation measures is an important part of the surveys, and is likely to yield not only insight into the physics of the radio galaxies, but into the origin of cosmic magnetism. There is a good chance that the intergalactic magnetic field will be detected and traced by these surveys.

\item Even without redshifts, several cosmological probes can use SPARCS survey data to measure cosmological parameters and test models of gravity,  with significantly more precision than current analyses. The combination of WODAN and EMU is likely to yield the best measurement yet of the Integrated Sachs-Wolfe (ISW) effect. The most  powerfully constraining probe of dark energy parameters turns out to be the source count correlations, while ISW provides the most stringent constraints on  modified gravity theories. If photometric redshifts can be obtained for a significant subset of the radio sources, then these cosmological tests become even more powerful.

\item Variable and transient sources are currently poorly studied, and the SPARCS surveys are likely to change this field significantly. In addition to constraining models of (e.g.) black hole accretion, these studies are also likely to result in unexpected discoveries.

\item While most SPARCS surveys are focussed on extragalactic science, they will also produce the best high-resolution data yet on the Galaxy, detecting essentially every SNR and UCHII region in the Galaxy, and vastly increasing the number of known radio stars. It is essential that this scientifically important aspect of the SPARCS surveys is not neglected in our drive to to answer the key questions of extragalactic astrophysics.

\item Because the SPARCS surveys are opening up a large area of unexplored parameter space, it is likely that they will make unexpected discoveries. It is important not to leave this process to chance, but to recognise it as a legitimate goal of large surveys, and to plan processes and software accordingly, to mine the deluge of data that these surveys will produce.

\end{itemize}

To achieve these exciting scientific goals, many technical challenges must be addressed by the survey instruments. Given the limited resources of the global radio-astronomical community, it is essential that we pool our skills and knowledge. We do not have sufficient resources to enjoy the luxury of re-inventing wheels. We face significant challenges in calibration, imaging, source extraction and measurement, classification and cross-identification, redshift determination, stacking, and data-intensive research. As the SPARCS instruments extend the observational parameters we will face further unexpected challenges in calibration, imaging, and interpretation. If we are to realise the full scientific potential of these expensive instruments, it is essential that we devote enough resources and careful study to understanding the instrumental effects and how they will affect the data. The prime role of SPARCS is to facilitate the process of doing so by ensuring we share resources and expertise across the projects.

\section*{Acknowledgments}

We are indebted to the Lorentz Center in Leiden for hosting and funding the workshop in February 2011 which gave rise to this paper.
Parts of this research were supported by the Australian Research Council Centre of Excellence for All-sky Astrophysics (CAASTRO), through project number CE110001020.
Part of the research described in this paper was carried out at the  
Jet Propulsion Laboratory, California Institute of Technology, under a  
contract with the National Aeronautics and Space Administration.
JA gratefully acknowledges support from the Science and Technology Foundation (FCT, Portugal) through the research grants//
 PTDC/FIS/100170/2008, PTDC/CTE-AST/105287/2008 and PEst-OE/FIS/UI2751/2011.
%
%
CF and AD acknowledge financial support by the {\it Agence Nationale de la Recherche through grant ANR-09-JCJC-0001-01}. AD acknowledges financial support from the joint PhD program of {\it Observatoire de la C\^ote d'Azur} and {\it Conseil r\'egional Provence-Alpes-C\^ote d'Azur}.

\section*{Authors Affiliations}
{\small \affil{1}\,CSIRO Astronomy \& Space Science, PO Box 76, Epping, NSW 1710, Australia}\\
{\small \affil{2}\,ARC Centre of Excellence for All-sky Astrophysics (CAASTRO)}\\ 
{\small \affil{3}\,Centro de Astronomia e Astrof\'{\i}sica da Universidade de Lisboa, Observat\'{o}rio Astron\'{o}mico de Lisboa, Tapada da Ajuda, 1349-018 Lisboa, Portugal}\\
{\small \affil{4}\,Institute of Cosmology and Gravitation, University of Portsmouth, Dennis Sciama Building, Burnaby Road, Portsmouth, PO1 3FX, UK}\\
{\small \affil{5}\,Max Planck Institut fur Radioastronomie, Auf dem Hugel 69, Bonn, Germany}\\ 
{\small \affil{6}\,School of Physics \& Astronomy, University of Southampton, Southampton, SO17 1BJ, UK}\\ 
{\small \affil{7}\,Jodrell Bank Centre for Astrophysics, School of Physics and Astronomy, The University of Manchester, Manchester, M13~9PL, UK}\\ 
{\small \affil{8}\,Institute for Astronomy, University of Edinburgh, Royal Observatory, Blackford Hill, Edinburgh, EH9 3HJ, UK}\\ 
{\small \affil{9}\,National Radio Astronomy Observatory, 520 Edgemont Road, Charlottesville, VA 22903, USA}\\ 
{\small \affil{10}\,Max Planck Institute for Plasma Physics, Boltzmannstr. 2 D-85748 Garching, Germany}\\ 
{\small \affil{11}\,INAF-IRA, Via P. Gobetti 101, 40129 Bologna, Italy}\\ 
{\small \affil{12}\,Dept.\ of Physics and Astronomy, The Johns Hopkins University, 3400 North Charles Street, Baltimore, MD 21218; USA}\\
{\small \affil{13}\,Physics Department, University of the Western Cape, Cape Town, 7535, South Africa}\\ 
{\small \affil{14}\,Laboratoire Lagrange, UMR 7293, Universit\'e de Nice Sophia-Antipolis, CNRS, Observatoire de la C\^ote d'Azur, 06300, Nice, France}\\
{\small \affil{15}\,Sydney Institute for Astronomy, School of Physics, The University of Sydney, NSW 2006, Australia}\\
{\small \affil{16}\,ASTRON, Postbus 2, 7990 AA Dwingeloo, The Netherlands}\\
{\small \affil{17}\,Astrophysics, Cosmology \& Gravity Centre, Department of Astronomy, University of Cape Town, Private Bag X3, Rondebosch 7701,, South Africa}\\
{\small \affil{18}\,Australian Astronomical Observatory, PO Box 296, Epping, NSW 1710, Australia}\\
{\small \affil{19}\, Centre for Astrophysics Research, Science \& Technology Research Institute, University of Hertfordshire, Hatfield, Herts., UK}\\
{\small \affil{20}\,Joint Institute for VLBI in Europe, Postbus 2, 7990 AA DWINGELOO, The Netherlands}\\
{\small \affil{21}\,Jet Propulsion Laboratory, California Institute of Technology,
Pasadena, CA 91109, USA}\\
{\small \affil{22}\,National Radio Astronomy Observatory, PO Box 0, Socorro, NM 87801, USA}\\
{\small \affil{23}\,School of Mathematics \& Physics, University of Tasmania, Private Bag 37, Hobart, 7001, Australia}\\ 
{\small \affil{24}\,Rhodes University, South Africa}\\
{\small \affil{25}\,Astronomisches Institut, Ruhr-Universit\"at Bochum, Universit\"atsstr. 150, 44801 Bochum, Germany}\\
{\small \affil{26}\,Carnegie Observatories, 813, santa Barbara St., Pasadena, CA 91101, USA}\\
{\small \affil{27}\,European Southern Observatory, Karl-Schwarzschild-Str. 2, D-85748 Garching bei M\"unchen, Germany}\\
{\small \affil{28}\,School of Physics \& Astronomy, University of Nottingham, Nottingham, NG7 2RD, UK}\\
{\small \affil{30}\,Leiden Observatory, Leiden University, P.O. Box 9513, 2300 RA Leiden,The Netherlands}\\
{\small \affil{31}\,INAF-Catania Astrophysical Observatory, Via S. Sofia 78, 95123 Catania, ITALY}\\
{\small \affil{32}\,National Research Council of Canada, National Science Infrastructure, Dominion Radio Astrophysical Observatory, P.O. Box 248, Penticton, BC V2A 6J9, Canada}\\
{\small \affil{33}\,California Institute of Technology, Pasadena, CA 91125,USA}\\
{\small \affil{34}\,Department of Astrophysics/IMAPP, Radboud University Nijmegen, P.O.Box 9010, 6500 GL Nijmegen, The Netherlands}\\
{\small \affil{35}\,School of Chemical \& Physical Sciences, Victoria University of Wellington, PO Box 600, Wellington 6140, New Zealand}\\
{\small \affil{36}\,Oxford Astrophysics, Denys Wilkinson Building, University of Oxford, Keble Rd, Oxford OX1 3RH}\\
{\small \affil{}\,Email: Ray.Norris@csiro.au}
School of Physics, Astronomy and Mathematics, University of Hertfordshire, College Lane, Hatfield AL10 9AB
Oxford Astrophysics, Denys Wilkinson Building, University of Oxford, Keble Rd, Oxford OX1 3RH 
Physics Department, University of the Western Cape, Cape Town, 7535, South Africa

%

\end{document}